\documentclass[11pt]{article}

\usepackage{jcappub} 
\usepackage[T1]{fontenc} % if needed
\usepackage{amsbsy}
\usepackage{color}
\usepackage{subcaption}

\usepackage{graphicx}
\usepackage[normalem]{ulem}
\usepackage{xurl}

\renewcommand{\H}{\mathcal H}
\newcommand{\I}{\mathcal I}

\newcommand{\D}{\mathcal D}
\newcommand{\Om}{\Omega_\mathrm{m}}

\title{\boldmath Relativistic matter bispectrum of cosmic structures on the light cone} 

\def\aap{\ref@jnl{A\&A}}
\def\jcap{\ref@jnl{J. Cosmology Astropart. Phys.}}

\usepackage{acro}

\DeclareAcronym{PNG}{
  short=PNG,
  long=Primordial Non-Gaussianity,
}

\DeclareAcronym{CDM}{
  short=CDM,
  long=Cold Dark Matter,
}

\DeclareAcronym{RE}{
  short=REs,
  long=Relativistic Effects,
}

\DeclareAcronym{CMB}{
  short=CMB,
  long=Cosmic Microwave Background,
}

\DeclareAcronym{LSS}{
  short=LSS,
  long=Large-Scale Structure,
}

\DeclareAcronym{IC}{
  short=ICs,
  long=Initial Conditions,
}

\DeclareAcronym{LPT}{
  short=LPT,
  long=Lagrangian Perturbation Theory,
}

\DeclareAcronym{CIC}{
  short=CIC,
  long=cloud-in-cell,
}

\author[a]{Thomas Montandon,}
\affiliation[a]{Department of Astrophysics, University of Vienna, Türkenschanzstraße 17, 1180 Vienna, Austria}

\author[b]{Julian Adamek,}
\affiliation[b]{Institute for Computational Science, Universit\"at Z\"urich, Winterthurerstrasse 190, 8057 Z\"urich, Switzerland}

\author[a,c]{Oliver Hahn,}
\affiliation[c]{Department of Mathematics, University of Vienna, Oskar-Morgenstern-Platz 1, 1090 Vienna, Austria}

\author[d]{Jorge Nore\~na,}
\affiliation[d]{Instituto de F\'isica, Pontificia Universidad Cat\'olica de Valpara\'iso, Casilla 4950, Valpara\'iso, Chile}

\author[a,c]{Cornelius Rampf,}

\author[e]{Cl\'ement Stahl,}
\affiliation[e]{Universit\'e de Strasbourg, CNRS, Observatoire astronomique de Strasbourg, UMR 7550, 67000 Strasbourg, France}

\author[f]{and Bartjan van Tent}
\affiliation[f]{Universit\'e Paris-Saclay, CNRS/IN2P3, IJCLab, 91405 Orsay, France}

\emailAdd{thomas.montandon@univie.ac.at}

\abstract{Upcoming surveys of cosmic structures will probe scales close to the cosmological horizon, which opens up new opportunities for testing the cosmological concordance model to high accuracy. In particular, constraints on the squeezed bispectrum could rule out the single-field hypothesis during inflation. However, the squeezed bispectrum is also sensitive to dynamical effects of general relativity as well as interactions of matter with residual radiation from the early Universe. In this paper, we present a relativistic simulation pipeline that includes these relativistic effects consistently. We produce light cones and calculate the observed number counts of cold dark matter for five redshift bins between $z=0.55$-$2.25$. We compare the relativistic results against reference Newtonian simulations by means of angular power- and bispectra. We find that the dynamical relativistic effects scale roughly inversely proportional to the multipole in the angular power spectrum, with a maximum amplitude of $10\%$ for $\ell \lesssim 5$.
By using a smoothing method applied to the binned bispectrum we detect the Newtonian bispectrum with very high significance. The purely relativistic part of the matter bispectrum, obtained by subtracting the Newtonian bispectrum from the relativistic one, is detected with a significance of $\sim 3\,\sigma$, mostly limited by cosmic variance. We find that the pure dynamical relativistic effects accounts for up to $3\%$ and $10\%$ of the total amplitude, respectively in the squeezed and equilateral limits. Our relativistic pipeline for modelling ultra-large scales yields gauge-independent results as we compute observables consistently on the past light cone, while the Newtonian treatment employs approximations that leave some residual gauge dependence. A gauge-invariant approach is required in order to meet the expected level of precision of forthcoming probes of cosmic structures on ultra-large scales.
}

\begin{document}

%\compress
%\toccontinuoustrue
\maketitle
% \flushbottom

\section{Introduction}\label{sec:Introduction}

With the current and upcoming \ac{LSS} surveys, such as DES \cite{DES:2021wwk}, DESI \cite{DESI:2016fyo}, Euclid \cite{Amendola:2016saw}, SPHEREx \cite{Dore:2014cca}, the Vera Rubin observatory \cite{Zhan:2017uwu} or SKA \cite{SKA:2018ckk}, the cosmological concordance model will be tested at the percent level for scales $L \gtrsim 1$ Mpc. This forthcoming wealth of data will allow us to probe the fundamental nature of dark energy and \ac{CDM}, the theory of general relativity, as well as various physical descriptions for the primordial Universe which, ultimately, provide initial data for cosmic structure formation. The most popular model for the early Universe is inflation. Amongst the vast number of different inflationary scenarios \cite{Martin:2013tda}, the most simple, compatible with all the current observations, is single-field slow-roll inflation. First introduced to explain the homogeneity, the isotropy and the flatness of the Universe, it was found later that it also predicts tiny primordial perturbations with very specific properties: they ought to be almost exactly Gaussian, adiabatic and nearly scale invariant. In particular, it was shown that an inflationary phase driven by a single scalar field can only produce a small amount of \ac{PNG}, hence rendering the single-field hypothesis testable. A clean probe of non-Gaussianity is the bispectrum which is the Fourier counterpart of the 3-point correlation function and vanishes for a Gaussian field.
If inflation was driven by only one single degree of freedom, Maldacena's consistency relation says that the amplitude of the bispectrum in the squeezed limit, i.e.\ the coupling between large and small scales parametrised by the amplitude of the local shape $f_{\rm NL}$, follows  $f_{\rm NL}= {}^5\!/_{\!12} (1-n_\mathrm{s})$ \cite{Maldacena:2002vr,Creminelli:2004yq}, where $n_\mathrm{s}$ is the spectral index of initial scalar perturbations. However, these computations were not performed in a fully gauge-invariant way, meaning that Maldacena's famous result is in fact gauge dependent. Later on, it was shown that Maldacena's value for $f_{\rm NL}$ is effectively zero, since it can be absorbed by a suitable coordinate transformation \cite{Tanaka:2011aj,Pajer:2013ana,Cabass:2016cgp}. Single-field inflation predicts however a small amount of \ac{PNG} \cite{Cabass:2016cgp}, while more than one active degrees of freedom produce a larger signal and can be characterised through the idea of the cosmological collider \cite{Arkani-Hamed:2015bza}. 

One of the most intriguing results of the Planck mission is the exclusion of the pure scale-invariant primordial power spectrum, with latest constraints giving $n_\mathrm{s} =  0.9665 \pm 0.0038$ \cite{Planck:2018vyg}. All popular shapes of bispectra that have been constrained by Planck remain compatible with a null hypothesis \cite{Planck:2019kim}. For example, the local $f_{\rm NL}$ constraint is given by $f_{\rm NL} = -0.9 \pm 5.1$.
However, forecasts for the upcoming \ac{LSS} surveys predict an improvement of the error bars to $\lesssim 1$ \cite{Desjacques:2016bnm,Dore:2014cca, Karagiannis:2019jjx}. This increase in sensitivity is particularly exciting as multi-field inflation predicts, without fine tuning of the parameters, $f_{\rm NL} \sim 1$ \cite{Achucarro:2022qrl}. However, as precision increases, we also need to account for several effects that could be safely neglected for previous studies. 
One prominent example of such effects, which is the subject of this paper, are the dynamical \ac{RE} in which we also include the early radiation effects (see below for definitions). 
\ac{RE} and \ac{PNG} contribute both to the bispectrum, and we attempt to disentangle them. \ac{RE} also allow to probe general relativity at cosmological scales~\cite{Bonvin:2018ckp}.

Perturbation theory is the analysis tool ubiquitously used to study Einstein's field equations at early time. It has been intensively studied in the Newtonian limit \cite{Bernardeau:2001qr}, and indeed, the volume probed by the previous \ac{LSS} surveys was small enough for this Newtonian limit to be applicable. However, the largest scales of the next generation of \ac{LSS} surveys will approach the cosmological horizon. Close to the horizon, the relativistic nature of our Universe starts to matter and the relativistic dynamics deviate from the classical Newtonian dynamics; these effects are called dynamical \ac{RE}. Many studies have been performed to account for these effects at second order \cite{Matsubara:1995kq,Matarrese:1997ay,Bruni:2013qta,Boubekeur:2008kn,Bartolo:2010rw,Pajer:2013ana,Villa:2014foa,Villa:2015ppa}, and even up to fourth order \cite{Castiblanco:2018qsd,Calles:2019prs} in perturbation theory. In particular, it was found that, unlike the Newtonian nonlinearities, some dynamical \ac{RE} contribute to the squeezed limit of the bispectrum and have the same time dependence as \ac{PNG}, hence polluting the local \ac{PNG}. Moreover, the squeezed limit involves both large scales, where \ac{RE} are important, and small scales where non-linearities (and possibly large \ac{PNG}  \cite{Stahl:2022did}) are important. These small scales have entered the horizon in the radiation-dominated era. The impact of this early radiation period on the squeezed bispectrum has been studied in Refs.~\cite{Fitzpatrick:2009ci, Tram:2016cpy, Pettinari:2014vja}, finding a contribution with the same time dependence as \ac{PNG}.

One further layer of complexity is that \ac{RE} can be gauge-dependent, which particularly obscures the direct comparison with Newtonian simulation outputs. Thus, any comparison should be performed at the level of gauge-invariant observables. To do so, we need to account for another type of \ac{RE}, namely the kinematical \ac{RE}. Indeed, in practice, the galaxies are observed in terms of angular positions in the sky and redshifts. The observed redshift of a galaxy is a combination of the cosmological redshift, the proper motion of galaxies causing a Doppler effect, a gravitational redshift, and an integrated Sachs-Wolfe/Rees-Sciama effect. In addition, the angular position and the observed shape of a galaxy are affected by lensing. 

 Since the pioneering work of Kaiser \cite{Kaiser:1987qv}, which included the proper motion of galaxies at the linear level in the 2-point statistics, many others have extended the work to all kinematical \ac{RE} and up to second order, see e.g.\  \cite{Yoo:2010ni, Bonvin:2011bg,Yoo:2014sfa}.
 The amplitude of this systematic effect close to the horizon, and even its actual existence, is still debated \cite{Castorina:2021xzs,Grimm:2020ays}. The theoretical prediction of the galaxy bispectrum including local \ac{PNG} and \ac{RE} is still under development, see e.g.~\cite{Bonvin:2011bg, DiDio:2015bua,DiDio:2018unb, Maartens:2020jzf,Yoo:2022klz}. However, we should note that in Ref.~\cite{Yoo:2022klz}, it was found that the kinematical \ac{RE} vanish in the squeezed limit of the bispectrum.

A number of analytical works by several groups have demonstrated the relevance of \ac{RE} for future LSS surveys. Moreover, perturbation theory breaks down for many nonlinear effects probed by LSS, and one needs to resort to other techniques such as N-body simulations. Since we are specifically interested in \ac{RE}, we employ the relativistic N-body simulation code \texttt{gevolution} \cite{Adamek:2016zes}. The public version allows to generate \ac{IC} using the linear propagation of the perturbations. But for accurate second-order dynamics, it should be initialised at second order. Recently, introducing \texttt{RELIC} \cite{Adamek:2021rot}, we solved this problem connecting the numerical Einstein-Boltzmann solver \texttt{SONG} \cite{Pettinari:2014vja} to \texttt{gevolution}. However, the use of \texttt{SONG} remains computationally expensive.

\begin{figure}
    \centering
    \includegraphics[scale=0.225]{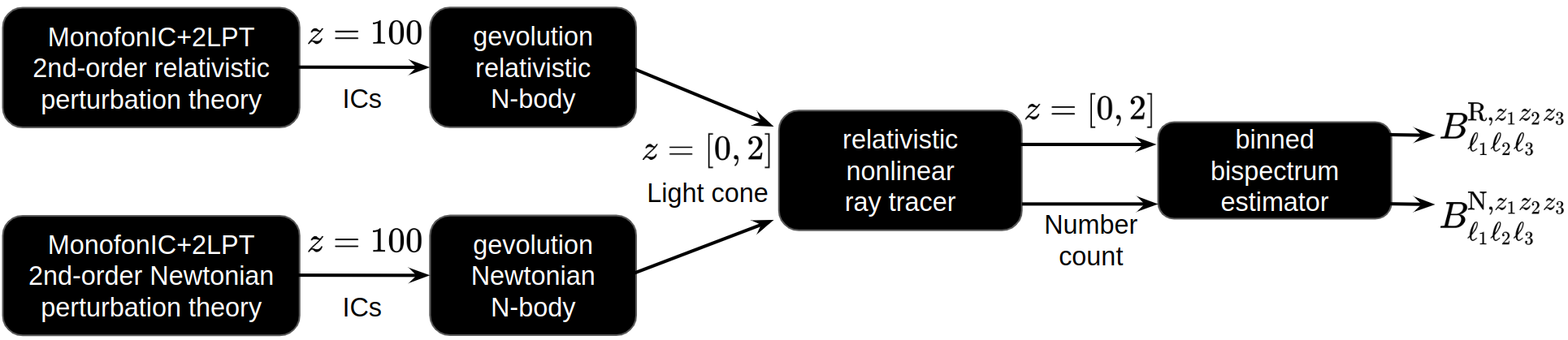}
    \caption{Simulation pipelines used to produce the relativistic ($B_{\ell_1\ell_2\ell_3}^{\mathrm{R}, z_1z_2z_3}$) and the Newtonian ($B_{\ell_1\ell_2\ell_3}^{\mathrm{N}, z_1z_2z_3}$) observable bispectra. We first use a modified version of \texttt{MonofonIC} and a discrete relativistic Lagrangian perturbation theory to produce the \ac{IC} up to second-order. Then, we use the N-body code \texttt{gevolution} to produce a light cone. The non-perturbative ray tracer implemented for \texttt{gevolution} is used to compute the \ac{CDM} number counts. Finally, the binned bispectrum estimator is used to compute the angular bispectrum of the maps of the number counts for various redshifts.} 
    \label{fig:my_label}
\end{figure}

In this paper, we perform a relativistic study on the light cone, which in particular includes non-Gaussian effects induced through nonlinearities at second order in relativistic perturbation theory. We also include the early radiation effects as they play a similar role as the second-order dynamics. Our method is based on an analytic result found in Refs.~\cite{Tram:2016cpy, Adamek:2021rot} that includes all second-order \ac{RE} (excluding massive neutrinos) and reproduces the numerical solution of \texttt{SONG} with a $\sim 1\%$ error. For convenience, we have implemented these results in the --- otherwise Newtonian --- \ac{IC} generator \texttt{MonofonIC} (the single-resolution surrogate for the upcoming~\texttt{MUSIC2}) \cite{Michaux:2020yis}. Our pipeline is then as follows (see also Fig.~\ref{fig:my_label}): first, we initialize \texttt{gevolution} using second-order perturbation theory. Then, we run the N-body simulation to $z=0$ and construct the light cones on which we apply a nonlinear numerical ray tracer \cite{Adamek:2018rru, Lepori:2020ifz}. Afterwards, we compute the binned bispectrum estimator \cite{Bucher:2009nm,Bucher:2015ura} on the sky maps of the number counts. Moreover, in order to single out the relativistic part, we perform a pair of simulations based on the same simulation pipeline. One of the simulations, the relativistic one, includes all radiation, dynamical, and kinematical \ac{RE}, while the other simulation, the Newtonian counterpart, uses Newtonian dynamics instead. Hence, the two simulations only differ by the dynamical/radiation \ac{RE}, which are both degenerate with local~\ac{PNG}.  

The paper is organised as follows. In Section~\ref{sec:pipeline}, we describe our simulation pipeline: we introduce the relativistic second-order perturbation theory and its implementation in the modified version of \texttt{MonofonIC}. We also review  and reformulate the discrete Lagrangian perturbation theory already introduced in Ref.~\cite{Adamek:2021rot}, discussing in particular the Newtonian limit. In Section~\ref{sec:numbercount}, we construct the observed number counts and introduce the binned bispectrum estimator. In Section~\ref{sec:result}, we first test the second-order dynamics of \texttt{gevolution}, and then present the angular power spectrum and bispectrum measurements of the \ac{CDM} number counts in the relativistic and Newtonian cases. Finally, we conclude in Section~\ref{sec:conclusion}.

\section{Simulation pipeline}\label{sec:pipeline}

In this section, we present the simulation pipeline. We first present in Section~\ref{sub:REIC} the second-order relativistic perturbation theory, followed by its implementation in \texttt{MonofonIC} in Section~\ref{sub:monofonic}. In Section~\ref{sub:gev}, we reformulate the discrete Lagrangian perturbation theory implemented in \texttt{gevolution}, and in Section~\ref{sub:Newton}, we discuss the Newtonian limit.

\subsection{Relativistic initial conditions} \label{sub:REIC}

In Poisson gauge and ignoring tensor perturbations, the line element can be written as
\begin{equation}
    \label{eq:generic}
    ds^2 = a^2 \left[-e^{2\psi}d\tau^2  + e^{-2\phi} \delta_{ij} (dx^i+\beta^id\tau)(dx^j+\beta^jd\tau) \right]\,,
\end{equation}
where $a$ is the cosmic scale factor, $\tau$ is conformal time, $\psi$ and $\phi$ are the two gravitational potentials, $\delta_{ij}$ is Kronecker's delta, and $\beta^i$ is the shift constrained by $\nabla_i \beta^i=0$, where here and in the following we imply summation over repeated indices. To describe perturbations at first order, we use the linear transfer functions $T_1$ in Poisson gauge, 
\begin{equation}
    \label{eq:first-transfer}
    \I_1(z,\boldsymbol{k}) = T^{\I}_{1}(z,k) \zeta(\boldsymbol{k})\,,
\end{equation}
where $\I$ can be any field (density, velocity, \ldots), $z$ is the redshift and $\zeta(\boldsymbol{k})$ is the primordial curvature perturbation in Fourier space. The transfer function is computed numerically by solving the linear Einstein-Boltzmann system. Since we initialise the simulations at the second order, the first-order quantities must be computed with a margin of error smaller than the second-order quantities. In this paper
we use \texttt{CLASS} \cite{Lesgourgues:2011re,Blas:2011rf}.

At second order we follow Refs.~\cite{Villa:2015ppa, Tram:2016cpy} and neglect the radiation in the background so in that case, we assume that the first Friedmann equation is of the form
\begin{equation}
    \label{eq:H}
    \H^2 = \H_0^2 (\Omega_{\mathrm{m}0} a^{-1} + \Omega_{\Lambda0} a^2 )\,,
\end{equation}
where $\mathcal{H} = d\ln a/d\tau$ and the subscript ``0'' indicates when a quantity is 
evaluated at the present time $\tau_0$, where specifically $\Omega_{\rm m0}$ and $\Omega_{\Lambda0}$ are the present-day values of the background density parameters for matter and for the cosmological constant.
%for the purpose of solving the second-order quantities.
The linear growth factor $\D$ is defined such that $\delta_1^\mathrm{N}(\tau,\boldsymbol{k}) = \D(\tau) \delta_1^\mathrm{N}(\tau_0, \boldsymbol{k})$ where $\delta_1^\mathrm{N}$ is the linear density contrast in the Newtonian limit and $\tau_0$ is the conformal time today. The linear growth factor is the growing solution of the standard ODE \cite{Meszaros:1974tb,Weinberg:2002kg} 
\begin{equation}
    \label{eq:D}
    \D'' + \H \D' = \frac{3}{2} \frac{\Omega_{\mathrm{m}0} \H_0^2}{a} \D\,,
\end{equation}
where the prime denotes a derivative with respect to conformal time $\tau$. The fastest-growing mode solution of Eq.~\eqref{eq:D} is $\propto \tau^2$ for an Einstein--de Sitter universe (EdS). The second-order growth factor ${\cal F}$ is the solution of a similar equation, but with an additional source term \cite{Villa:2015ppa},
\begin{equation}
    \label{eq:F}
    {\cal F}'' + \H {\cal F}' = \frac{3}{2} \frac{\Omega_{\mathrm{m}0} \H_0^2}{a} ({\cal F}+ \D^2) \,.
\end{equation}
For an EdS universe, the fastest growing-mode solution is ${\cal F} = -{}^3\!/_{\!7} \mathcal{D}^2$, and we note that an explicit analytic solution valid for $\Lambda$CDM can be found in Ref.\,\cite{Matsubara:1995kq} (see also \cite{Rampf:2022tpg}).
Following the notation of Ref.~\cite{Tram:2016cpy}\footnote{Note that the $\Om$ defined in Ref.~\cite{Tram:2016cpy} corresponds to our $\Omega_{\mathrm{m}0}$, while the variable $u$ defined in Ref.~\cite{Tram:2016cpy} is our time-dependent $\Om$.} we define 
\begin{equation}
    \label{eq:def}
  f = \frac{\D'}{\H\D}, \qquad v = \frac{7 {\cal F}}{3\D^2}, \qquad w = \frac{7  {\cal F}'}{6 \H \D^2} \,,
\end{equation}
which all reduce to unity in the EdS limit. Since radiation is not accounted for in the background Eq.~\eqref{eq:H}, the linear and second-order growth factors are the standard $\Lambda$CDM solutions. We stress that in order to compute our linear fields, we do not use $\D$, but the full transfer functions following Eq.~\eqref{eq:first-transfer} which include radiation. Here, the $\Lambda$CDM linear growth factor is only used to compute the second-order fields, through Eq.~\eqref{eq:def}. However, radiation will nevertheless be accounted for thanks to additional terms derived in Ref.~\cite{Tram:2016cpy}.

The second-order density field can be expressed in Fourier space as the convolution integral
\begin{equation}
    \label{eq:second-order-stuff}
    \delta_2 (\tau,\boldsymbol{k}) = \left( \frac{2 }{3\H^2\Om}\right)^2  \int \frac{d^3k_1d^3k_2}{(2\pi)^3} \delta^{(3)}_{\rm D}(\boldsymbol{k} - \boldsymbol{k}_1 -\boldsymbol{k}_2) k_1^2 k_2^2 \mathcal K(k_1,k_2,k) \phi_1(\boldsymbol{k}_1) \phi_1(\boldsymbol{k}_2)\,,
\end{equation}
where $\phi_1$ is the gravitational potential defined in Eq.~\eqref{eq:generic} evaluated at $\tau$ (we suppress the time dependence of $\phi_1$ for brevity), $\delta_\mathrm{D}^{(3)}$ denotes the three-dimensional Dirac delta function, and $k = |\boldsymbol k|$.
The kernel $\mathcal K$ can be computed numerically for the density by solving the second-order Einstein-Boltzmann system \cite{Pettinari:2013ieg}. In \cite{Adamek:2021rot}, we have used the numerical code \texttt{SONG} to evaluate Eq.~\eqref{eq:second-order-stuff}, neglecting the coupling between small scales, i.e.\ all terms $\phi(\boldsymbol{k}_1) \phi(\boldsymbol{k}_2)$ if both $k_1$ and $k_2$ are smaller than a cut-off scale. This approximation was appropriate for the study of \ac{RE} since the coupling of small scales is dominated by Newtonian effects. This method allowed us to evaluate Eq.~\eqref{eq:second-order-stuff} within $\sim 100$ days for $1024$ modes on a single processor, instead of $\sim 1000$ years for the full integral. In this paper, we instead follow Ref.~\cite{Tram:2016cpy} who found an accurate analytical approximation of the second-order density transfer function by accounting for all the \ac{RE} and the early radiation, see Appendix~\ref{app:2ps}. They argue that radiation effects matter only for small scales. But in the small-scale equilateral configuration the Newtonian nonlinearities dominate, such that the early radiation effects only matter in the squeezed limit. By using a separate-universe approach, they found an analytical expression that is in agreement with the numerical solution computed by \texttt{SONG} at the percent level. Here, we want to exploit the fact that this expression is separable, even though it is not local. 

For convenience, we can write the full second-order density as
\begin{align}\label{eq:5.54}
   \delta^{\rm fl}_2 =   \delta_2^{\rm N} + \delta_2^{\rm R1} + \delta_2^{\rm R2} \,.
\end{align}
The subscript ``fl'' stands for fluid, meaning that the density perturbation is measured in the rest frame of the fluid, see Appendix~\ref{app:2EE} for more details. The first term on the right-hand side stands for the Newtonian contribution. It takes the form \cite{Bernardeau:2001qr}
\begin{equation}
    \label{eq:newtonian}
        \delta^\mathrm{N}_2 = \left(\frac{2}{3 \H^2 \Om} \right)^{2} \left(\frac{1}{2} \left[ 1+\frac{3}{7}v \right] (\Delta\phi_1)^2 +  \phi_1^{,l} \Delta\phi_{1,l} + \frac{1}{2} \left[ 1-\frac{3}{7}v \right] \phi_1^{,lm}\phi_{1,lm} \right)\,. 
\end{equation}
The second term on the right-hand side of Eq.~\eqref{eq:5.54} contains all relativistic contributions proportional to~$\H^2$. It can be written as  
\begin{multline}
    \label{eq:R1}
        \left(\frac{2}{3 \H^2 \Om} \right)^{-2} \frac{\delta^{\rm R1}_2}{\H^{2}} =  -\left[ \frac{3}{4} \Om+ 2 f \right] (\nabla \phi_1)^2
        + \frac{9}{7} w \Delta^{-1} \left[ \phi_1^{,lm}\phi_{1,lm} -(\Delta\phi_1)^2\right]  \\
         + \left[ 3 \Om - f + f^2\right] \phi_1\Delta\phi_1 \\
        -\frac{1}{2} \left(f+\frac{3}{2}\Om\right) \mathcal F^{-1}_{\boldsymbol x}\left[ \frac{1}{k^2} \frac{\partial \log T^{\phi}_{1}}{\partial \log k}  \mathcal F_{\boldsymbol{k}} \left[ 2 \phi_1 \Delta^2  \phi_1 -  2(\Delta\phi_1)^2 \right] \right]\,,
\end{multline}
where $\mathcal F_{\boldsymbol{k}}$ denotes a forward Fourier transform and $\mathcal F^{-1}_{\boldsymbol x}$ its inverse. The last term on the right-hand side of Eq.~\eqref{eq:5.54} contains all relativistic contributions proportional to $\H^4$ and reads
\begin{multline}
    \label{eq:R2}
    %\begin{split}
        \left(\frac{2}{3 \H^2 \Om} \right)^{-2} \frac{\delta^{\rm R2}_2}{\H^{4}} 
        =  \left[ \frac{9}{4}\Omega_m f^2-9 \Om f + \frac{9}{2}f+\frac{27}{4} \Om \right] \phi_1^2 \\ 
         - 27 \Om f \Delta^{-1}\left[\frac{1}{2} \Delta^{-1}\left[\phi_1^{,lm}\phi_{1,lm} - (\Delta\phi_1)^2\right] - \frac{1}{3}\phi_1^{,l}\phi_{1,l} \right] \\
        -\frac{1}{2} \left(f+\frac{3}{2}\Om\right) \mathcal F^{-1}_{\boldsymbol x}\left[\frac{3 f}{k^4} \frac{\partial \log T^{\phi}_{1}}{\partial \log k}  \mathcal F_{\boldsymbol{k}} \left[ 2 \phi_1 \Delta^2  \phi_1 -  2(\Delta\phi_1)^2 \right] \right]\,.
    %\end{split}
\end{multline}
The first two lines of Eq.~\eqref{eq:R1} and Eq.~\eqref{eq:R2} correspond to the second-order relativistic corrections computed in a $\Lambda$CDM universe by neglecting the early radiation. They were first computed in Ref.~\cite{Bartolo:2010rw}. In both equations, the last line is the additional correction found in Ref.~\cite{Tram:2016cpy} due to the presence of radiation in the early Universe. As we can see, these last terms have no local expressions but are separable in the sense that they do not involve any convolution integrals.

In a matter-dominated universe we find
\begin{itemize}
    \item in the super-horizon limit $k \ll \H$: $\delta^{\rm fl}_2 \simeq \delta_2^{\rm R2} \propto a^0$
    \item at the horizon scale $k \sim \H$: $\delta^{\rm fl}_2 \simeq \delta_2^{\rm R1} \propto a^1$
    \item in the sub-horizon limit $k \gg \H$: $\delta^{\rm fl}_2 \simeq \delta_2^{\rm N} \propto a^2$.
\end{itemize}
Note that the linear density in the Poisson gauge also has two limiting cases:
\begin{itemize}
    \item in the super-horizon limit $k \ll \H$: $\delta^{\rm fl}_1 \propto a^0$
    \item in the sub-horizon limit $k \gg \H$: $\delta^{\rm fl}_1 \propto a^1$.
\end{itemize}

To compute the canonical momentum\footnote{In our previous work \cite{Adamek:2021rot} we used the symbol $q_i$ to denote the canonical momentum of the particles. Here we change our notation in order to avoid confusion with the Lagrangian coordinate $\boldsymbol{q}$.} $\boldsymbol{p}$ and the gravitational potential at second order, which are necessary to initialize the particles, we follow Ref.~\cite{Adamek:2021rot}, and expand Einstein's equations at second order. For convenience we define a ``momentum potential'' $Q$ such that $\nabla\cdot\boldsymbol{p} / m = a \Delta Q$. Note also that $Q_1 = V_1$, where $V_1$ is the velocity potential at first order. The relation becomes more complicated at second order. See Appendix~\ref{app:2EE} for more details.

\subsection{Implementation in \texttt{MonofonIC}}\label{sub:monofonic}
We use a modified version of the \texttt{MonofonIC} software package\footnote{The official code is available at \url{https://bitbucket.org/ohahn/monofonic/}. Our modified branch can be found at \url{https://bitbucket.org/tomamtd/monofonic/src/relic/}.} \cite{Hahn:2011uy,Michaux:2020yis}. \texttt{MonofonIC} (\texttt{MUSIC2}) is an \ac{IC} generator for Newtonian simulations with \ac{CDM} and baryons based on Lagrangian perturbation theory up to third order. It is written in C++ and supports hybrid parallelisation with MPI and OpenMP/threads. It also accounts for numerical errors that can be generated by aliasing when taking products of fields that are discretised on a mesh. To de-alias the fields, \texttt{MonofonIC} makes use of ``Orszag’s ${}^3\!/_{\!2}$ rule,'' see Ref.~\cite{Michaux:2020yis} for more details.

\begin{figure}
    \centering
    \includegraphics[scale=0.5]{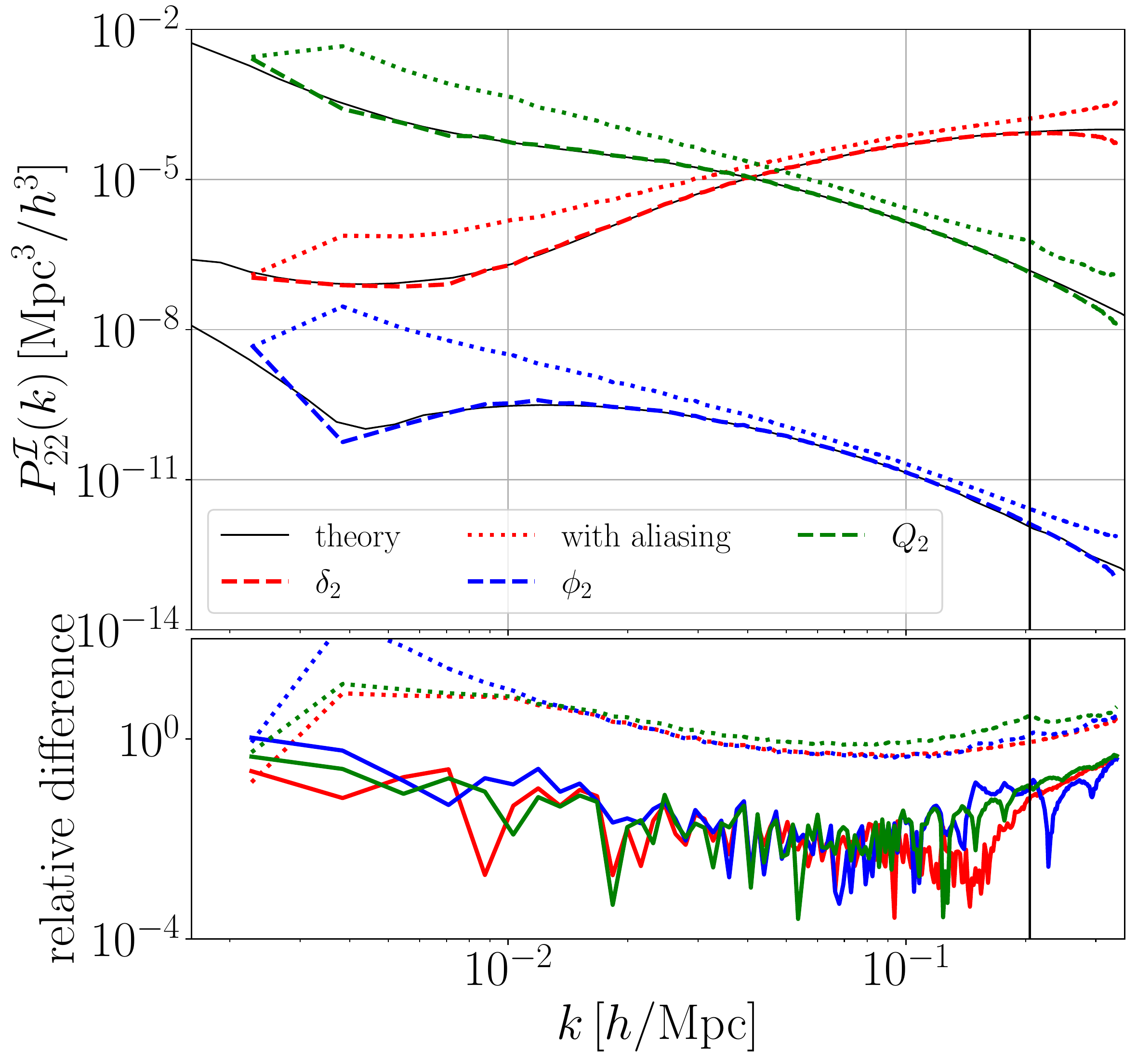}
    \caption{Power spectra of the second-order fields computed by \texttt{MonofonIC} in the relativistic case at redshift $z=100$. The dashed lines show the actual measurements of the fields after de-aliasing, which are in good agreement with the corresponding theoretical computations indicated as black lines (see Appendix~\ref{app:2ps}). The dotted lines show the power spectra of the fields without de-aliasing. In the lower panel, we show the relative difference between the measurements and the theoretical power spectra. Here we use a box with fundamental mode $1.6\times 10^{-3}$ $h/$Mpc and $256$ grid points, and the vertical black line indicates the Nyquist wavenumber of the box.
    }
    \label{fig:Pk2ini}
\end{figure}

\begin{figure}
    \centering
    \includegraphics[scale=0.32]{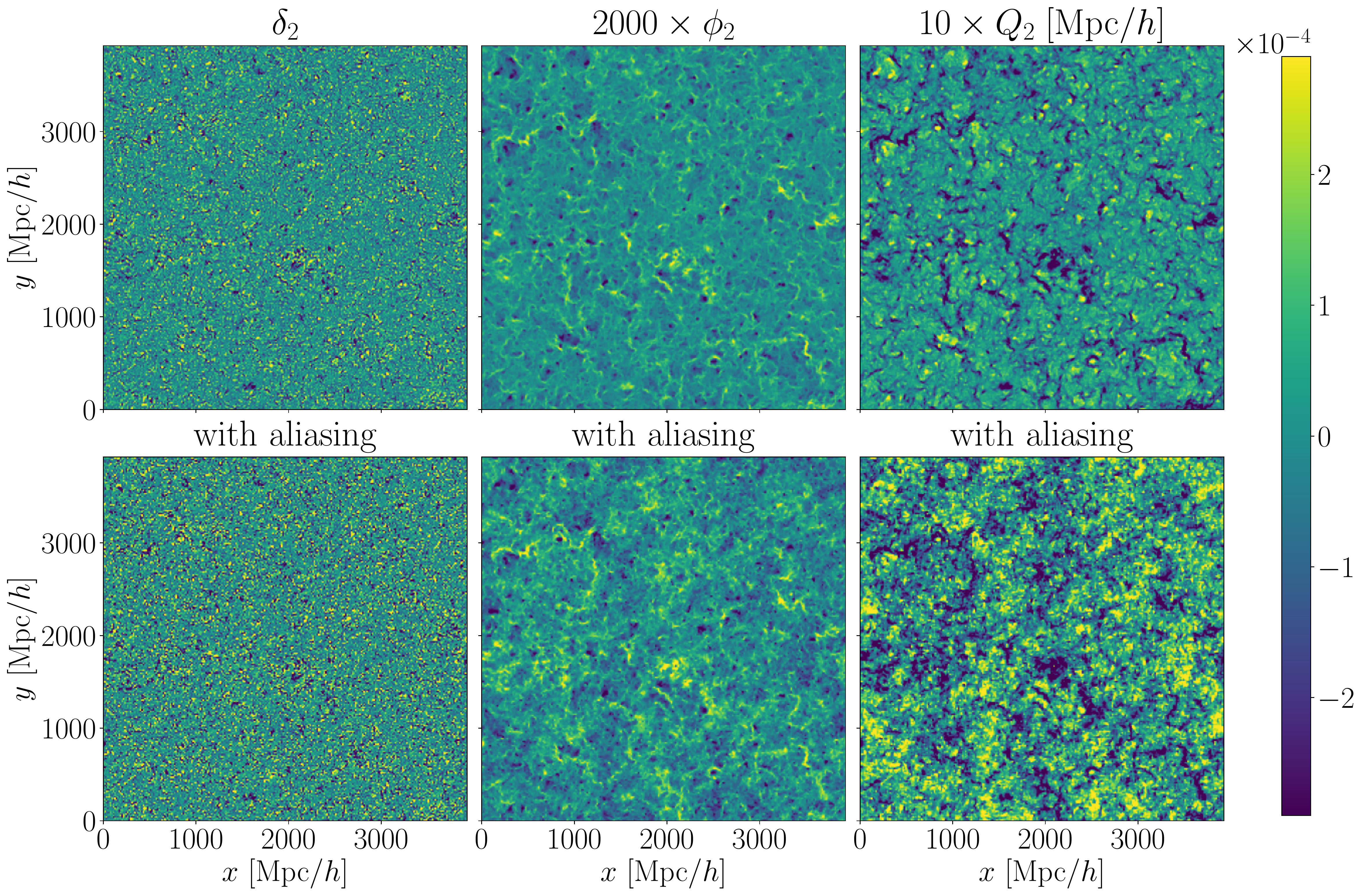}
    \caption{Slices of the three second-order fields computed by \texttt{MonofonIC}, $\delta_2$, $\phi_2$ and $Q_2$ (from left to right). The last two are multiplied by factors of $2000$ and $10$, respectively, in order to accommodate a common colour scale. The first row shows the fields computed with de-aliasing, while the second row shows the result of the computation without de-aliasing. We use the same resolution as in Fig.~\ref{fig:Pk2ini}. }
    \label{fig:imshow}
\end{figure}

In Figs.~\ref{fig:Pk2ini} and~\ref{fig:imshow}, we illustrate the impact of aliasing on the second-order fields. The black solid lines in Fig.~\ref{fig:Pk2ini} show the theoretical predictions for the power spectra of $\delta_2$, $\phi_2$ and $Q_2$, see Appendix~\ref{app:2ps} for the details of the computation. The dotted lines show the respective  power spectra if the numerical evaluation employs products of discretised fields in a naive way. As we can see, aliasing generates considerable errors at all scales. These errors are mostly generated by terms containing many derivatives because aliasing is more important for products of fields with blue spectra. The power spectra in dashed lines are obtained with de-aliasing. In Fig.~\ref{fig:imshow}, we can see the large visual impact that the aliasing has on the fields. The first (second) row shows the second-order fields computed with (without) de-aliasing. Except for the density, where the effect is difficult to see because the power on small scales is large, we clearly see numerical effects on $\phi_2$ and $Q_2$. To evaluate the various density contributions given in Eq.\,\eqref{eq:5.54}, as well as the expressions for $Q_{2}$ and $\phi_{2}$ (see also Appendix~\ref{app:2EE}), we use \texttt{MonofonIC} as a library, taking advantage of its parallelisation and efficient de-aliasing method. The first-order transfer functions are computed with the Einstein--Boltzmann code \texttt{CLASS}~\cite{Blas:2011rf}. In our analysis, we treat baryons and \ac{CDM} as a single fluid. At first order, we use the weighted sum of the \ac{CDM} and the baryon density and velocity fields. At second order, we approximate the matter density and the velocity by the corresponding quantities of \ac{CDM}.

\texttt{MonofonIC} also contains a numerical ODE solver, used to solve Eq.~\eqref{eq:D}. To compute $f$, $v$, and $w$, see Eq.~\eqref{eq:def}, we have extended this solver to second-order, see Eq.~\eqref{eq:F}. Recall that at second order, radiation is not accounted for in the background but only at the perturbation level. Hence, the growth factors are initialised in an EdS universe 
\begin{equation}
    \label{eq:ICD}
\D_0 = a_0, \qquad \D_0' = 2 \H a_0, \qquad {\cal F}_0 = \frac{3}{7} a_0^2, \qquad {\cal F}_0' = \frac{12}{7} \H a_0^{2} \,.
\end{equation}
Our relativistic version of \texttt{MonofonIC} can generate all the fields ($\delta_2, \phi_2, Q_2$, as well as the auxiliary fields $\chi_2 = \phi_2 - \psi_2$ and $\phi_2'$) within $10$ hours on a single processor for a $N=1024$ grid. The fields generated are used as input for the relativistic N-body code \texttt{gevolution}\footnote{The public version is available at \url{https://github.com/gevolution-code/gevolution-1.2}; our branch, able to initialise particles up to any order, can be found in \url{https://github.com/TomaMTD/gevolution-1.2}.}.

\subsection{Discrete Lagrangian perturbation theory}\label{sub:gev}
To initialise the particles, we use the same method and implementation as described in Ref.~\cite{Adamek:2021rot}. This iterative method can initialise the particles up to any order in perturbation theory, as long as one provides the fields $\delta, Q$ and $\phi$ accurately at the considered order. Here, we want to discuss the connection to Lagrangian perturbation theory, first introduced in Ref.~\cite{Zeldovich:1969sb}.

In the Newtonian case, the whole dynamical information of the collisionless fluid is encapsulated in the Lagrangian displacement field $\boldsymbol \xi$, defined through the mapping $\boldsymbol q \mapsto \boldsymbol x(\boldsymbol q,t) = \boldsymbol q + \boldsymbol \xi(\boldsymbol q,t)$ from initial  position $\boldsymbol q$ to Eulerian position~$\boldsymbol x$ at time~$t$. The main goal of \ac{LPT} is to compute the displacement field $\boldsymbol \xi$, which is expanded as a Taylor series in powers of the linear growth function. Once the displacement is determined up to a certain perturbation order, the Eulerian density can be reconstructed using the law of mass conservation $\rho(\boldsymbol x) d^3x = \bar \rho d^3q$, as well as the relation $d^3 x = {\cal J} d^3 q$ which defines the Jacobian determinant~${\cal J}$ of the transformation. All-order recursion relations were first found in Refs.~\cite{Rampf:2012up,Zheligovsky:2013eca} and implemented up to third order in \texttt{MonofonIC} \cite{Michaux:2020yis}.

In general relativity, the issue of Eulerian and Lagrangian gauge was discussed in \cite{Villa:2015ppa}. They show that the unique Lagrangian frame is the synchronous comoving gauge, while the Poisson gauge is an Eulerian frame. Let us derive here the link between the displacement field and the density. For an ensemble of point-like particles with masses $m_p$, the full stress-energy tensor takes the form \cite{Adamek:2016zes}
\begin{equation}
    \label{eq:Tmunu_def}
    T^{\mu\nu} =   \sum_p \delta_\mathrm{D}^{(3)}(\boldsymbol x - \boldsymbol x_p ) \frac{m_p v^\mu v^\nu}{\sqrt{g g_{\kappa\lambda} v^\kappa v^\lambda}}
    =   \sum_p \delta_\mathrm{D}^{(3)}(\boldsymbol x - \boldsymbol x_p ) \frac{\epsilon_p}{n_0^2} v^\mu v^\nu  
    \,,
\end{equation}
where the above relation defines the variable $\epsilon_p$, $n_0$ is the $0$th component of the unit normal of the equal-time hypersurface of the Poisson gauge $n_\mu = \partial \tau / \partial x^\mu$,
$v^\mu = dx^\mu / d\tau$ is the coordinate four-velocity (not to be confused with the covariant four-velocity $u^\mu$), and $p$ denotes the particle index. Taking the continuum limit of Eq.~\eqref{eq:Tmunu_def}, we can write the continuum stress-energy tensor as the integral over the Lagrangian coordinates which now replace the discrete particle index
\begin{equation}
    \label{eq:Tmunu_def2}
    T^{\mu\nu} = \int \bar \rho a^3  d^3 \boldsymbol{q}\, \delta_\mathrm{D}^{(3)}(\boldsymbol x - \boldsymbol q - \boldsymbol \xi ) \frac{v^\mu v^\nu}{\sqrt{g g_{\kappa\lambda} v^\kappa v^\lambda}}  \,.
\end{equation}
By change of variables to Eulerian coordinates we find
\begin{equation}
    \label{eq:Tmunu_def3}
    T^{\mu\nu} =   \frac{\bar \rho a^3 \mathcal J^{-1} v^\mu v^\nu}{\sqrt{g g_{\kappa\lambda} v^\kappa v^\lambda}} \,.
\end{equation}
This equation is the equivalent of the mass conservation in the Newtonian case. Indeed, one way of defining the density in the context of N-body simulations is by projecting the stress-energy tensor on $n_\mu$,
\begin{equation}
    \label{eq:Tmunu_def4}
    \rho(\boldsymbol x)= \bar \rho (1+\delta)= n_0^2 T^{00}=   \frac{\bar \rho a^3 n_0^2 \mathcal J^{-1}}{\sqrt{g g_{\mu\nu} v^\mu v^\nu}}  \,.
\end{equation}
This relation is gauge-independent and non-perturbative, and expresses the relation between the Lagrangian displacement field $\boldsymbol{\xi}$ and the density in Eulerian coordinates. The iterative algorithm presented in Ref.~\cite{Adamek:2021rot} is effectively a method to reconstruct $\boldsymbol{\xi}$ in a discrete setup, with some subtleties that we now explain.

In a particle-mesh scheme the stress-energy density of the particle ensemble can be convolved with a projection kernel to get a quasi-continuous distribution. For this we employ the \ac{CIC} kernel, yielding  %This yields 
\begin{equation}
    \label{eq:density_tot}
    \rho(\boldsymbol x_g) = \sum_{p} w_{\text{CIC}}(\boldsymbol x_g - \boldsymbol x_p )\epsilon_p \,,
\end{equation}
where $\boldsymbol x_g$ denotes the coordinates of a mesh point.
We can now perturb the particle position with respect to the initial position on a homogeneous template. In the language of \ac{LPT}, this means using the transformation $\boldsymbol x_p = \boldsymbol q_p + \boldsymbol \xi (\boldsymbol q_p)$ and expanding Eq.~\eqref{eq:density_tot} with respect to the displacement field. Assuming that the displacement can be written as a discrete gradient of a displacement field potential~$\xi$,
\begin{equation}\label{eq:ansatz}
    \boldsymbol \xi(\boldsymbol{x}) = \sum_{\boldsymbol{x}'_g} \boldsymbol{w}_\mathrm{grad}(\boldsymbol{x}'_g - \boldsymbol{x}) \xi(\boldsymbol{x}'_g)\,,
\end{equation}
where $\boldsymbol{w}_\mathrm{grad}$ is a discrete gradient kernel, we can solve for $\xi$ iteratively order by order, following Ref.~\cite{Adamek:2021rot}. The result approximates the displacement computed with \ac{LPT} with the following modifications:
\begin{itemize}
    \item Order by order, the discrete displacement also cancels discretisation errors that are not present in the continuum formulation. This means, for example, that it is not really important which type of gradient kernel is used in eq.~\eqref{eq:ansatz}. In practice we use for all simulations the gradient interpolation described in Appendix B of Ref.~\cite{Adamek:2016zes}.
    \item Even in the continuum limit, each order of the reconstructed displacement does not directly correspond to the Lagrangian displacement at the same order of \ac{LPT}. This is partly due to the fact that orders of \ac{LPT} mix in different iterations of the reconstruction and only the full displacement should be compared.
    \item The reconstruction is insensitive to Lagrangian displacements that do not affect the Jacobian at the present order. In other words, the reconstructed displacement and the \ac{LPT} displacement may differ by a volume-preserving transformation that can be ignored for the purpose of setting initial conditions in the Eulerian frame. However, at low perturbation order, this difference is also negligible for a standard cosmology. 
\end{itemize}
It should be stressed that we compute the Eulerian velocity field (or equivalently the canonical momentum) using the divergence of the momentum constraint in Poisson gauge rather than \ac{LPT}. The particle momenta are then obtained by interpolating to the Eulerian particle positions. 

In Poisson gauge, the variable $\epsilon_p$ takes the form
\begin{equation}\label{eq:t_poisson}
\epsilon_p = \frac{m_p}{a^3}e^{3\phi} \sqrt{1+e^{2\phi} (\nabla Q)^2}\,,
\end{equation}
where the four-velocity has been expressed in terms of the canonical momentum.

\subsection{Newtonian limit} \label{sub:Newton}

To isolate relativistic effects, we construct the same considered observables also from a Newtonian simulation.
The code \texttt{gevolution} allows us to run a Newtonian simulation by employing the so-called N-body gauge \cite{Fidler:2015npa}, and using the same numerical setup has the advantage that systematic effects due to the specific implementation are essentially kept fixed in the comparison. In this gauge all the linear relativistic corrections vanish except for the ones related to radiation. Therefore, it is a natural gauge for interpreting the linear perturbations of a Newtonian simulation in a relativistic framework. 

In the terminology of Ref.~\cite{Angulo:2021kes} the pipeline of setting \ac{IC} in the relativistic simulations uses a \textit{forward} method for the first-order fields and adds the second-order (Newtonian and relativistic) terms in the \ac{IC} following a \textit{back-scaling} approach. The philosophy behind this mixed scheme is to recover the correct solution in Poisson gauge at all redshifts. To achieve this, we need to account for the linear effects of radiation on matter in \texttt{gevolution} by evaluating the radiation perturbations with \texttt{CLASS} at each time step. It is sufficient to only account for the linear radiation perturbations in the simulation as for redshifts $z \lesssim 100$ their magnitude is similar to the second-order matter perturbations. This also means that the first-order dynamics will be incorrect by an amount similar to the second order if the linear radiation is not accounted for in the simulations. 

We cannot follow this approach for the Newtonian simulation. Indeed, to be consistent, we would need a Newtonian equivalent of the last lines of Eqs.~\eqref{eq:R1} and \eqref{eq:R2}. However, radiation is a pure relativistic effect in the sense that it cannot be incorporated within a Newtonian theory of gravity, unlike the linear matter dynamics \cite{Fidler:2015npa, Fidler:2017ebh}. Turning off \ac{RE} is however possible thanks to the N-body \textit{back-scaling} method \cite{Angulo:2021kes}, which allows us to take into account the radiation at first order with pure \ac{CDM} dynamics \cite{Fidler:2017ebh}. The transfer function, which accounts for all effects at the linear level, is first evaluated at redshift $z=0$. Then, we ``scale it back'' by using the linear growth factor $\D$ which ignores the radiation, i.e.~Eqs.~\eqref{eq:H} and \eqref{eq:D} where one sets $\Omega_{\Lambda0} = 1 - \Omega_{\mathrm{m}0}$; the small amount of radiation present today leads to an adjustment of $\Omega_{\Lambda0}$
%the cosmological constant 
in the fourth significant digit, the impact of which is diminished even further by the fact that CDM dominates the gravitational evolution at early times. Thus, the first-order density and the potential are computed as 
\begin{equation}
    \label{eq:first-transfer_back}
    \delta^{\rm Nb}_1(z,\boldsymbol{k}) = \D(z) T^{\rm Nb,\delta}_{1}(z=0,k) \zeta(\boldsymbol{k})\,,
\end{equation}
where the superscript ``Nb'' means that we are now evaluating the fields in the N-body gauge. Recall that the potential can be back-scaled in the same way with $\D/a$. The velocity takes the usual Newtonian form 
\begin{equation}
    \label{eq:vel_back}
    V^{\rm Nb}_1 = -\frac{2 f }{3 \H\Omega_\mathrm{m} } \phi_1^{\rm Nb} \,.
\end{equation}

At second order, we can take the Newtonian limit of Eq.~\eqref{eq:5.54} and substitute the potential by its back-scaled counterpart,
\begin{equation}
    \label{eq:5.54_newton}
        \left(\frac{2}{3 \H^2 \Omega_\mathrm{m}} \right)^{-2} \delta^{\rm fl,Nb}_2 = \frac{1}{2}(1+\frac{3}{7}v)(\Delta\phi_1^{\rm Nb})^2 +  \phi_1^{{\rm Nb},l} \Delta\phi^{\rm Nb}_{1,l} + \frac{1}{2} (1-\frac{3}{7}v) \phi_1^{{\rm Nb},lm}\phi^{\rm Nb}_{1,lm} \,.
\end{equation}
Finally, the second-order canonical momentum in the Newtonian limit takes the form \cite{Villa:2015ppa}
\begin{equation}
    \label{eq:Newtonian_vel}
      Q_2^{\rm Nb} =  V_2^{\rm Nb} = \frac{\H}{2} \left(\frac{2 }{3 \H^2 \Om} \right)^2 \left(- f \phi^{\rm Nb}_{1,i}\phi_1^{{\rm Nb},i} + \frac{6w}{7} \Delta^{-1}\left[ \phi_1^{{\rm Nb}, lm} \phi^{\rm Nb}_{1,l m} - (\Delta\phi_1^{\rm Nb})^2\right]  \right)\,.
\end{equation}

Once the density and velocity are computed, the Newtonian displacement field can be obtained by following the procedure explained in Section \ref{sub:gev} and by replacing 
\begin{equation}
    \label{eq:Nbody_gauge_t}
    \epsilon^{\rm Nb}_p = \frac{m_p}{a^3}\,.
\end{equation}
Note that, unlike in the relativistic case where $\phi$ should also be provided to \texttt{gevolution} to initialise the particles due to its appearance in Eq.~\eqref{eq:t_poisson}, it is not needed in the Newtonian case. However, it is used internally by \texttt{MonofonIC} to evaluate Eqs.~\eqref{eq:5.54_newton} and \eqref{eq:Newtonian_vel}.

\section{Number-count bispectrum}\label{sec:numbercount}

In order to compare the Newtonian and the relativistic simulations, we construct an observable which, by definition, is gauge invariant. To this end, we employ a method developed in Ref.~\cite{Lepori:2021lck} and construct the relativistic particle number counts for matter on the light cone. A gauge-invariant signature of non-Gaussianity can then be extracted from the angular bispectrum. Note that the particle number count is not, strictly speaking, an observable as particles representing CDM cannot be observed directly. But assuming they could be observed, the number count is manifestly gauge invariant. In this work we avoid the complications of using astrophysical objects like galaxies as tracers of the underlying matter distribution and work with matter counting statistics directly. The inclusion of biased tracers of matter is left to future work.

Many theoretical developments have been made on the angular power spectrum and bispectrum in \ac{LSS}, see e.g.\ \cite{Bonvin:2011bg, DiDio:2015bua}. The linear angular power spectrum of the CDM number counts was implemented in \texttt{CLASS} \cite{DiDio:2013bqa}. It includes dynamical and kinematical relativistic effects and is therefore suitable to be compared with our simulations. 

For the bispectrum, no complete theoretical expression has been derived so far, as the large number of effects to be accounted for up to second order makes it extremely tedious. In Ref.~\cite{DiDio:2018unb} however, one of the dominant terms of the bispectrum was explicitly computed and implemented in the public code \texttt{Byspectrum}\footnote{\url{https://gitlab.com/montanari/byspectrum}}. Yet we cannot directly compare our results with \texttt{Byspectrum} as it does not feature any redshift window functions and extracting number counts for infinitesimal redshift bins from simulations is impractical. Moreover, one of the dominant Newtonian terms, namely the redshift-space distortion, as well as a host of relativistic effects (such as those induced through Eqs.\,\ref{eq:R1}--\ref{eq:R2}), have not been implemented in \texttt{Byspectrum}; we leave such issues for future work.

The binned bispectrum estimator \cite{Bucher:2009nm,Bucher:2015ura}, developed in the context of \ac{CMB} experiments, provides the full bispectrum in observed angular multipoles. In this paper, we propose its first application to \ac{LSS} experiments. Unlike bispectrum estimators commonly used in \ac{LSS} \cite{Scoccimarro:2015bla,Sefusatti:2015aex}, we employ an estimator based on spherical harmonics since, for \ac{RE}, one needs to probe the largest scales possible. Galaxy surveys are naturally provided on the light cone in angular/redshift space. This can make it challenging to interpret the large-scale results of an estimator based on 3-dimensional Fourier transforms, see e.g.\ \cite{Grimm:2020ays,Castorina:2021xzs}. The use of an angular bispectrum estimator avoids the complexity of wide-angle effects as it takes the pure observable as direct input. Note also that we want to compare two simulations performed in different gauges: Poisson and N-body gauge. It is therefore crucial to construct gauge-invariant quantities.

We want to simulate a full-sky survey for tracers of matter, taken to be particles for simplicity. With \texttt{gevolution}, we produce the particle light cone for an observer in the center of the simulation box. Different implementations of relativistic ray tracers have been applied to N-body codes in the past, e.g.\ \cite{Adamek:2018rru, Breton:2021htu}. We will use the nonlinear relativistic ray tracer developed for \texttt{gevolution} \cite{Adamek:2018rru,Lepori:2020ifz} to compute the null geodesics connecting each particle world line to the observer. We use the same relativistic ray tracer for the relativistic and the Newtonian simulations. Hence, in the Newtonian pipeline, the ray tracer finds the particle positions in the wrong gauge \cite{Adamek:2019aad}. Later on we estimate the impact of the gauge correction by means of the angular power spectrum, using a modified version of \texttt{CLASS} (see Fig.\,\ref{fig:Cl}). For the bispectrum, however, this effect remains to be accounted for if we want a fair estimation of the second-order \ac{RE}. Any differences we find at the level of the bispectrum can yet be a combination of genuine \ac{RE} and a known shortcoming of the Newtonian approximation that could be rectified in principle (see Ref.~\cite{Adamek:2019aad} for a worked example).

From the ray tracer, we obtain the observed angular positions $\hat {\boldsymbol n}$ and redshifts $z$ for N-body particles on the light cone with which we can compute the particle number-count perturbation $\Delta(\hat {\boldsymbol n}, z)$, defined as 
\begin{equation}\label{eq:numbercount}
    \Delta(\hat {\boldsymbol n}, z) = \frac{N(\hat {\boldsymbol n}, z)}{\left <N\right>(z)} -1\,,
\end{equation}
where $N(\hat {\boldsymbol n}, z)$ is the number of particles in a given solid angle $d \hat {\boldsymbol n}$ and redshift bin $dz$. Note that the particle number count is an unbiased tracer of the matter density. The redshift dimension is split into bins such that for a given redshift bin, the number-count perturbation $\Delta(\hat {\boldsymbol n}, z)$ is a 2-dimensional spherical map that can be decomposed using spherical harmonics. 
Let us follow \cite{Bucher:2015ura} and define the filtered map $\Delta_\ell^z$
\begin{equation}\label{eq:Delta_ellm}
    \Delta_\ell^z(\hat {\boldsymbol n}) = \sum_{m = -\ell}^\ell a_{\ell m}^z  Y_{\ell m}(\hat {\boldsymbol n})\,,
\end{equation}
where $Y_{\ell m}$ are the usual spherical harmonics and where $a_{\ell m}^z$ can be obtained with
\begin{equation}
    a_{\ell m}^z = \int d\hat {\boldsymbol n} \, \Delta(\hat {\boldsymbol n}, z) Y^*_{\ell m}(\hat {\boldsymbol n})\,.
\end{equation}
The angular power spectrum is commonly defined as 
\begin{equation}
     \left < a_{\ell_1 m_1}^{z_1} a_{\ell_2 m_2}^{z_2} \right > = \delta_{\ell_1 \ell_2}\delta_{m_1 m_2} C_{\ell_1}^{z_1z_2}\,,
\end{equation}
and where $\left< \right>$ denotes the ensemble average, which is replaced by an average over the $m$'s for the angular power spectrum estimator. The \textit{angle averaged} bispectrum is then obtained by
\begin{equation}\label{eq:angleaverage_bisp}
    B^{z_1z_2z_3}_{\ell_1\ell_2\ell_3} = \left< \int d\hat{\boldsymbol n} \, \Delta_{\ell_1}^{z_1} \Delta_{\ell_2}^{z_2} \Delta_{\ell_3}^{z_3} \right> \,.
\end{equation}
One can show that $B^{z_1z_2z_3}_{\ell_1\ell_2\ell_3}$ can be obtained by averaging the full bispectrum over the $m$'s: 
\begin{equation}
    B^{z_1z_2z_3}_{\ell_1\ell_2\ell_3} = 
 \sqrt{\frac{(2\ell_1+1)(2\ell_2+1)(2\ell_3+1)}{4\pi}}
    \begin{pmatrix}
\ell_1 & \ell_2 & \ell_3\\
0 & 0 & 0
\end{pmatrix}
    \sum_{m_1m_2m_2} 
    \begin{pmatrix}
\ell_1 & \ell_2 & \ell_3\\
m_1 & m_2 & m_3
\end{pmatrix}
    \left< a_{\ell_1 m_1}^{z_1} a_{\ell_2 m_2}^{z_2} a_{\ell_3 m_3}^{z_3}\right>
\end{equation}
Statistical isotropy ensures that the angle-averaged bispectrum contains all the physical information. 

Even if working with the angle-averaged bispectrum means we do not have to consider the dependence on $m_1,m_2,m_3$, it is still computationally unfeasible for current or future experiments to compute the bispectrum for each multipole combination $\ell_1, \ell_2$, $\ell_3$ \cite{Bucher:2015ura}. One way of reducing the computational cost is to consider instead the bispectrum of the binned filtered maps, i.e.
\begin{equation}\label{eq:}
    \Delta_i^z(\hat {\boldsymbol n}) = \sum_{\ell =\ell^{\rm min}_i}^{\ell_i^{\rm max}} \sum_{m = -\ell}^\ell a_{\ell m}^z  Y_{\ell m}(\hat {\boldsymbol n})\,,
\end{equation}
where $[\ell_i^{\rm min}, \ell_i^{\rm max}]$ is the bin interval of the $i^{\rm th}$ bin. In the analysis of \ac{CMB} anisotropies the binning has been adapted to the study of three relevant theoretical templates: the local, equilateral and orthogonal shapes. Thanks to the fact that they vary slowly with $\ell$, an optimal binning choice (with less than $1\%$ loss of information) was found with only 57 bins, to be compared with the roughly $2500$ multipoles measured by Planck. For the purpose of studying LSS the criterion of binning might need to be revisited, but we leave this to future work. In the following, in order to have an accurate sampling for all scales, we use a logarithmic binning for $\ell< 120$ and a linear binning with $\ell_i^{\rm max} - \ell_i^{\rm min} = 69$ for $120 \leq \ell < 750$. In total, we have $15$ bins in the range $ [2, 749]$. Note that the upper limit $\ell = 750$ is chosen to be slightly above the multipole corresponding to the Nyquist frequency of the highest-redshift bin.

Given the lack of a complete theoretical prediction to guide us, we study the full binned bispectrum numerically.
In \cite{Bucher:2015ura}, a statistical method has been developed to search for any non-Gaussian signal in the bispectrum. To this end the bispectrum is convolved with a Gaussian such that
\begin{equation}\label{eq:smooth}
    S\left[ B^{z_1 z_2 z_3}_{i_1 i_2 i_3}\right] =  
    (2\pi\sigma^2_{\rm bin})^{-3/2}\sum_{i_1',i_2',i_3'} \exp{\left[-\frac{1}{2} \frac{ (i_1 - i_1')^2+(i_2 - i_2')^2+(i_3-i_3')^2 }{\sigma_{\rm bin}^2} \right]} 
    B^{z_1z_2z_3}_{i_1'i_2'i_3'}\,,
\end{equation}
where $\sigma_{\rm bin}$ is a parameter to adjust. We will use the same value as in the Planck analysis \cite{Planck:2019kim}, i.e.\ $\sigma_{\rm bin} = 2$. This ``smoothing'' allows us to reduce the noise affecting the bins. Hence, it is well adapted for the search of any broad non-Gaussian features like the ones expected for the \ac{RE}.

\section{Results}\label{sec:result}
In this section, we test the relativistic and Newtonian simulation pipelines and then compare them by calculating the gauge-invariant particle number counts on the light cone. First, we present our simulation setting in Section~\ref{sub:setting}. In Section~\ref{sub:evolution}, we validate the second-order evolution by comparing the power spectrum (in Section~\ref{subsub:power}) and the bispectrum (in Section~\ref{subsub:bisp}) of the second-order density on snapshots produced by \texttt{gevolution} with the respective theory calculation implemented in \texttt{MonofonIC}. In Section~\ref{sub:nbc}, we calculate the particle number counts for five redshift bins using a nonlinear ray tracer. Finally, we compare the angular power spectrum (in Section~\ref{subsub:angle_ps}) and the angular bispectrum (in Section~\ref{subsub:angle_bs}) of the relativistic and Newtonian number counts.

\subsection{Simulation settings} \label{sub:setting}
We run two types of simulations: relativistic simulations including radiation in the evolution, and Newtonian simulations where radiation is only accounted for through proper back-scaling in the \ac{IC} but otherwise does not appear in the evolution. To observe a significant difference between the relativistic and the Newtonian simulations, we use a large simulation volume with fundamental mode $k_\mathrm{f} = 8\times 10^{-4} \;h/$Mpc, i.e.\ a box size of $7854$ Mpc$/h$. We use $N=1024$, where $N$ is the number of grid points in each space dimension, and $2048^3$ particles. This gives a resolution of $7.67$~Mpc$/h$. The primordial power spectrum is a nearly scale-invariant power law with amplitude $A_\mathrm{s} = 2.215 \times 10^{-9}$, defined at the pivot scale $k_\mathrm{p} = 0.05$ Mpc$^{-1}$, and with tilt $n_\mathrm{s} = 0.9619$. These values are based on the Planck 2013 cosmological fit \cite{Planck:2013pxb} and we set $h = 0.67556$, $\Omega_\mathrm{b} = 0.048275$ and $\Omega_\mathrm{c} = 0.26377$ which are also compatible with that fit.

\subsection{Second-order evolution}\label{sub:evolution}
To validate our implementation of the second-order dynamics we start at a fairly early redshift $z=100$, even though our \ac{IC} based on second-order \ac{LPT} can be used at much later redshift \cite{Michaux:2020yis}. We then check the consistency between \texttt{MonofonIC} and \texttt{gevolution} by analysing density snapshots, hereafter $\hat \delta$, at redshifts $100, 90, 70, 50, 30$ and $10$. At these high redshifts, perturbation theory holds for the scales $k < 0.4\,h/\mathrm{Mpc}$ considered here, and we can compare the power spectra and bispectra of the density snapshots with the theory. The power spectrum $P_\I$ of any random field $\I$ is defined through
\begin{equation}\label{eq:ps_def}
     \left\langle \I(\boldsymbol k_1)\I(\boldsymbol k_2) \right\rangle = (2\pi)^3 \delta_\mathrm{D}^{(3)}(\boldsymbol k_1+\boldsymbol k_2) P_\I(k_1) \,,
\end{equation}
and the bispectrum $B_\I$ of the field $\I$ through
\begin{equation}\label{eq:bs_def}
     \left< \I(\boldsymbol k_1 )\I(\boldsymbol k_2)\I(\boldsymbol k_3 ) \right> = (2\pi)^3 \delta_\mathrm{D}^{(3)} (\boldsymbol k_1+\boldsymbol k_2 +\boldsymbol k_3 ) B_\I(k_1,k_2,k_3) \,.
\end{equation}
To estimate the constant-time hypersurface power spectra and the bispectra of the simulation snapshots, we use the estimator provided by the code package \texttt{Pylians3} \cite{Pylians}.

\begin{figure}
    \centering
    \includegraphics[scale=0.38]{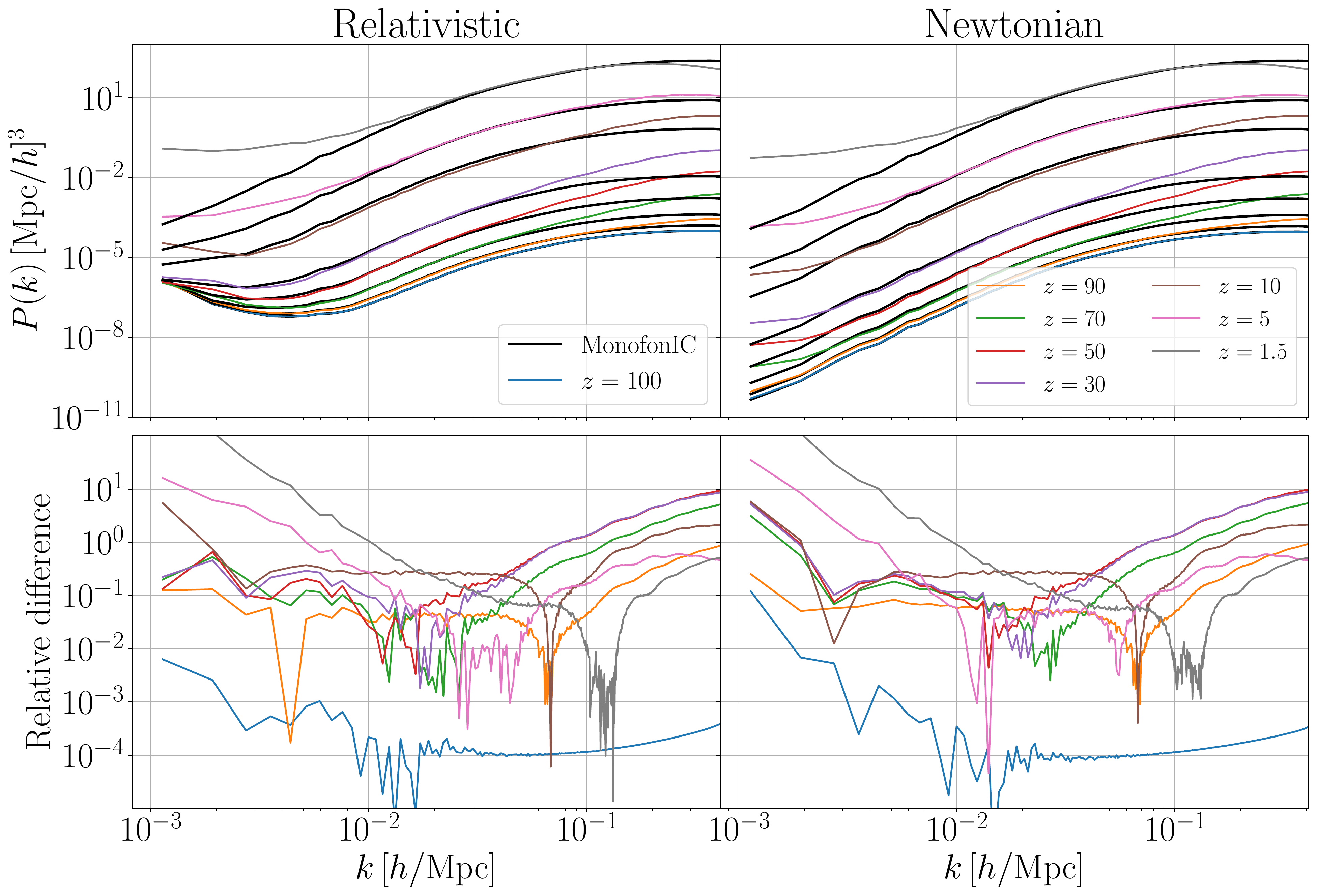}
    \caption{The top panels show the second-order power spectra computed on the \texttt{MonofonIC} realisations in black and on the \texttt{gevolution} snapshots in colours. For the snapshots of \texttt{gevolution}, we have subtracted the linear field evaluated at the corresponding redshift with \texttt{MonofonIC}, i.e.\ the correlations contain everything apart from the fiducial first-order contribution in this case (see text for details). 
    The power is increasing with time. The initial redshift is $z=100$, where \texttt{gevolution} matches \texttt{MonofonIC} to very high precision. The lower panels show the relative difference between \texttt{gevolution} and \texttt{MonofonIC} for each redshift. The left panels correspond to the relativistic simulation, including radiation, while the right panels correspond to the Newtonian simulation, without radiation. 
    }
    \label{fig:Pk_at_all_z}
\end{figure}

\subsubsection{Power spectrum}\label{subsub:power}
As long as perturbation theory holds, the density of the snapshots $\hat \delta$ contains the linear evolution of the \ac{IC}, as well as additional nonlinear (at late times even non-perturbative) terms sourced by the evolution. In the perturbative regime we may write
\begin{equation}\label{eq:gev_density}
    \hat \delta(z \leq z_{\mathrm {ini}}) = \delta_1(z) + \delta_2(z) + \ldots \,.
\end{equation}
The first-order density $\delta_1(z)$ can be evaluated at any redshift by \texttt{MonofonIC} and subtracted from the nonlinear $\hat \delta$. If the cancellation is exact, the power spectrum of the remaining fields, $\hat P_{\hat \delta-\delta_1}$, takes the form 
\begin{equation}\label{eq:gev_density_ps}
\begin{split}
    \hat P_{\hat \delta-\delta_1}  = P_{22} + \Delta P_{\rm NL} \,,
\end{split}
\end{equation}
where $P_{22}$ is the power spectrum of $\delta_2$,
while $\Delta P_{\rm NL}$ stands for the nonlinear residual which is assumed to be perturbative at sufficiently early times (in which case it contains $P_{33}$ and $P_{24}$ at next-to-leading order).

In the top panels of Figure~\ref{fig:Pk_at_all_z}, we plot in black the power spectra of the second-order realisations evaluated at the given redshift by \texttt{MonofonIC} (the agreement between \texttt{MonofonIC} and the theory was established in Section~\ref{sub:monofonic}). The left and right panels are, respectively, the results of the relativistic and Newtonian simulations. The power spectra have increasing amplitude with decreasing redshift. In colours, we show the power spectra $\hat P_{\hat \delta-\delta_1}$ obtained by subtracting the first-order field $\delta_1$ of \texttt{MonofonIC} from the \texttt{gevolution} snapshot. The lower panels show the relative difference between $\hat P_{\hat \delta-\delta_1}$ and $P_{22}$. 

Initially, except for the large scales, we have $\hat P_{\hat \delta-\delta_1} = P_{22}$ with a relative error of order $10^{-4}$. This shows the efficiency of the iterative algorithm introduced in Ref.~\cite{Adamek:2021rot} to compute the initial particle displacements. For $k \lesssim 3\times 10^{-3}\; h/$Mpc, the relative difference increases to reach $1\%$ for the relativistic case, and $10\%$ for the Newtonian case. This may be partly due to the fact that the amplitude itself is very small at large scales (it tends to zero in the Newtonian case). With the evolution, we observe three different contributions, affecting the small scales $k \gtrsim 10^{-1}\; h/$Mpc, the intermediate scales $ 3 \times 10^{-3}\; h/\mathrm{Mpc} \lesssim k \lesssim 10^{-1}\; h/$Mpc and the large scales $k \lesssim 3 \times 10^{-3}\; h/$Mpc.

At small scales, we observe an increasing discrepancy in both the relativistic and Newtonian simulations, which finally decays at redshift $z \leq 5$. This increasing power arises from numerical effects due to discretisation and can be mitigated by starting the simulation at a later time \cite{Michaux:2020yis}. We have checked that this small-scale power arises from the nonlinear evolution of the first-order \ac{IC}, by computing similarly $\hat P_{\hat \delta-\delta_1}$ on a simulation initialised at first order. 
In the intermediate scales, we observe a general agreement at a level between $1\%$ and $10\%$, which is consistent with Ref.~\cite{Adamek:2021rot}. Note that these scales are dominated by Newtonian nonlinearities. The growth of the second-order part seems to be slightly smaller than expected for the highest redshifts $z>5$, but then catches up with the theoretical prediction. This hints at the presence of a decaying mode.
Finally, at large scales, the initial large-scale inaccuracy of the discrete N-body realisation appears to propagate through the simulation, causing systematically larger power at large scales. For the lowest redshifts $z \leq 5$, the measurements deviate significantly from the theory up to $k\sim 10^{-2}\; h/\mathrm{Mpc}$.

\begin{figure}
    \centering
    \includegraphics[scale=0.38]{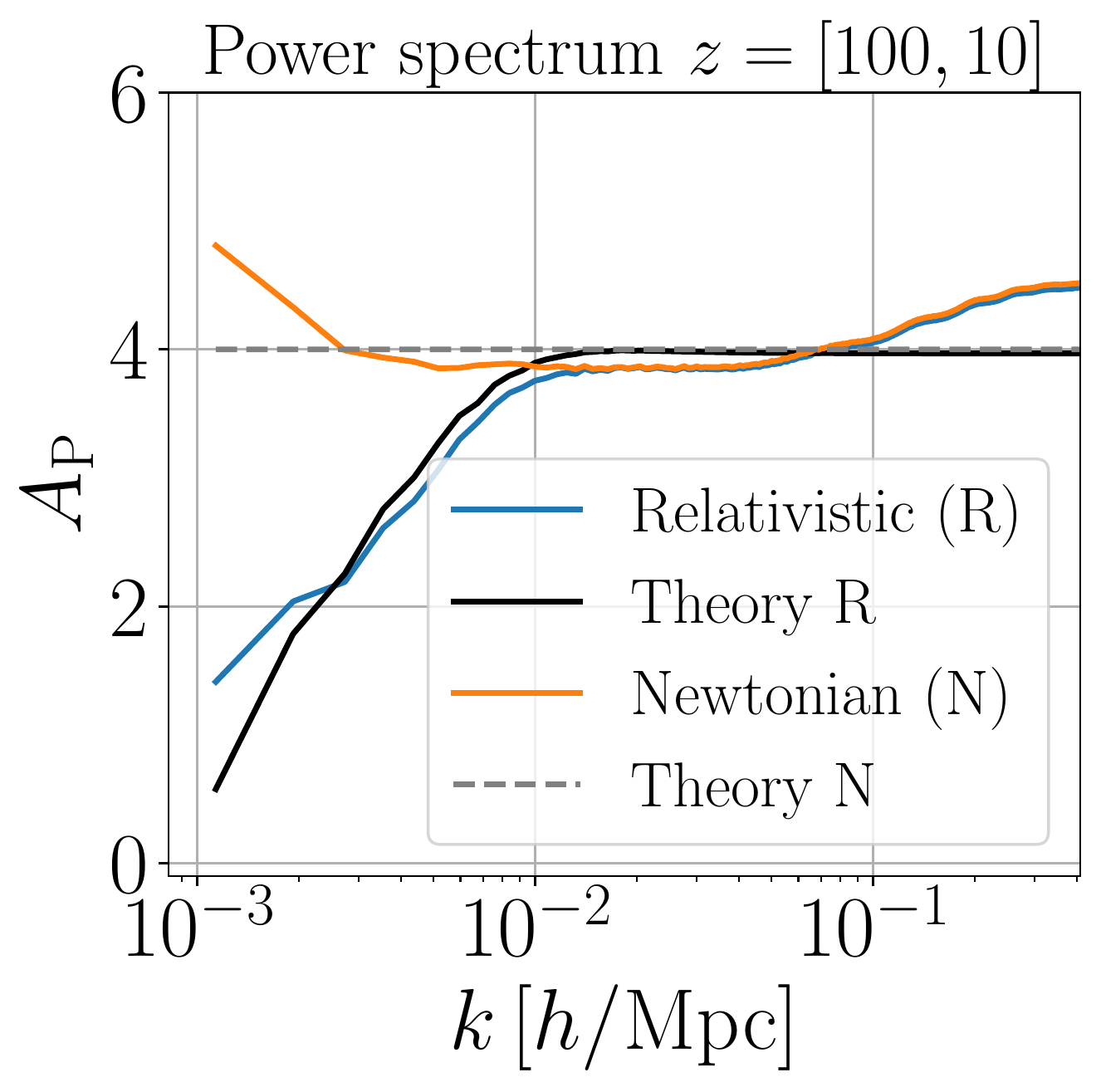}
    \includegraphics[scale=0.38]{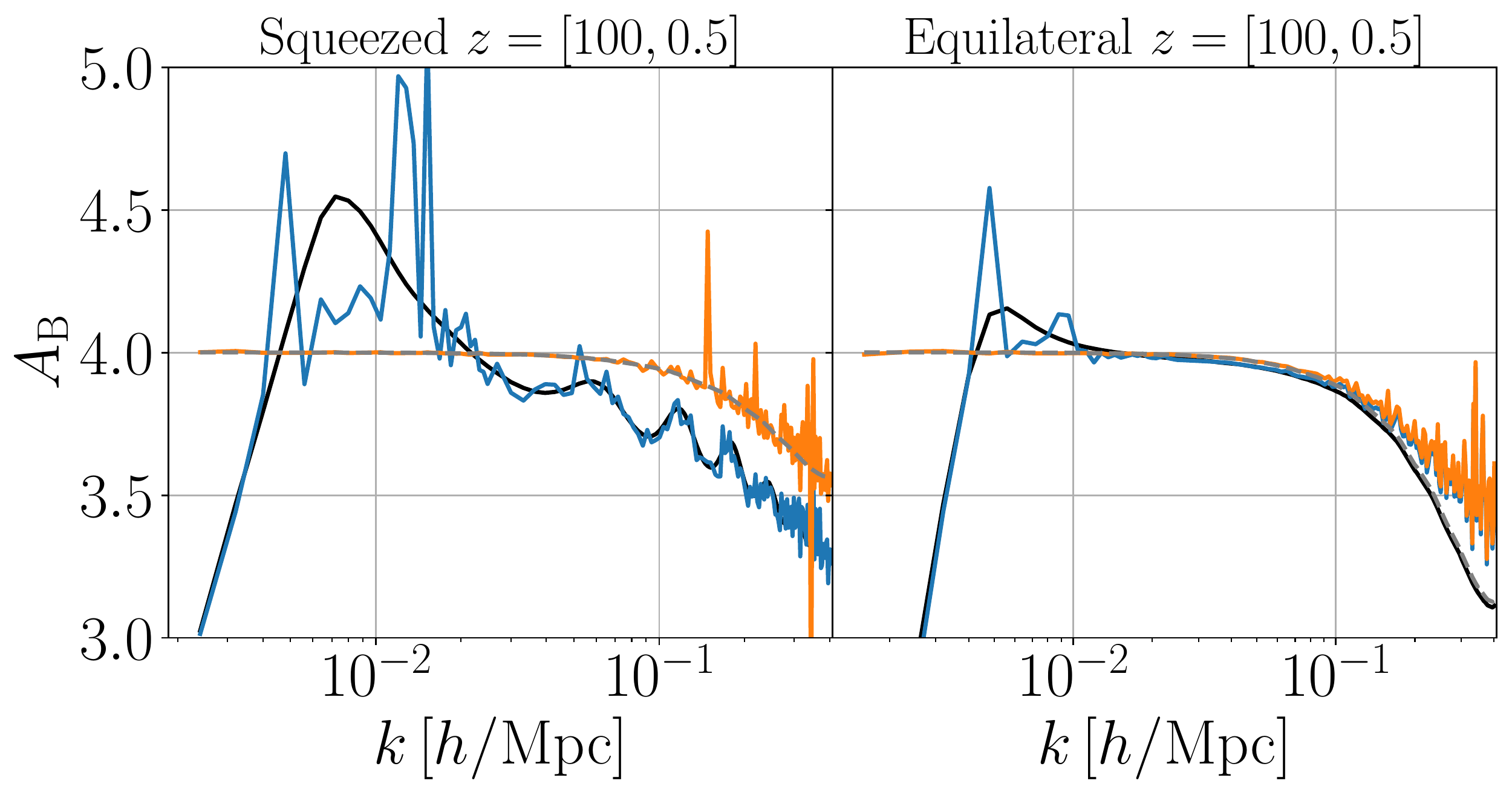}
    \caption{Mean logarithmic growth rate $A_{\mathrm{P}}$ and $A_{\mathrm{B}}$ over the redshift range $100 < z < 10$ for $A_{\mathrm{P}}$ and $100 < z < 0.5$ for for $A_{\mathrm{B}}$ as a function of $k$. The left panel shows the results for the power spectrum $P_{\hat \delta-\delta_1} = P_{22}$, while the middle and right panels show the results for the bispectrum in the squeezed limit and in the equilateral configuration, respectively. The logarithmic growth rate is defined in \eqref{eq:fit_power} for the power spectrum and in \eqref{eq:fit_bispectre} for the bispectrum. In solid black (dashed grey), we show the relativistic (Newtonian) theory.
    }
    \label{fig:fct_of_a}
\end{figure}

In order to test the time evolution of each mode in a more accurate way, we fit for each $k$ the linear relation 
\begin{equation}\label{eq:fit_power}
 \ln{\hat P_{\hat \delta-\delta_1}(a; k)} = A_{\mathrm{P}}(k) \ln{a} + C_{\mathrm{P}}(k)\,,    
\end{equation}
where $A_\mathrm{P}$ is the logarithmic growth rate and $C_\mathrm{P}$ is a constant in time. If we fit the relation between two redshifts, the coefficient $A_\mathrm{P}$ can be interpreted as the averaged logarithmic growth rate of the power spectrum between these redshifts. In our case, we use $z=100$ and $z=10$ since the nonlinearities have a significant impact for $z<10$, see Fig.~\ref{fig:Pk_at_all_z}. In the left panel of Fig.~\ref{fig:fct_of_a}, the logarithmic growth rate is plotted as a function of $k$. The theoretical logarithmic growth rate, shown in solid black (dashed grey) lines for the relativistic (Newtonian) case, is obtained by fitting $P_{22}$ computed directly from \texttt{MonofonIC}. As explained in Section~\ref{sub:REIC}, the Newtonian second-order part of the density $\delta^{\rm N}_2$ is expected to grow like $a^2$. Hence, we expect a logarithmic growth rate of $A_\mathrm{P} = 4$ for $P_{22}$. In agreement with Fig.~\ref{fig:Pk_at_all_z}, we recover a slightly smaller growth rate in the Newtonian simulation for the modes smaller than $\sim 5\times 10^{-2} \;h/$Mpc. The relativistic simulation also has a plateau close to $A_\mathrm{P} \simeq 4$ between $10^{-2}\;h/\mathrm{Mpc}$ and $5\times 10^{-2} \;h/\mathrm{Mpc}$. As explained in Section~\ref{sub:REIC}, for scales of order $\H$, $\delta_2^{\rm R1}$ is dominant and one observes a transition from $A_\mathrm{P} \simeq 4$ to $A_\mathrm{P} \simeq 2$, and eventually to $A_\mathrm{P} \simeq 0$ for super-horizon scales where $\delta_2^{\rm R2}$ finally dominates. In the left panel of Fig.~\ref{fig:Pk_at_all_z}, we see that this first point is constant until redshift $30$, while it starts to grow at $z = 10$ as the horizon expands in time. Thus, the averaged logarithmic growth rate between $z=100$ and $z=10$ is around $A_\mathrm{P} \simeq 1$ for the first point. 
Overall, the underestimation of the growth disappears if we would have chosen $z\leq 5$. But, as we can see in Fig.~\ref{fig:Pk_at_all_z}, the large-scale discrepancy would overestimate the growth.

\subsubsection{Bispectrum}\label{subsub:bisp}

\begin{figure}
    \centering
    \includegraphics[scale=0.43]{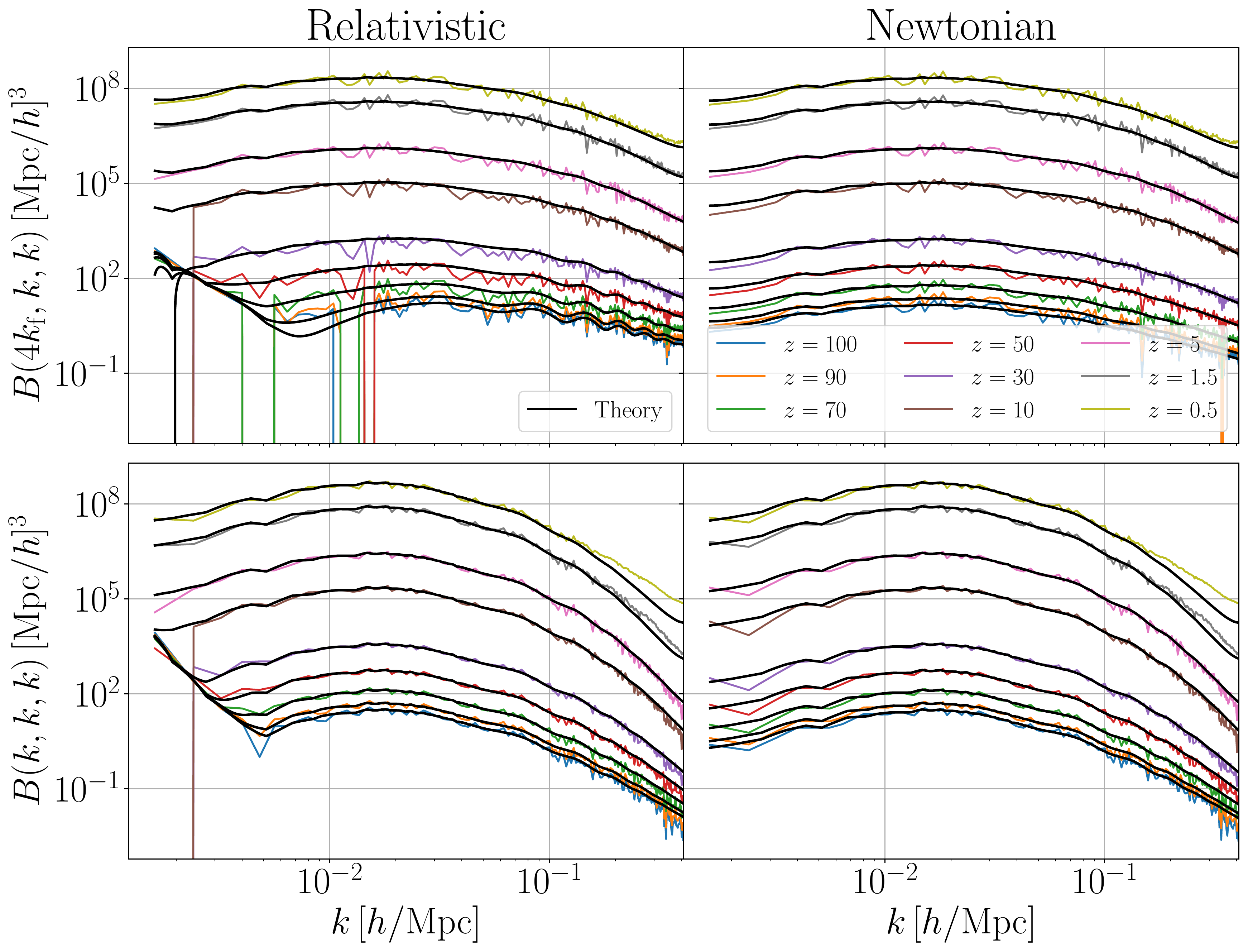}
    \caption{Relativistic (left panels) and Newtonian (right panels) bispectra measured on different redshift snapshots (the bispectrum increases with time). In the top panels, the first mode is fixed to $4$ times the fundamental mode $k_\mathrm{f}$, and we vary the last two modes which are equal. Therefore, the leftmost point of the plot represents a large-scale equilateral configuration, while the rightmost point represents a squeezed configuration. In the bottom panels all modes are taken equal so that we have equilateral configurations. We use the ``pairing'' method to reduce the cosmic variance, following Ref.~\cite{Angulo:2016hjd}, and we bin the $k$-space such that there are four $k$-values per bin. The black lines show the theoretical bispectrum where, in addition to the pairing, we have substituted the theoretical power spectrum by the measured one in order to introduce the cosmic variance in the theory. 
    }
    \label{fig:Bk_at_all_z}
\end{figure}

The bispectrum is to some extent a cleaner probe of the nonlinear evolution as one does not need to subtract the linear part of the field as we did in Section~\ref{subsub:power}. However, the variance is much more of a concern here as it is in general much larger than the signal. In Ref.~\cite{Adamek:2021rot}, we artificially amplified the second-order signal by a factor of $100$ and used the ensemble average over $20$ realisations in order to beat the cosmic variance in the bispectrum. An amplification was fine to test \ac{IC}, but cannot be used for the evolution as it does not describe a valid solution to the dynamical problem. Here we will use a method introduced in Refs.~\cite{Pontzen:2015eoh, Angulo:2016hjd}, called ``pairing'', which drastically reduces the variance due to phase correlation in the nonlinear regime. The idea consists of averaging the bispectra obtained by two simulations with exactly opposite primordial Gaussian curvature perturbations. However, this method does not cancel the large-scale cosmic variance. We could use the method called ``fixing'' to cancel it \cite{Pontzen:2015eoh}, but this also introduces non-Gaussianities which we do not want to interfere with our \ac{RE}. Instead of cancelling the cosmic variance in the power spectrum, we will account for it in the theory. To do so, we substitute the theoretical power spectrum by the measured one, $P_{11} \rightarrow \hat P_{\hat \delta}$, in the computation of the theoretical bispectrum. Hence, we compute the theoretical bispectrum as
\begin{equation}
\label{eq:semi-th}
    B^{\mathrm{th}}_{\delta}(k_1,k_2,k_3) = 2 \left[\hat P_{\hat \delta}(k_1) \hat P_{\hat \delta}(k_2) \mathcal K(k_1,k_2,k_3) +\text{2 perm.}\right]\,,
\end{equation}
where $\mathcal K$ is the kernel from Eq.~\eqref{eq:second-order-stuff}. 

In Fig.~\ref{fig:Bk_at_all_z}, we plot the bispectra of the relativistic simulations in the left panels, and of the Newtonian simulations in the right panels. The top panels show an increasingly squeezed isosceles configuration where one mode is fixed to $4\,k_\mathrm{f} = 3.2\times 10^{-3} \; h/$Mpc and the two others are equal and plotted on the horizontal axis. The first point corresponds to a folded configuration with $k=2k_{\rm f}$, the third point to an equilateral configuration with all modes equal to $4\,k_\mathrm{f}$, and the last point on the right is a maximally squeezed configuration. The lower panels show an equilateral configuration where all modes are equal. Similarly to Fig.~\ref{fig:Pk_at_all_z}, each panel contains the measurements for different redshifts, with increasing amplitude for decreasing redshift. From the lowest to the largest bispectrum (blue, orange, green, red, violet, and brown), we see the bispectra for the \ac{IC} at $z=100$, and for the redshifts $z=90, 70, 50, 30, 10, 5, 1.5, 0.5$. For a cleaner plot, we have binned the $k$-range in a linear way with $4$ modes per bin for the modes larger than $2.5\times 10^{-2}\; h/$Mpc. Shown as black curves, the theoretical predictions for the bispectrum follow Eq.~\eqref{eq:semi-th}. 

In all cases, a good agreement is observed between the theory and the measurements from the \ac{IC} to redshift $z=0.5$. The relativistic simulation in the squeezed limit seems to agree less at high redshift and close to the horizon as it changes sign. This is actually an artifact of the logarithmic axis. We can see that for super-horizon scales, all curves, including the \ac{IC}, agree up to redshift $50$. Again for the relativistic squeezed limit, the bispectrum amplitude seems to be slightly smaller than expected for the redshift range $70 \geq z \geq 10$. Similarly to the power spectrum, the measured power catches up with the theory at lower redshift $z \leq 10$. In the equilateral configuration, we observe a nonlinear growth at low redshift and small scales which is not as significant in the squeezed limit.

In Fig.~\ref{fig:fct_of_a}, we present in the middle and right panels the time evolution of each mode of the bispectrum, respectively in the squeezed limit and in the equilateral configuration. Similarly to Eq.~\eqref{eq:fit_power}, we fit for each mode the linear relation 

\begin{equation}\label{eq:fit_bispectre}
\ln{B(a; k_1, k_2, k_3)} = A_{\mathrm{B}}(k_1, k_2, k_3) \ln{a} + C_{\mathrm{B}}(k_1, k_2, k_3)\,,    
\end{equation}
    where $A_\mathrm{B}$ is the logarithmic growth rate of the bispectrum and $C_{\mathrm{B}}$ is again a constant in time. The theoretical curves are obtained by fitting the theoretical curves of Fig.~\ref{fig:Bk_at_all_z}. Since the agreement between the simulation and the theory holds until $z=0.5$, except for the smallest scales in the equilateral configuration, we choose to fit the relation \eqref{eq:fit_bispectre} between $z=100$ and $z=0.5$. The parameter $A_\mathrm{B}$ represents the average logarithmic growth between these redshifts.

The Newtonian logarithmic growth rate ($B \propto a^4$ except for small scales) is recovered well in both simulations in the equilateral configurations and in the squeezed configuration for the Newtonian simulation. Except for the small-scale equilateral configuration, we have a good agreement between the simulation and the theory.

Given the discussion of Section~\ref{sub:REIC} and the relation \eqref{eq:semi-th}, in the super-horizon limit and for the horizon-scale equilateral configuration, one expects the relativistic bispectrum to grow like $a^0$ and $a^1$, respectively. This limit is recovered well in the relativistic simulations. In both the squeezed and the equilateral configurations, we observe a larger logarithmic growth rate than the Newtonian one around scale $5\times 10^{-3}\; h/$Mpc confirmed by the theory curves. The squeezed limit is obtained by coupling a large relativistic mode with two small Newtonian modes. We therefore expect to converge to $a^3$, which is recovered well in the middle panel of Fig~\ref{fig:fct_of_a}.

\subsection{Number counts}\label{sub:nbc}
For all our simulations, light cones containing approximately half a billion particles are produced with respect to an observer in the center of the box, observing a full sky with a radius $3900$ Mpc$/h$, i.e.\ close to half of the box size. This corresponds to a full-sky survey from redshift $z=0$ to $z=2.3$ and the number of observed objects is roughly commensurate with the size of large galaxy catalogues in this redshift window, although we remind the reader that we use unbiased tracers of matter for the purpose of this work. After ray tracing, we compute the particle number counts following Eq.~\eqref{eq:numbercount} and using a \texttt{Healpix} \cite{Gorski:2004by} pixelisation of the sky. We recall here that the number count is observed on the past light cone, and therefore does not depend on the gauge choice made for the relativistic simulation. As explained in Section~\ref{subsub:bisp}, we run ``paired'' simulations to obtain better measurements of the ensemble mean of the bispectrum. We run $N_{\rm r}=10$ paired realisations for each case (relativistic and Newtonian), i.e.\ $40$ simulations in total. Thus, defining $X_+$ and $X_-$ as two paired quantities (power spectrum or bispectrum in our case), the mean $\hat X$ reads
\begin{equation}\label{eq:mean}
    \hat X = \frac{1}{N_{\rm r}} \left(\sum_{i=1}^{N_{\rm r}} \frac{X_{i,+} + X_{i,-} }{2} \right) \,.
\end{equation}
Note that pairing and averaging are two operations that commute. However, this is not the case for the standard deviation. In this article, we are going to discuss two types of error bars. First, the ``observational'' error bars: the standard deviation $\sigma_\mathrm{STD}$ of the realisations that ignore pairing:
\begin{equation}\label{eq:std}
    \sigma_\mathrm{STD}^2 = \frac{1}{2 N_{\rm r}} \sum_{i=1}^{N_{\rm r}} \left(X_{i,+} -  \hat X \right)^2 + \frac{1}{2N_{\rm r}} \sum_{i=1}^{N_{\rm r}} \left(X_{i,-} - \hat X\right)^2 \,.
\end{equation}
For $N_\mathrm{r} \rightarrow \infty$, $\sigma_\mathrm{STD}$ is a measure of the cosmic variance, which means that it quantifies our ability to detect $X$ in real observations. However, thanks to the pairing and the multiple realisations, our knowledge of the mean in our simulations is much more accurate than the one estimated with Eq.\,\eqref{eq:std}. The accuracy in this case is given by the standard error on the mean, $\sigma_\mathrm{STE}$, after the pairing operation. It reads
\begin{equation}\label{eq:ste}
    \sigma_\mathrm{STE}^2 = \frac{1}{(N_{\rm r}-1) N_{\rm r}}  \sum_{i=1}^{N_{\rm r}} \left(\left[\frac{X_{i,+}+X_{i,-}}{2}\right] - \hat X\right)^2  \,.
\end{equation}
Thus, even in cases where we do not expect a detection in real observations because of the cosmic variance given by Eq.\,\eqref{eq:std}, the fact that we do have an accurate determination after pairing and averaging over ten realisations allows us to discuss the theoretical shape of the relativistic power spectrum and bispectrum.

\subsubsection{Angular power spectrum}\label{subsub:angle_ps}

\begin{figure}
    \centering
    \includegraphics[scale=0.4]{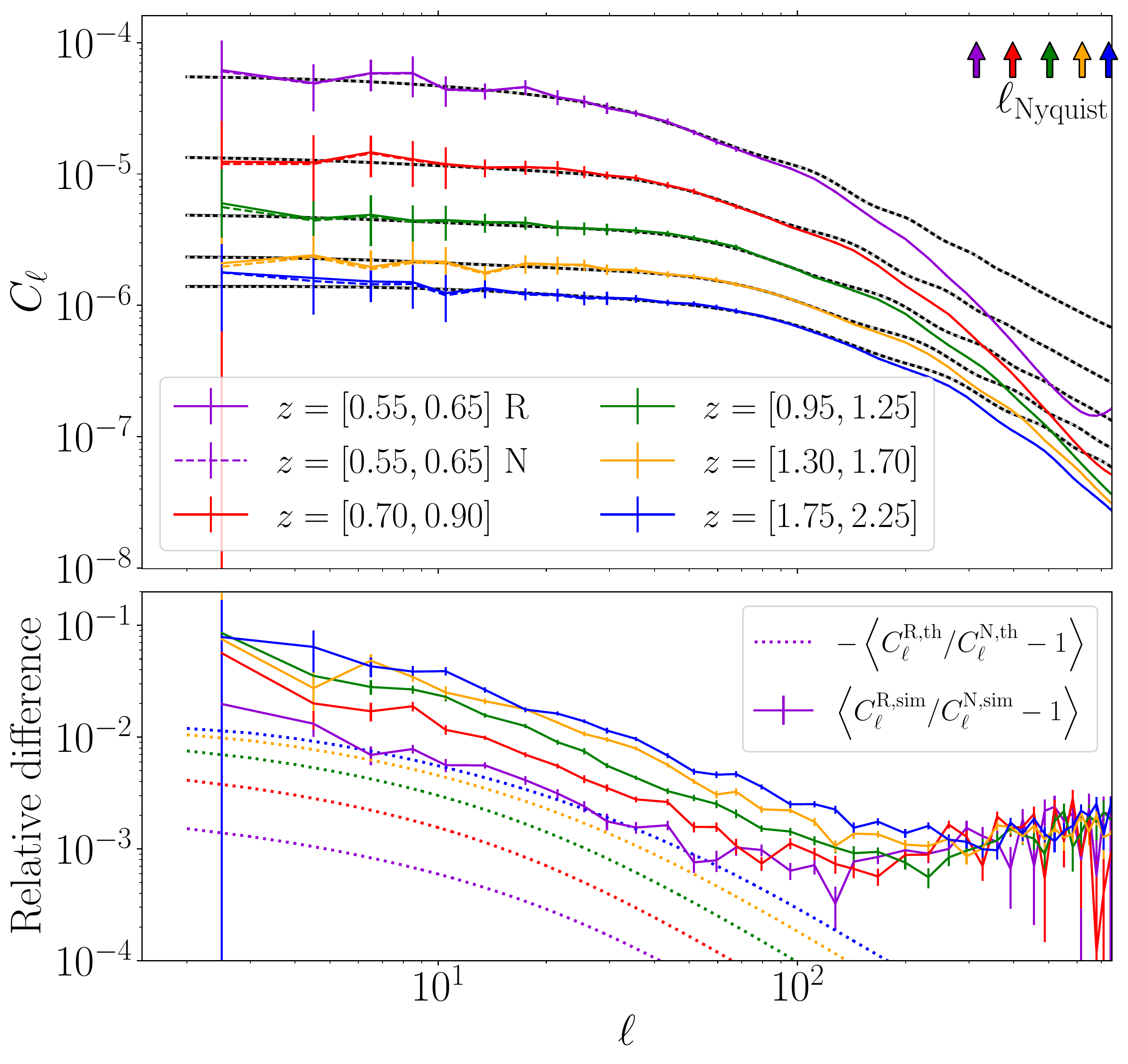}
    \caption{In the top panel, we plot in black solid lines the linear relativistic theoretical prediction with the Newtonian prediction in dotted grey lines superimposed (the power increases with time). In coloured lines, we plot the averaged angular power spectra of the \ac{CDM} particle number count for different redshift bins as indicated by the legend. The power spectra are binned for readability and to increase the signal-to-noise. The error bars are the result of the combination of the cosmic variance $\sigma_{\rm STD}$, see Eq.~\eqref{eq:std}, of different $\ell$'s in a bin. The dashed (solid) lines correspond to the Newtonian (relativistic) cases. 
    %In the bottom panel we plot in solid (dotted) lines the relative difference between the relativistic and the Newtonian simulations (theories). 
    In the bottom panel we plot in solid lines the relative difference between the relativistic and the Newtonian simulations.
    By contrast, the dotted lines in the bottom panel show the relative difference between general relativity and Newtonian theory; this difference is due to an inconsistent interpretation of particle position and gauge choice within the (too simplistic) Newtonian picture.
    The error bars in the bottom panel involve only the standard error on the mean $\sigma_{\rm STE}$ (since cosmic variance is not applicable for actual observations). Theoretical predictions are computed with \texttt{CLASS} at first order. The arrows indicate the multipoles corresponding to the Nyquist frequency for each redshift. }
    \label{fig:Cl}
\end{figure}

Similarly to Ref.~\cite{Lepori:2021lck}, we compare the nonlinear angular power spectra of our simulations with the linear prediction of the number-count angular power spectrum provided by \texttt{CLASS}. For this purpose, we use the same top-hat filter as used to construct the maps of the number counts. The linear bias and magnification bias are set to $1$ and $0$, respectively, consistent with the fact that we consider matter particles directly and that the survey has no detection threshold. In Fig.~\ref{fig:Cl}, we plot the ensemble mean of the angular power spectra for five redshift bins. The dashed (solid) lines correspond to the Newtonian (relativistic) simulations. Note that the dashed and solid lines are almost perfectly aligned. Also, we have binned the power spectra to increase the signal-to-noise and for readability. The error bars show the combination of the cosmic variance for different $\ell$'s in a bin, such that the error of a bin reads
\begin{equation}
 \sigma_{\rm bin} = \frac{1}{N_{\rm bin}\sqrt{N_{\rm bin}-1}} \sum_{i=1}^{N_{\rm bin}} \sigma^i_{\rm STD} \,,
\end{equation}
where $N_{\rm bin}$ is the number of $\ell$'s in the considered bin and where $\sigma^i_{\rm STD}$ is the cosmic variance defined in Eq.~\eqref{eq:std} of the $i^{\rm th}$ $\ell$. The black lines are the result of \texttt{CLASS} which are relativistic and accurate to first order.

As explained in Section~\ref{sec:numbercount}, we use the relativistic nonlinear ray tracer for all simulations. Hence, even in the Newtonian case, the ray tracer assumes that the positions of the particles are in Poisson gauge, as is common practice for lensing calculations. As mentioned earlier, at very large scales this introduces a subtle gauge artifact. To account for this error, we swap out the spatially flat gauge density fluctuation for the comoving gauge density fluctuation in \texttt{CLASS}, see Eqs.~(2.7) and (2.12) in Ref.~\cite{DiDio:2013bqa}. The result, called ``Newtonian theory'' is shown in Fig.~\ref{fig:Cl} as dotted grey lines. Similarly as with the simulation results, the linear Newtonian (dotted grey lines) and relativistic predictions (black solid lines) are almost perfectly aligned. The coloured arrows in Fig.~\ref{fig:Cl} indicate the multipoles corresponding to the Nyquist frequency of the simulation mesh projected on the light cone for each redshift which are estimated as follows. The transverse distance $L$ corresponding to the angle $\theta$ at radial distance $r$ is $L = \theta r$. By using the relations $\ell \simeq \pi/\theta$ and $k = 2\pi/L$ we have
\begin{equation}\label{eq:lNyquist}
    \ell(z) \simeq \frac{r(z) {k}}{2}\,.
\end{equation}
To estimate the ``Nyquist multipole'' $\ell_{\mathrm{Nyquist}}(z)$, we substitute $k$ by the Nyquist frequency $k_\mathrm{f} \times N/2$. For simplicity, we use the center of the redshift bins in Eq.~\eqref{eq:lNyquist}.

As already shown in Ref.~\cite{Lepori:2021lck}, the nonlinear ray tracer applied to \texttt{gevolution} recovers the linear theory of \texttt{CLASS} at large physical scales. Given our resolution, our simulations are compatible with the theory for $\ell< 100$. For $\ell > 100$, the multipole modes start to be sensitive to the finite resolution of the simulations, which causes a lack of power in the angular power spectrum. Note that resolution effects cancel out at leading order when we look at the ratio of two spectra, which means that ratios remain significantly more robust as one approaches the Nyquist scale. Results for the multipoles beyond the Nyquist scale, especially for the lowest redshift which has the smallest $\ell_{\rm Nyquist}$, should be discarded.

In the bottom panel of Fig.~\ref{fig:Cl}, we show in coloured solid lines the relative difference between the Newtonian and the relativistic simulations. The error bars in the bottom panel only include the contribution from the standard error on the mean $\sigma_{\rm STE}$, since cosmic variance is not applicable to this measurement which cannot be replicated in real observations. As we can see, except for the first bin, these relative differences are well-detected thanks to the combination of multiple realisations. We observe an increasing difference as we go to small $\ell$. This difference scales like $\ell^{-1}$. From  multipoles $\ell=1$ to $\ell=10$, we observe an effect between $10\%$, for the smallest $\ell$, and $1\%$ for $\ell \gtrsim 10$. As expected, the effect decreases with decreasing redshift, since a fixed angular scale corresponds to ever smaller physical distance scales. The observed relative difference is a combination of \ac{RE} and of the fact that the ray tracer finds the particle positions in the wrong gauge when using the Newtonian simulations. To estimate the amplitude of this second effect, we plot in dotted lines the relative difference between the relativistic and the Newtonian theories. As we can see, the error induced by the gauge artifact is roughly one order of magnitude smaller than the full relative difference between the simulations. Moreover, its sign is reversed. We can therefore argue that, up to an error of $10\%$, the relative difference between the relativistic and the Newtonian simulations is dominated by the second-order dynamical \ac{RE}, with a scaling of $\ell^{-1}$. However, the impact of \ac{RE} remains very small, well below cosmic variance for each individual multipole. Combining many multipoles across many different maps in a tomographic measurement could conceivably raise the signal to noise above the detection threshold, but we leave such an analysis to future work.

\begin{figure}
    \centering
    \includegraphics[scale=0.38]{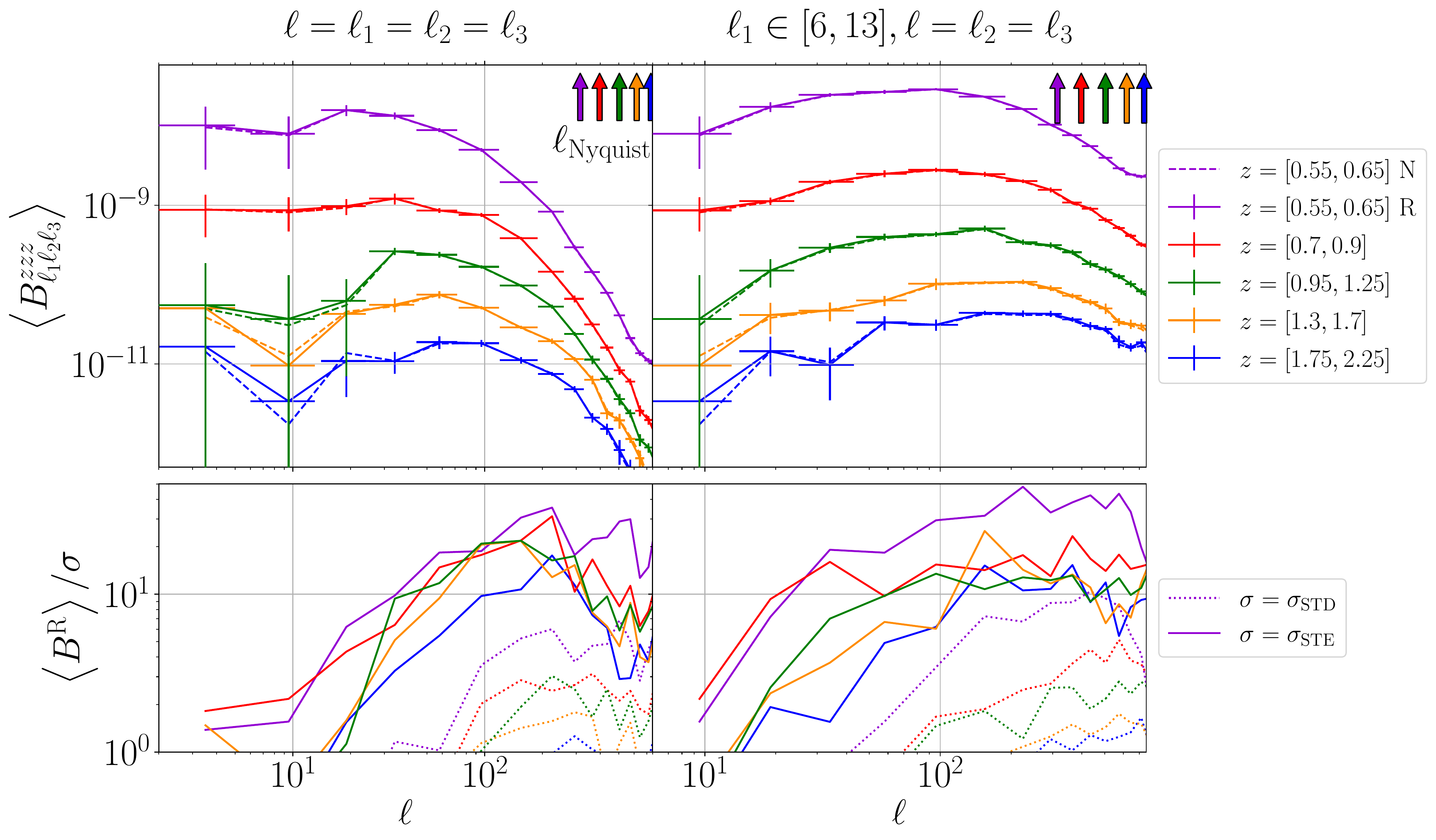}
    \caption{The first row shows the auto-bispectrum $(z = z_1 = z_2 = z_3)$ in two different configurations. In the left panels, we show %the auto-bispectrum $(z = z_1 = z_2 = z_3)$ in 
    the equilateral configuration $\ell_1 = \ell_2 = \ell_3$ for the $5$ same redshift bins as in Fig.~\ref{fig:Cl}. In the right panels, we plot the squeezed limit by fixing one of the multipole bins to $[6, 13]$ and by varying $\ell=\ell_2 = \ell_3$. Solid and dashed lines, even though they are almost exactly superposed, show the relativistic and Newtonian bispectra, respectively. The error bars show $\sigma_{\rm STE}$ defined in Eq.~\eqref{eq:ste}.
    In the second row, we plot the auto-bispectrum divided by $\sigma_{\rm STE}$ or $\sigma_\mathrm{STD}$ as solid or dotted lines, respectively. 
    }
    \label{fig:Bl_equi}
\end{figure}

\begin{figure}
    \centering
    $S[ \left< B^{\rm N} \right>]$
\begin{tabular}{c c c}
      \includegraphics[width=50mm, trim={4cm 0 6.5cm 0},clip]{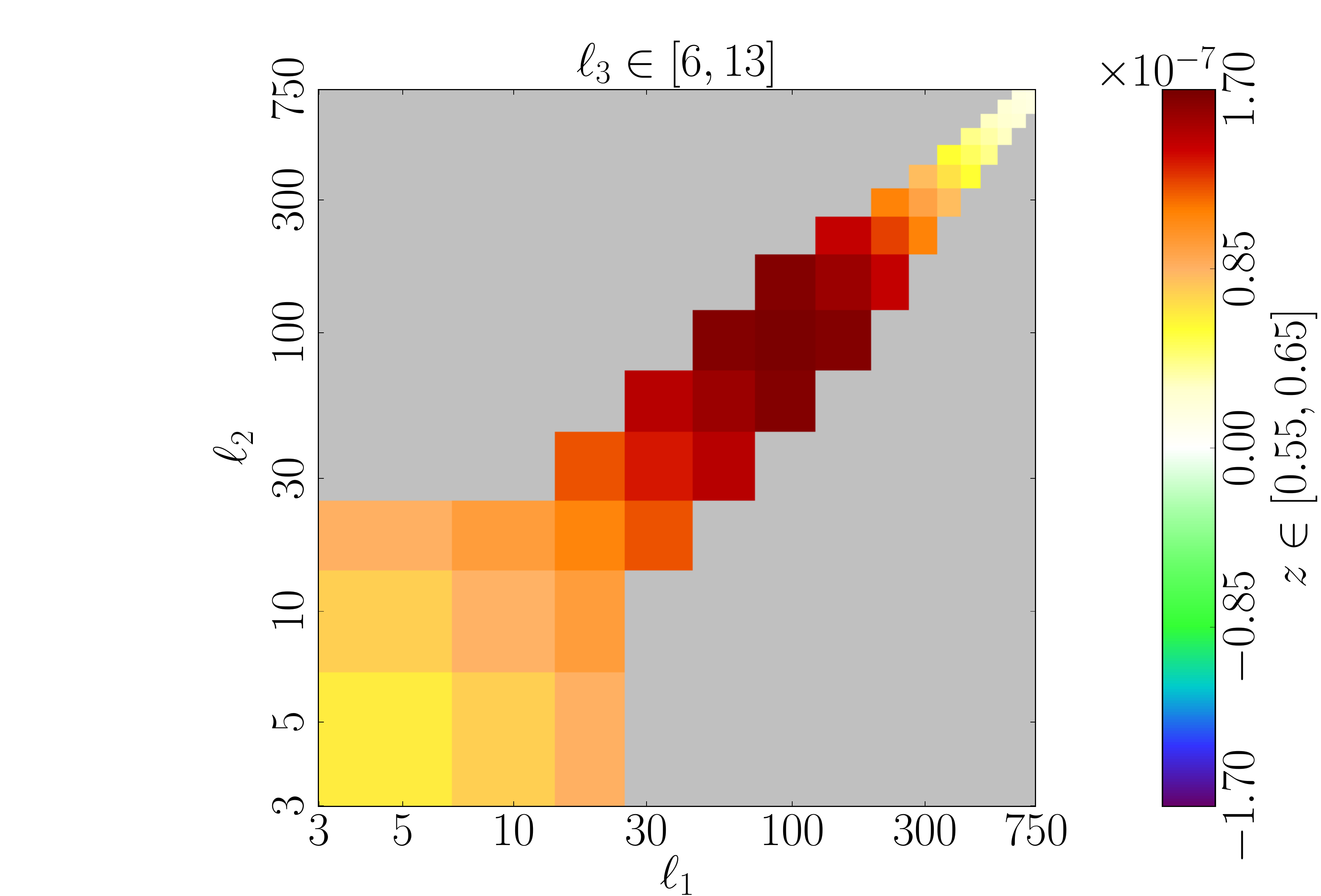} &
      \hspace{-0.45cm}\includegraphics[width=44mm, trim={7cm 0 6.5cm 0},clip]{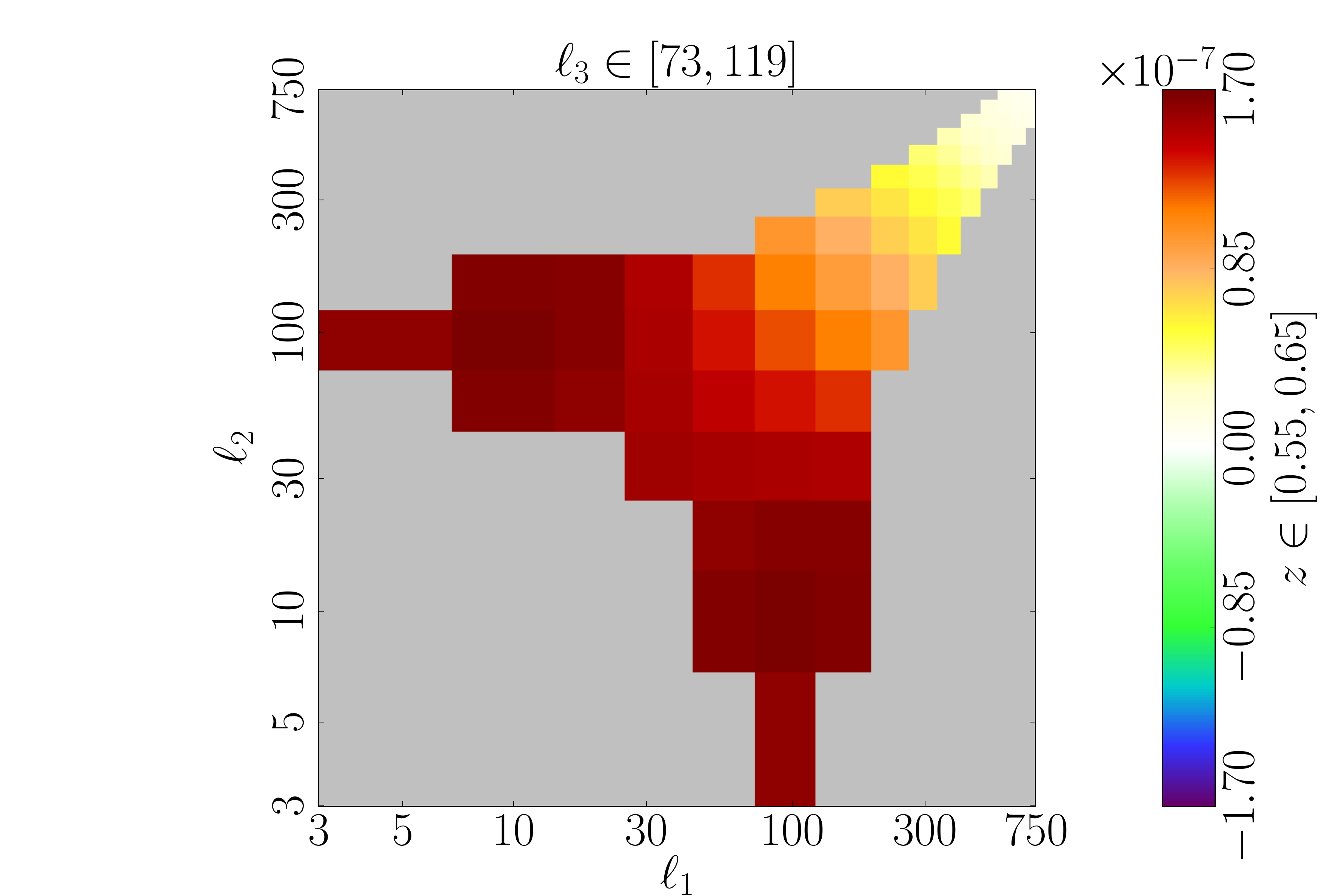} &
      \hspace{-0.45cm}\includegraphics[width=57mm, trim={7cm 0 0cm 0},clip]{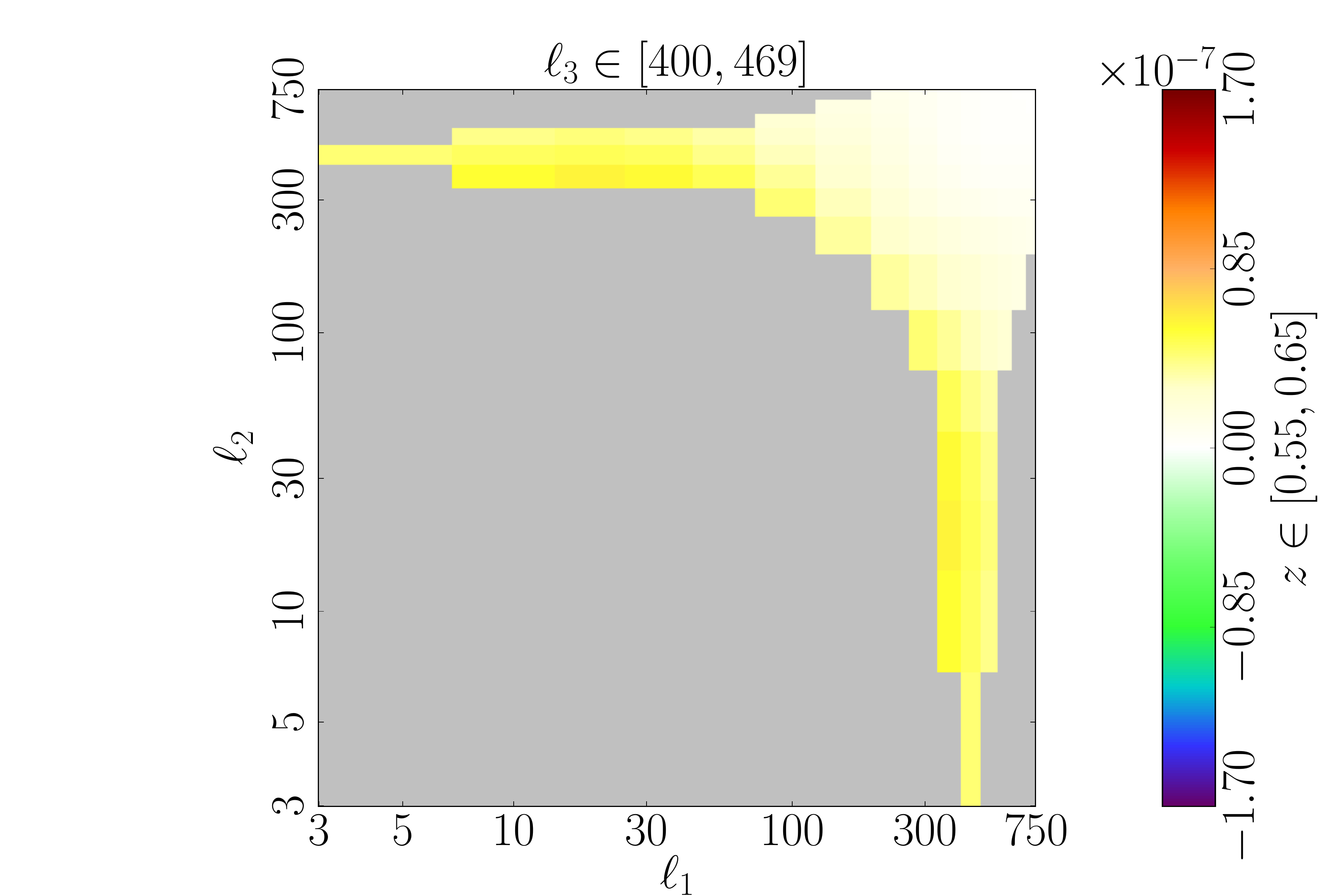}\\
      \includegraphics[width=50mm, trim={4cm 0 6.5cm 0},clip]{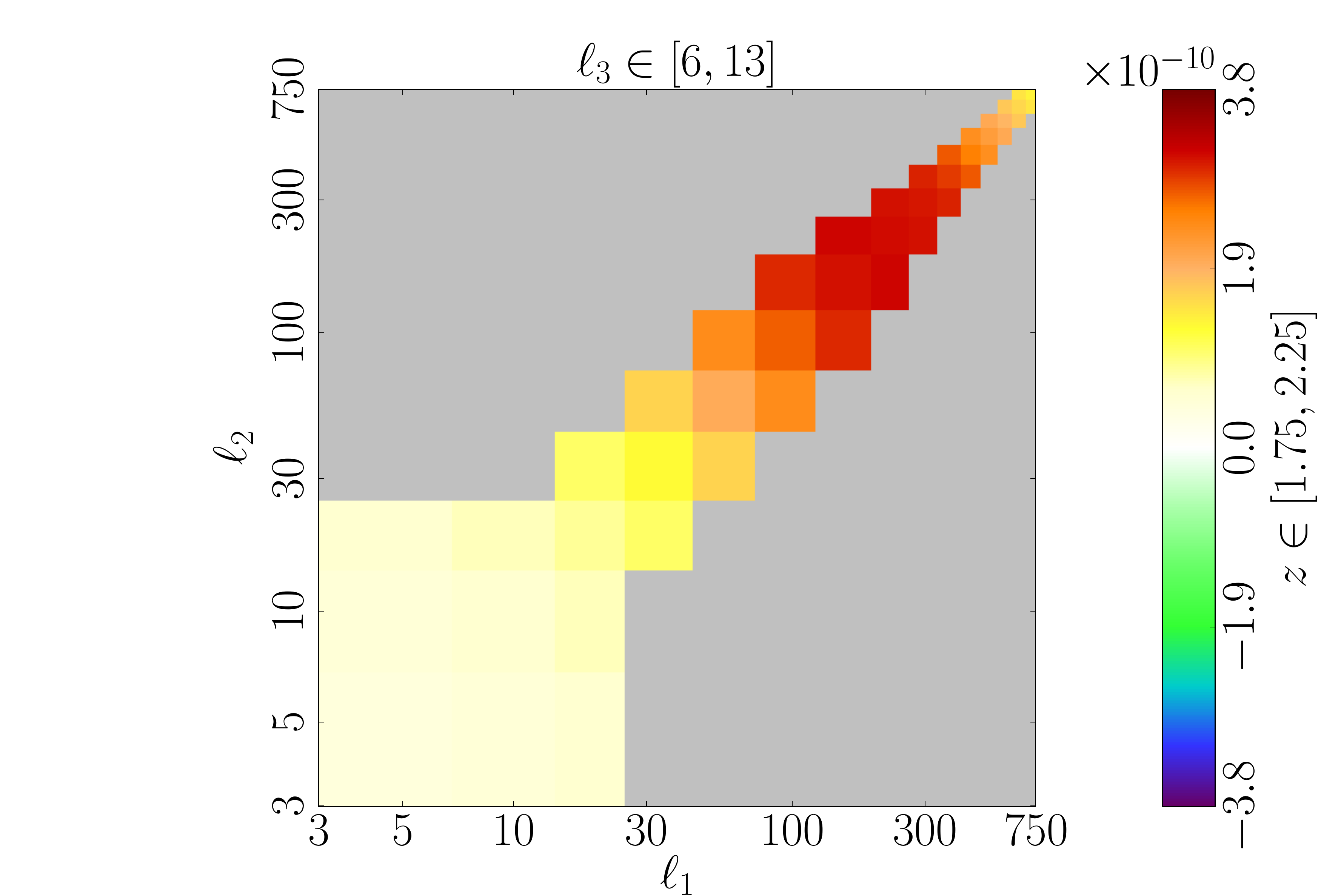} &
      \hspace{-0.45cm}\includegraphics[width=44mm, trim={7cm 0 6.5cm 0},clip]{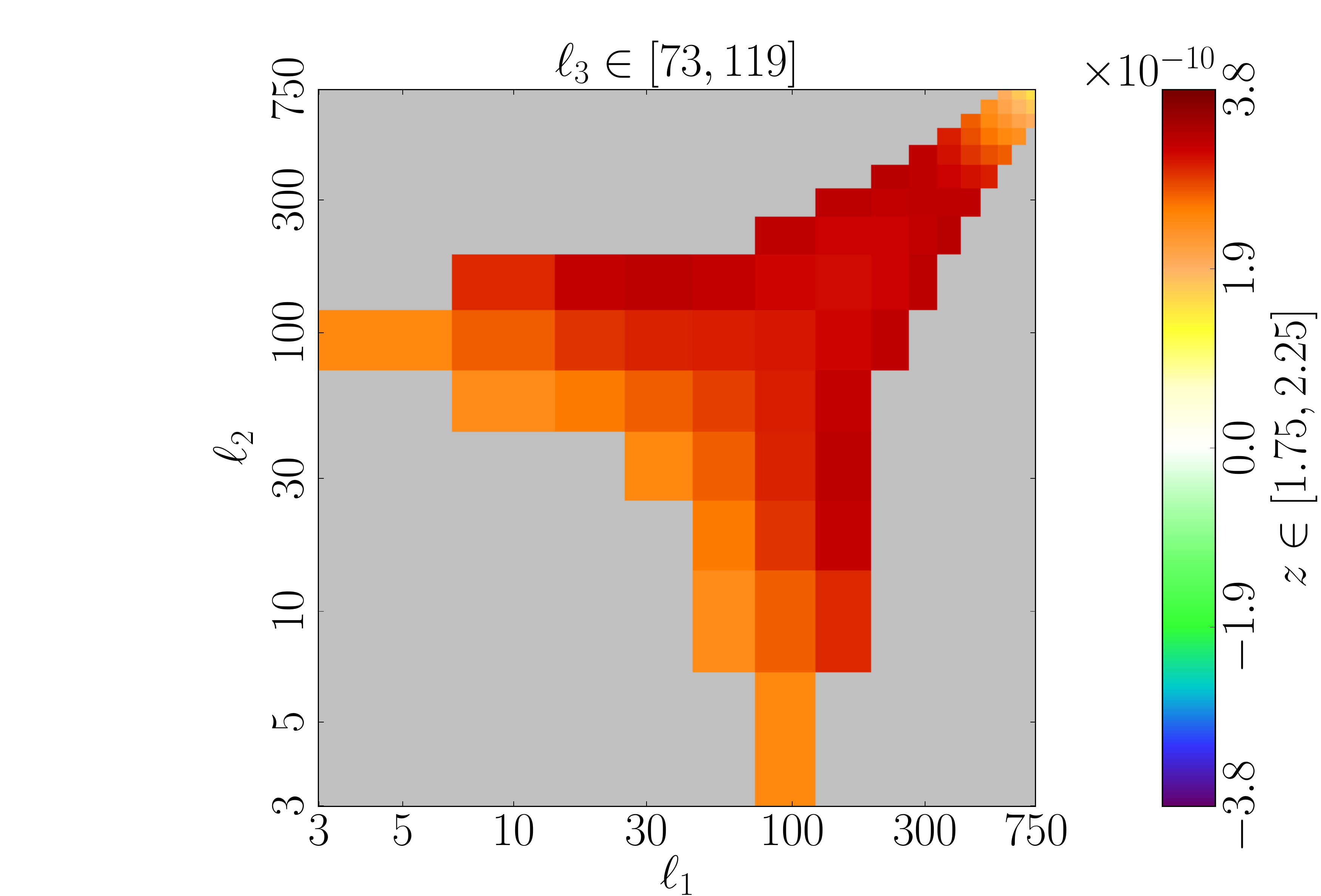} &
      \hspace{-0.45cm}\includegraphics[width=57mm, trim={7cm 0 0cm 0},clip]{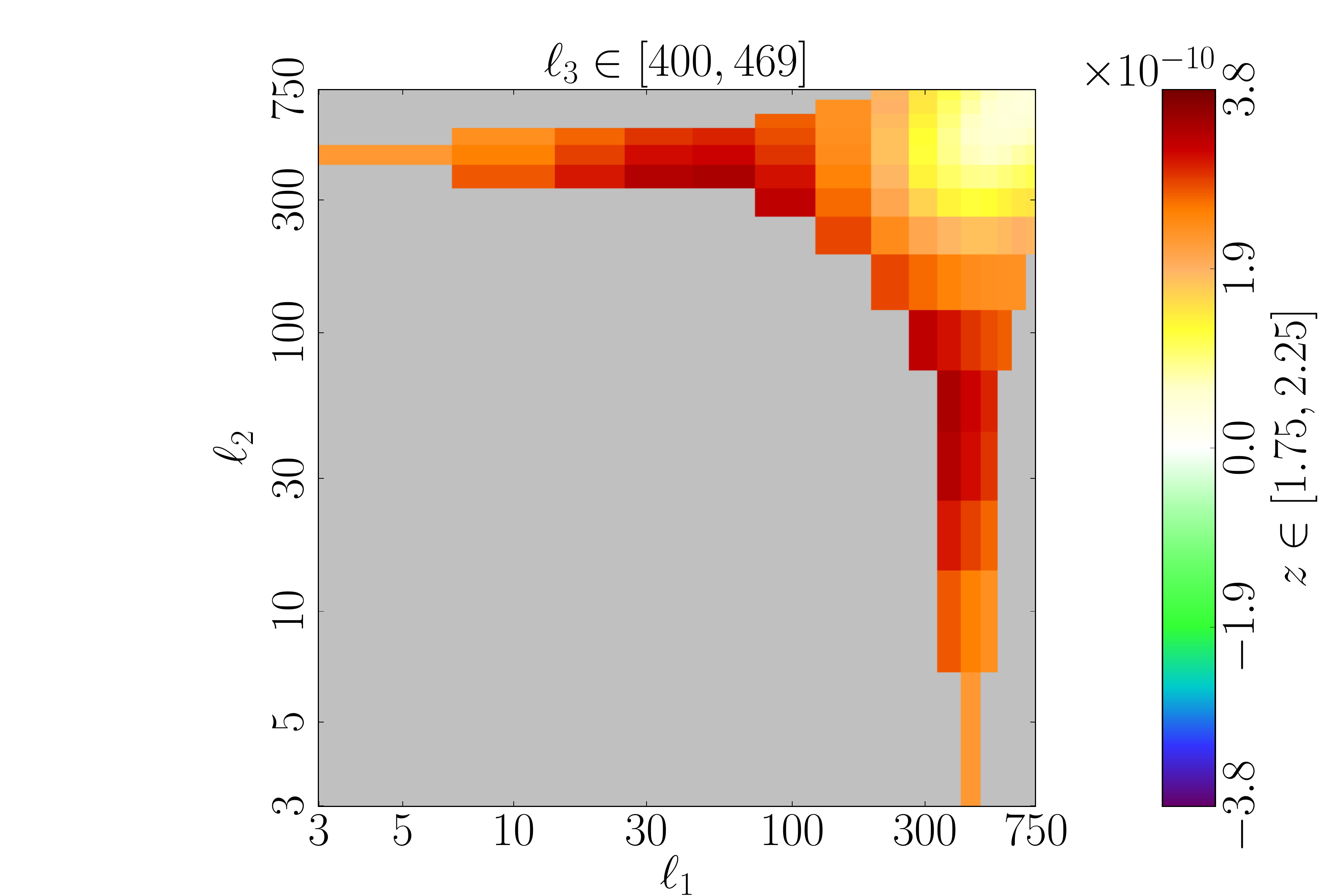}\\
\end{tabular}

    \caption{Newtonian bispectrum using the smoothing procedure~\eqref{eq:smooth}, averaged over the 10 simulations, as a function of $\ell_1$ and $\ell_2$. The different columns represent different values of $\ell_3$ as indicated. The first (second) row corresponds to the lowest (highest) redshift bin $z = [0.55, 0.65]$ ($z = [1.75, 2.25]$). Note the difference in colour scale used for the two rows. See Appendix~\ref{app:bispectre} for results at other redshifts.
    }
    \label{fig:Newtonian_bisp}
\end{figure}

\begin{figure}
    \centering
    $S[ \left< \mathcal B^{\rm N}_{\rm STD} \right>]$
\begin{tabular}{c c c}
      \includegraphics[width=50mm, trim={4cm 0 6.5cm 0},clip]{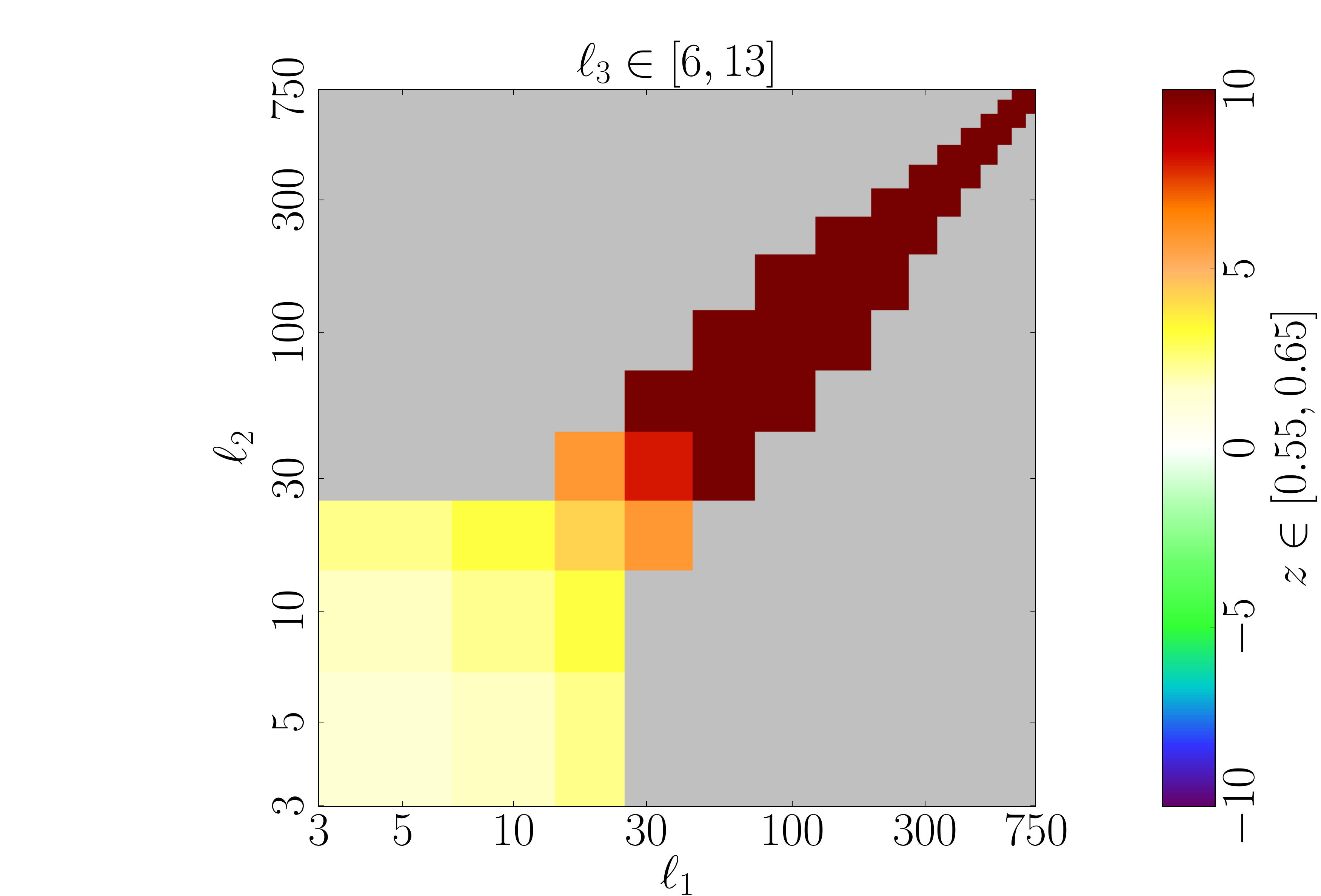} &
      \hspace{-0.45cm}\includegraphics[width=44mm, trim={7cm 0 6.5cm 0},clip]{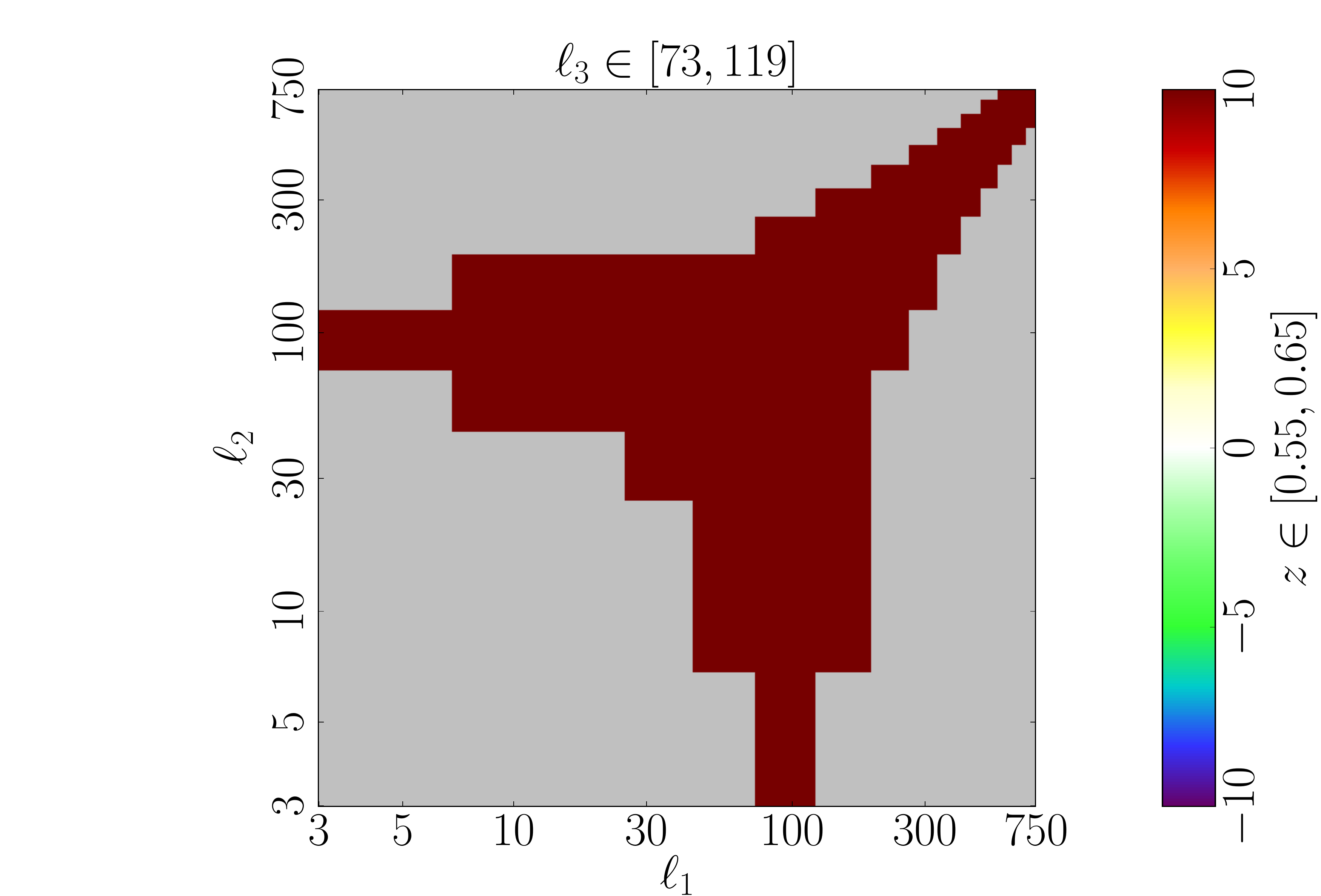} &
      \hspace{-0.45cm}\includegraphics[width=57mm, trim={7cm 0 0cm 0},clip]{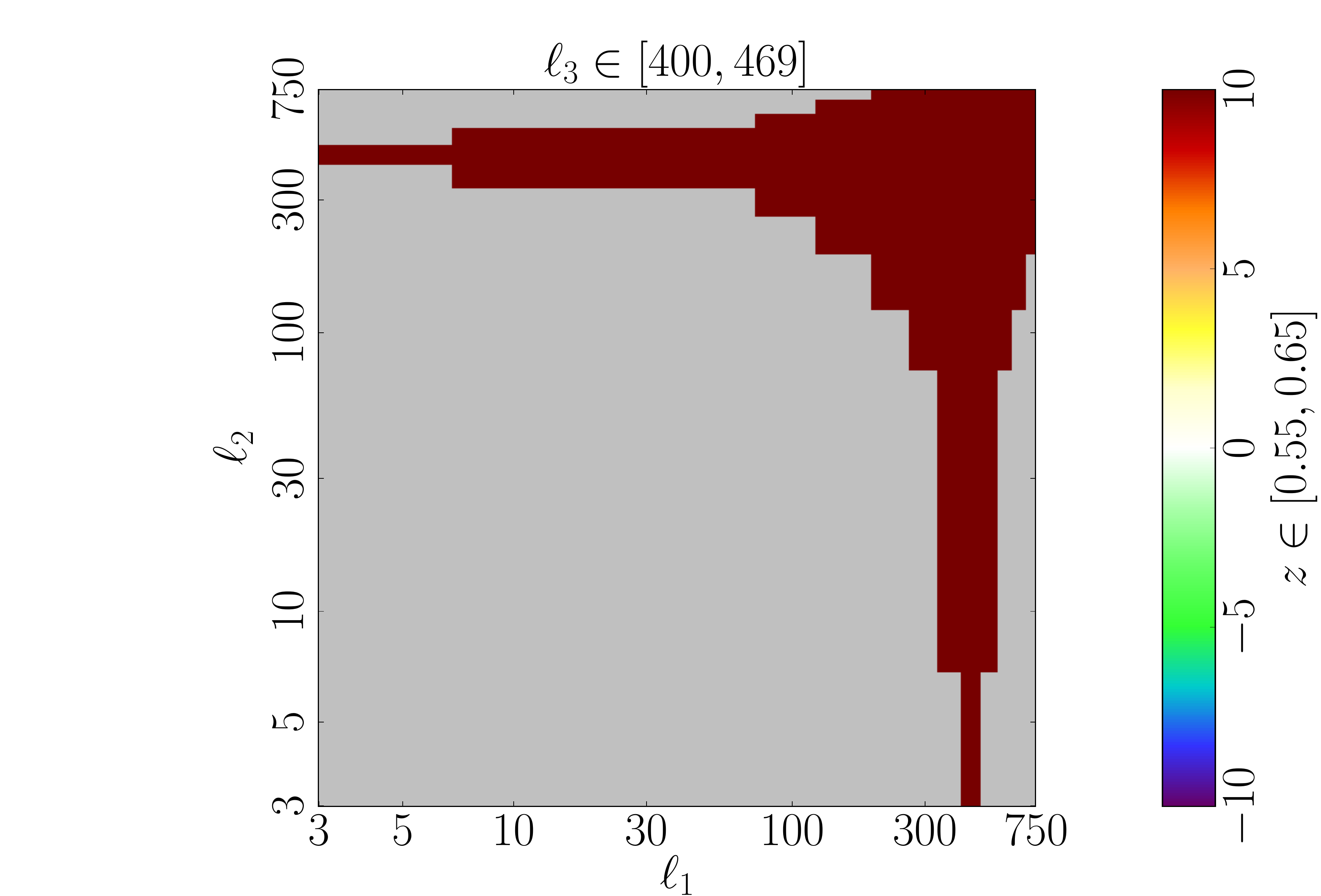}\\
      \includegraphics[width=50mm, trim={4cm 0 6.5cm 0},clip]{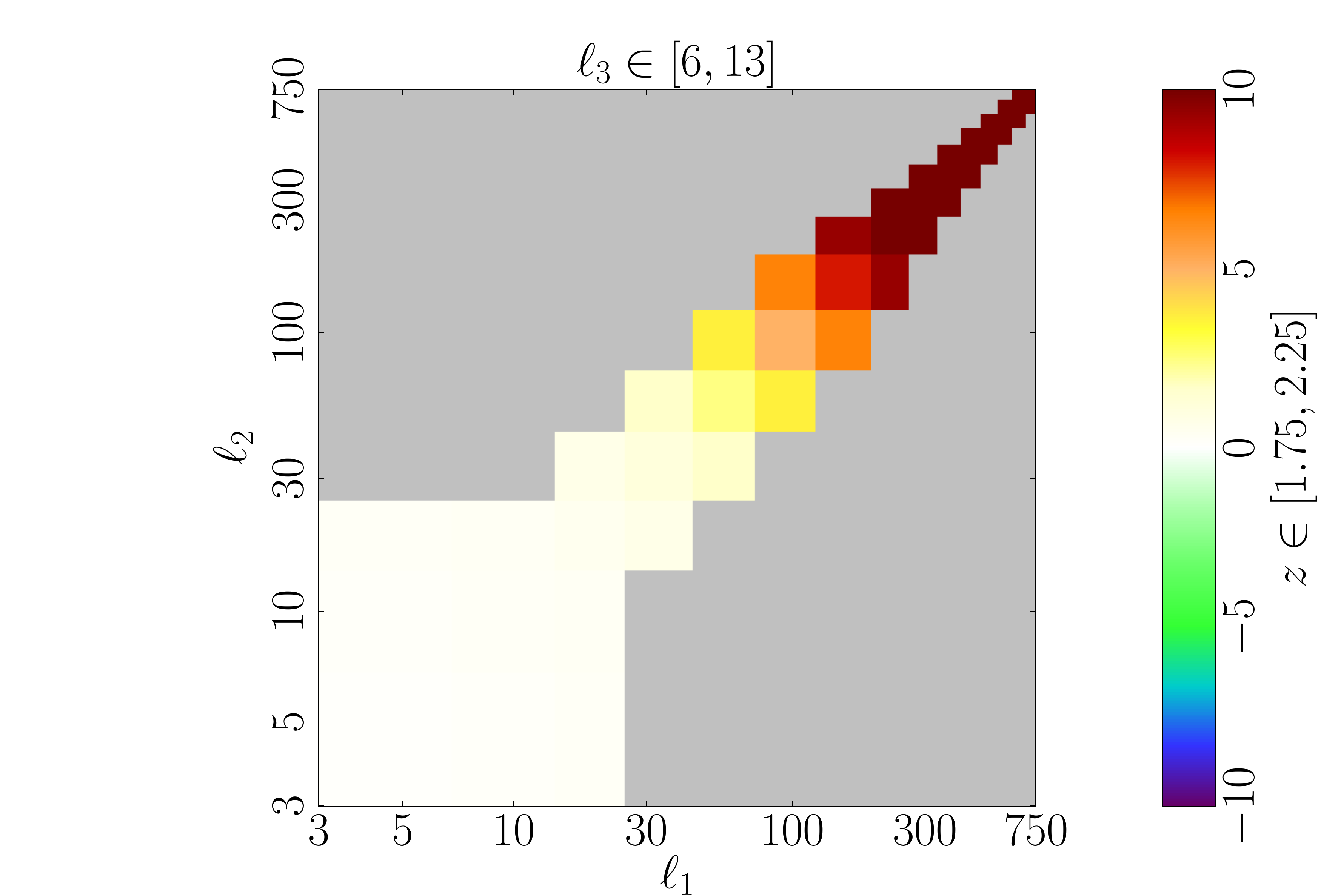} &
      \hspace{-0.45cm}\includegraphics[width=44mm, trim={7cm 0 6.5cm 0},clip]{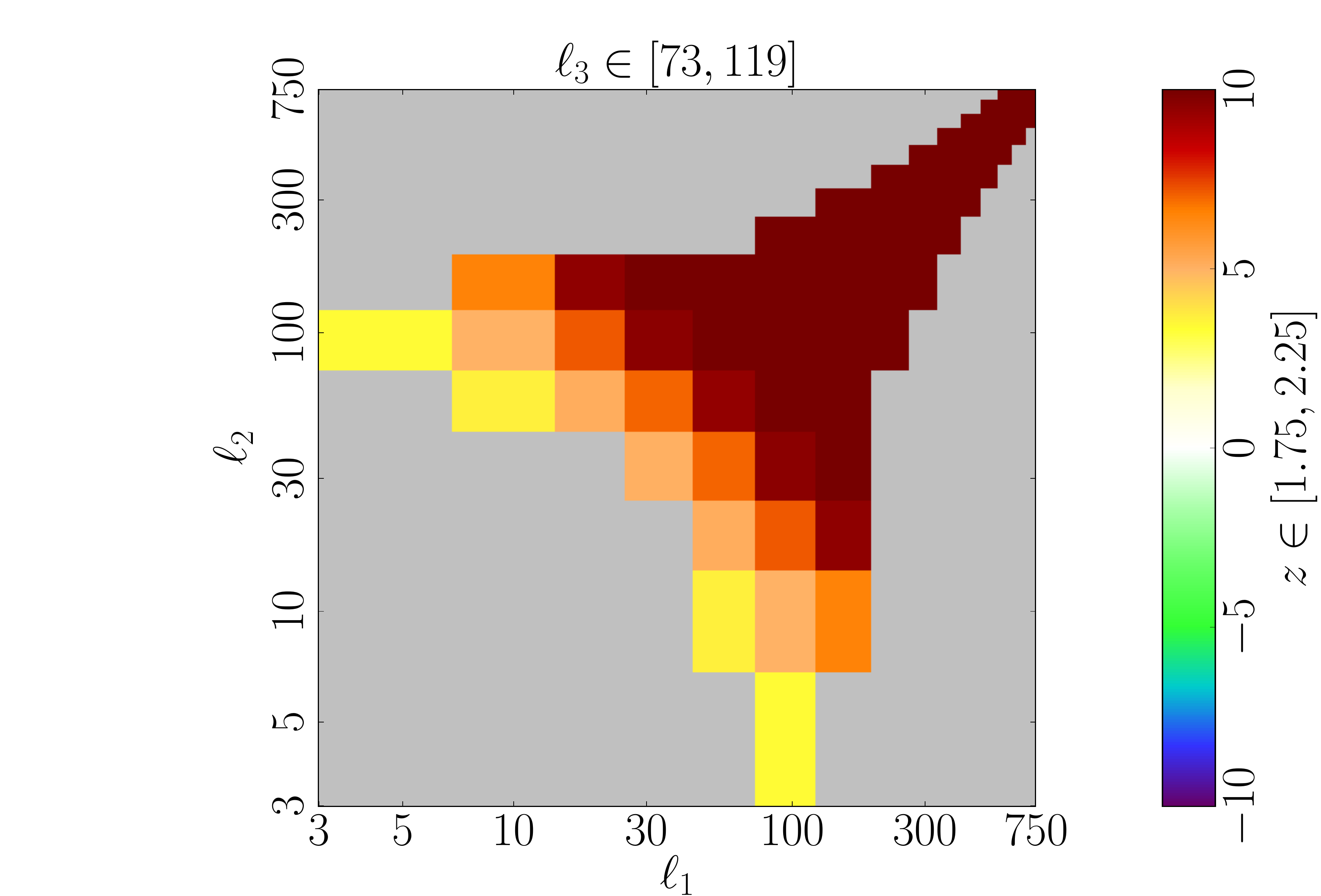} &
      \hspace{-0.45cm}\includegraphics[width=57mm, trim={7cm 0 0cm 0},clip]{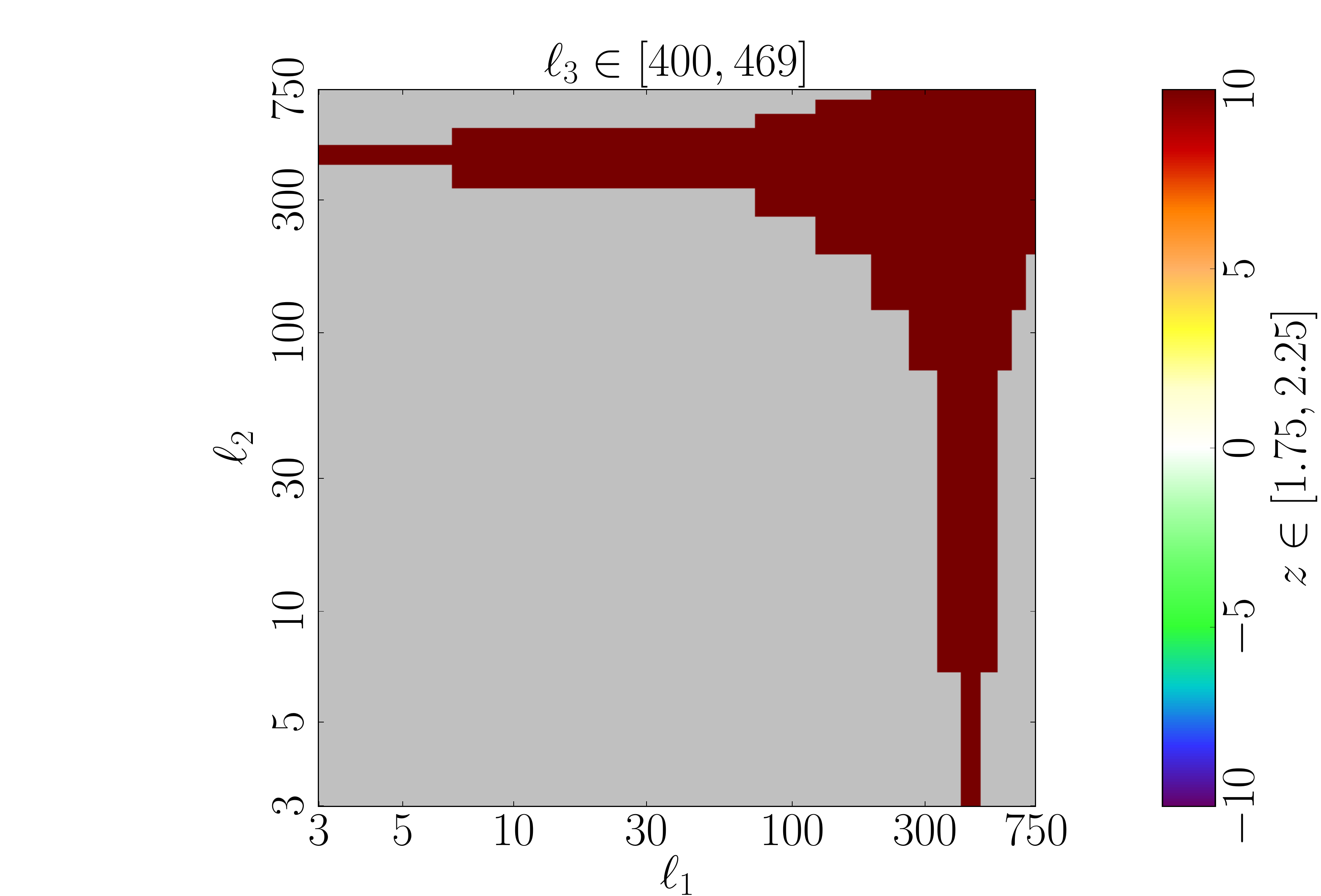}\\
\end{tabular}
    \caption{Similar to Fig.~\ref{fig:Newtonian_bisp}, but with the smoothed Newtonian bispectrum divided by $\sigma_{\rm STD}$.} 
    \label{fig:Newtonian_detection_bisp}
\end{figure}

\subsubsection{Angular bispectrum}\label{subsub:angle_bs}

\begin{figure}
    \centering
  $S \left[ \left<B^{\mathrm R - \mathrm N} \right>\right]$  
\begin{tabular}{c c c}
      \includegraphics[width=50mm, trim={4cm 0 6.5cm 0},clip]{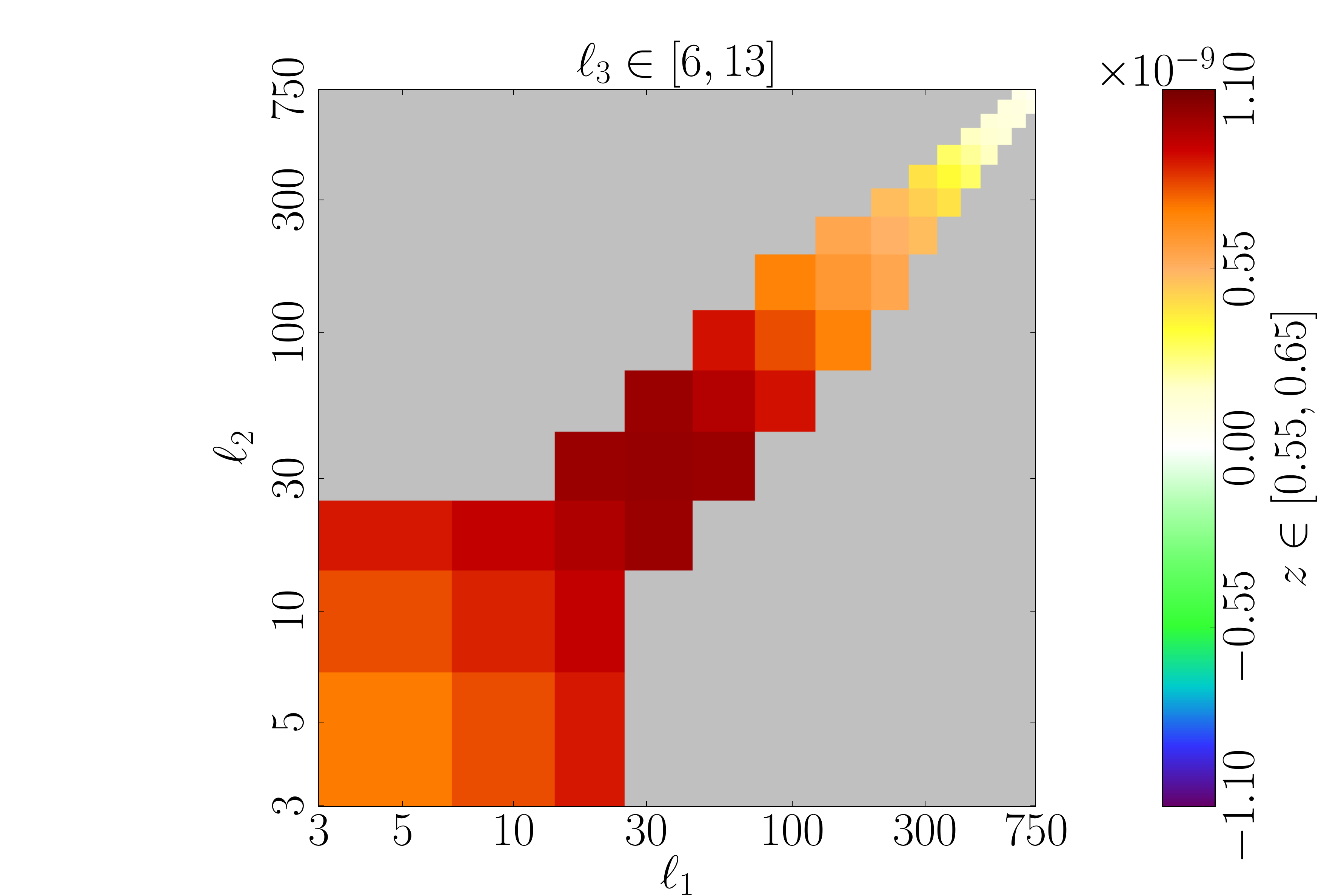} &
      \hspace{-0.45cm}\includegraphics[width=44mm, trim={7cm 0 6.5cm 0},clip]{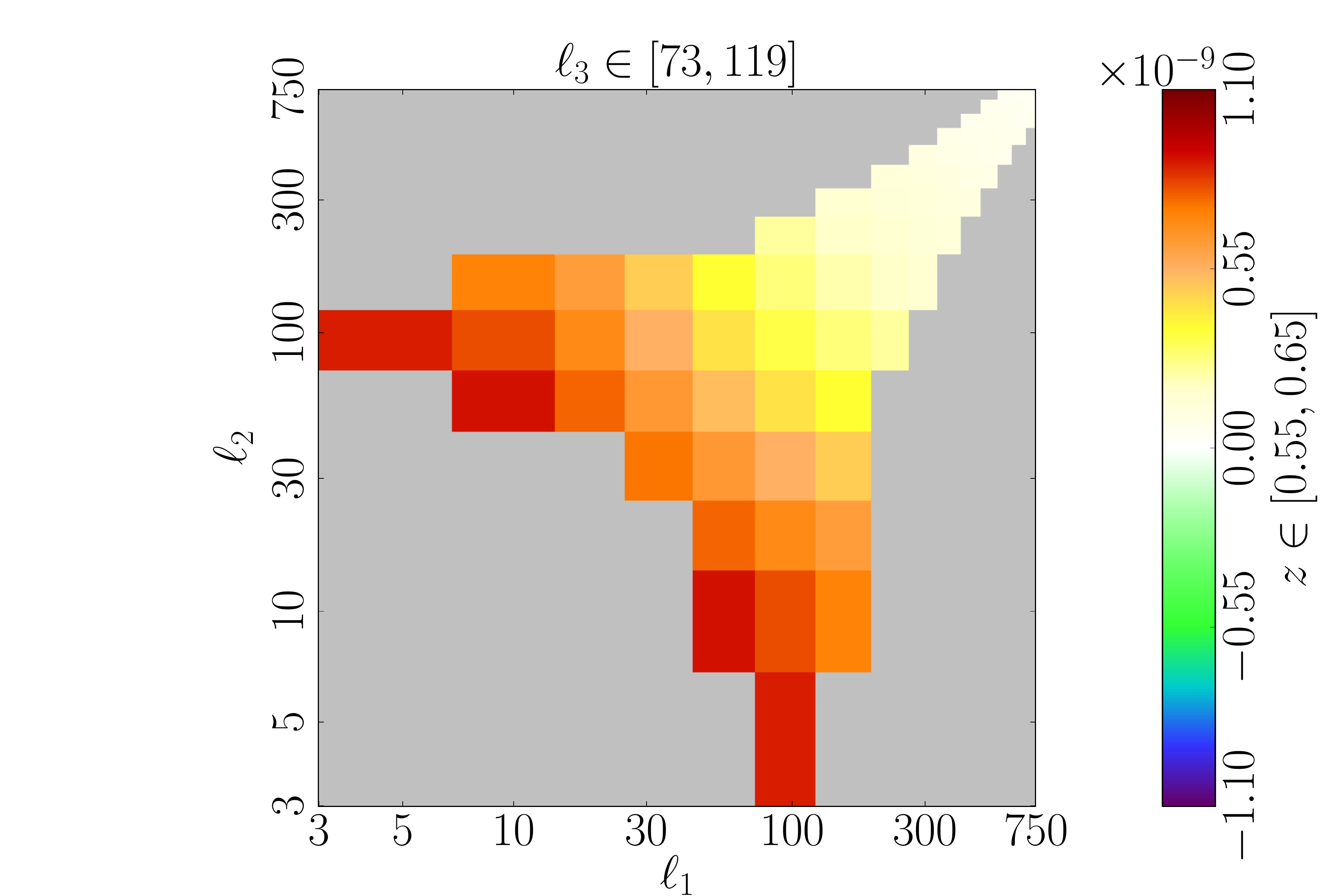} &
      \hspace{-0.45cm}\includegraphics[width=57mm, trim={7cm 0 0cm 0},clip]{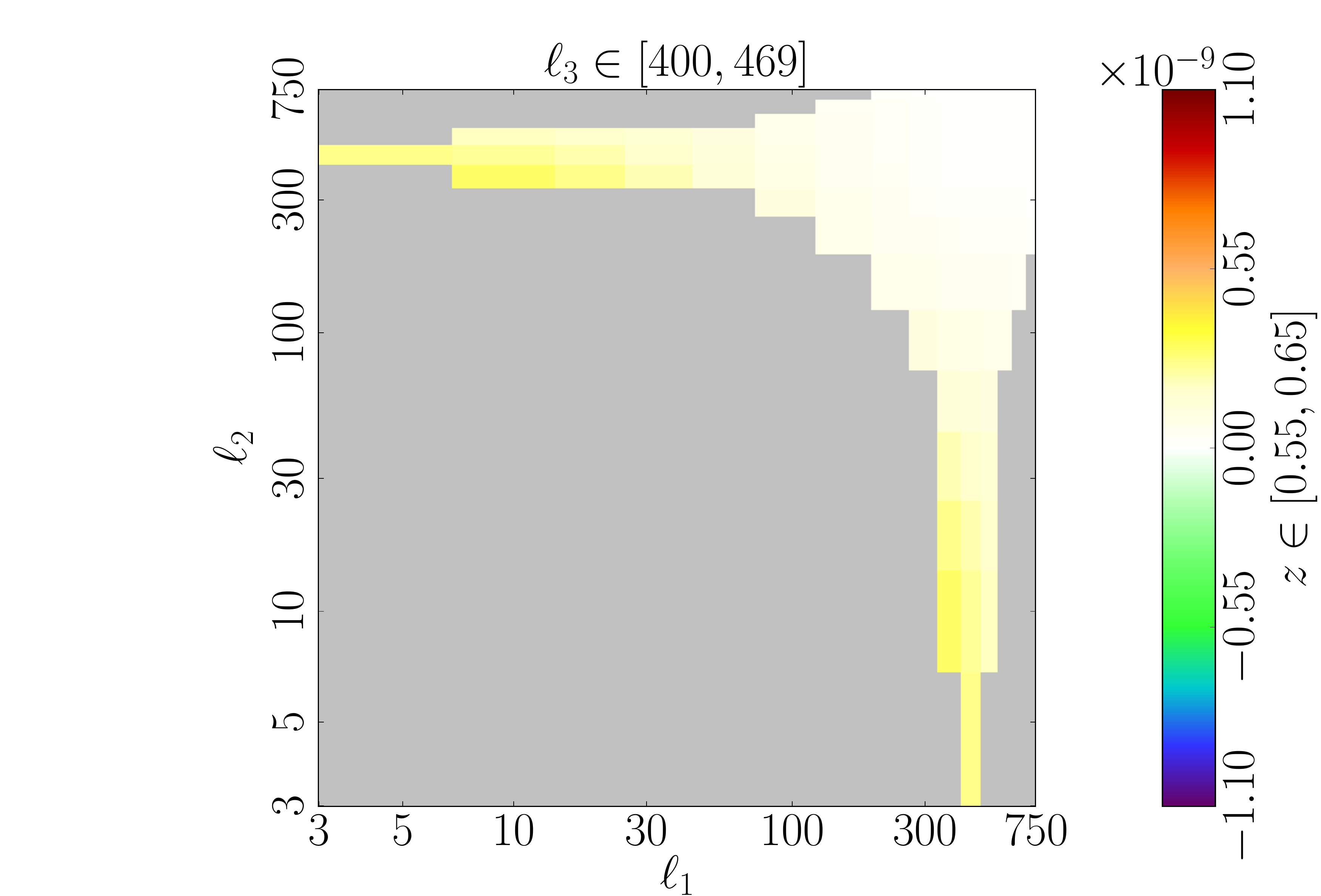}\\
      \includegraphics[width=50mm, trim={4cm 0 6.5cm 0},clip]{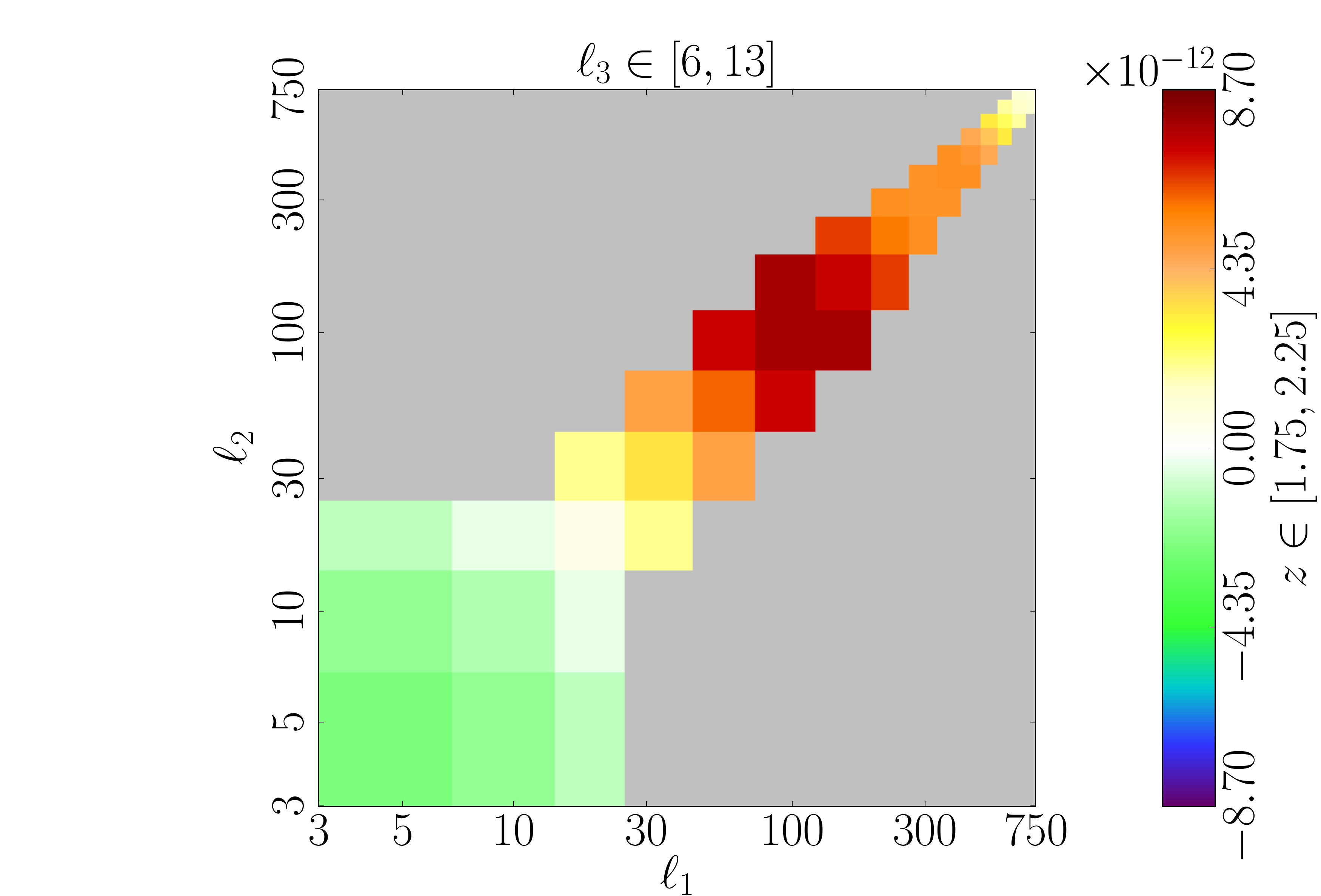} &
      \hspace{-0.45cm}\includegraphics[width=44mm, trim={7cm 0 6.5cm 0},clip]{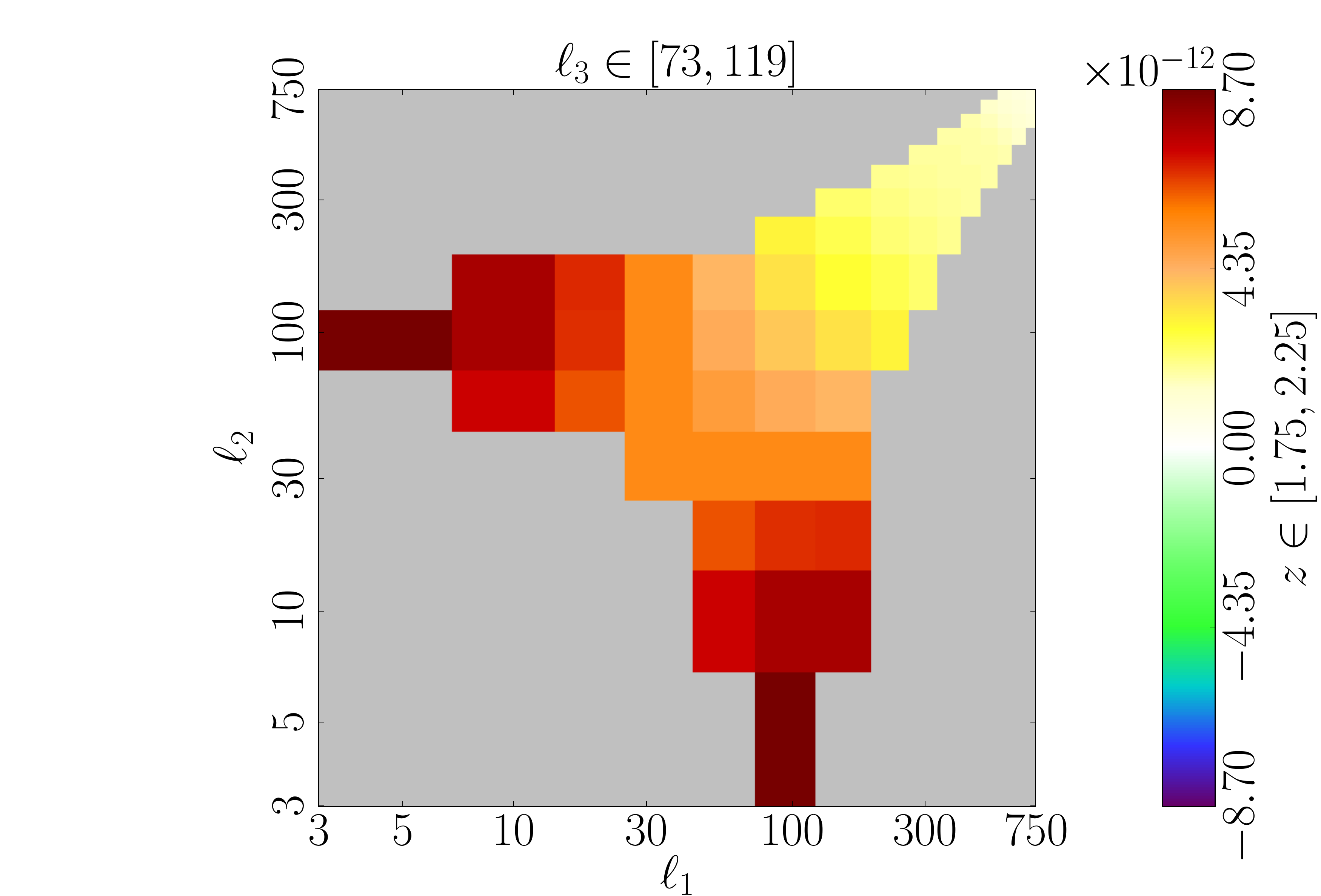} &
      \hspace{-0.45cm}\includegraphics[width=57mm, trim={7cm 0 0cm 0},clip]{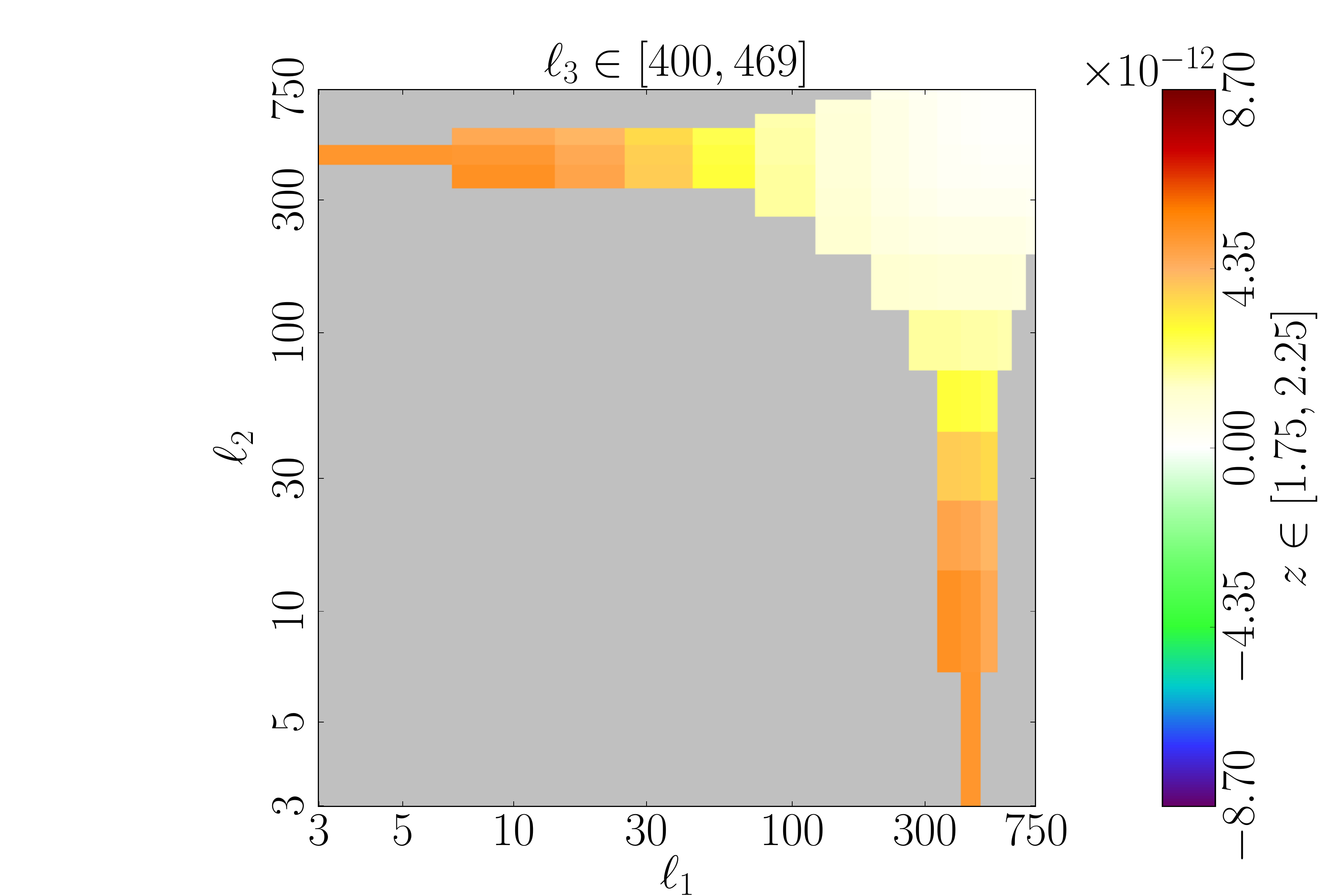}\\
\end{tabular}
    \caption{Similar to Fig.~\ref{fig:Newtonian_bisp}, but with the smoothed difference between the relativistic and the Newtonian bispectra. Note the difference in colour scale used for the two rows. See Appendix~\ref{app:bispectre} for results at other redshifts.}
    \label{fig:rRmN}
\end{figure}

\begin{figure}
    \centering
    $S\left[ \left<\mathcal B^{\mathrm R - \mathrm N}_{\rm STD} \right>\right]$
\begin{tabular}{c c c}
      \includegraphics[width=50mm, trim={4cm 0 6.5cm 0},clip]{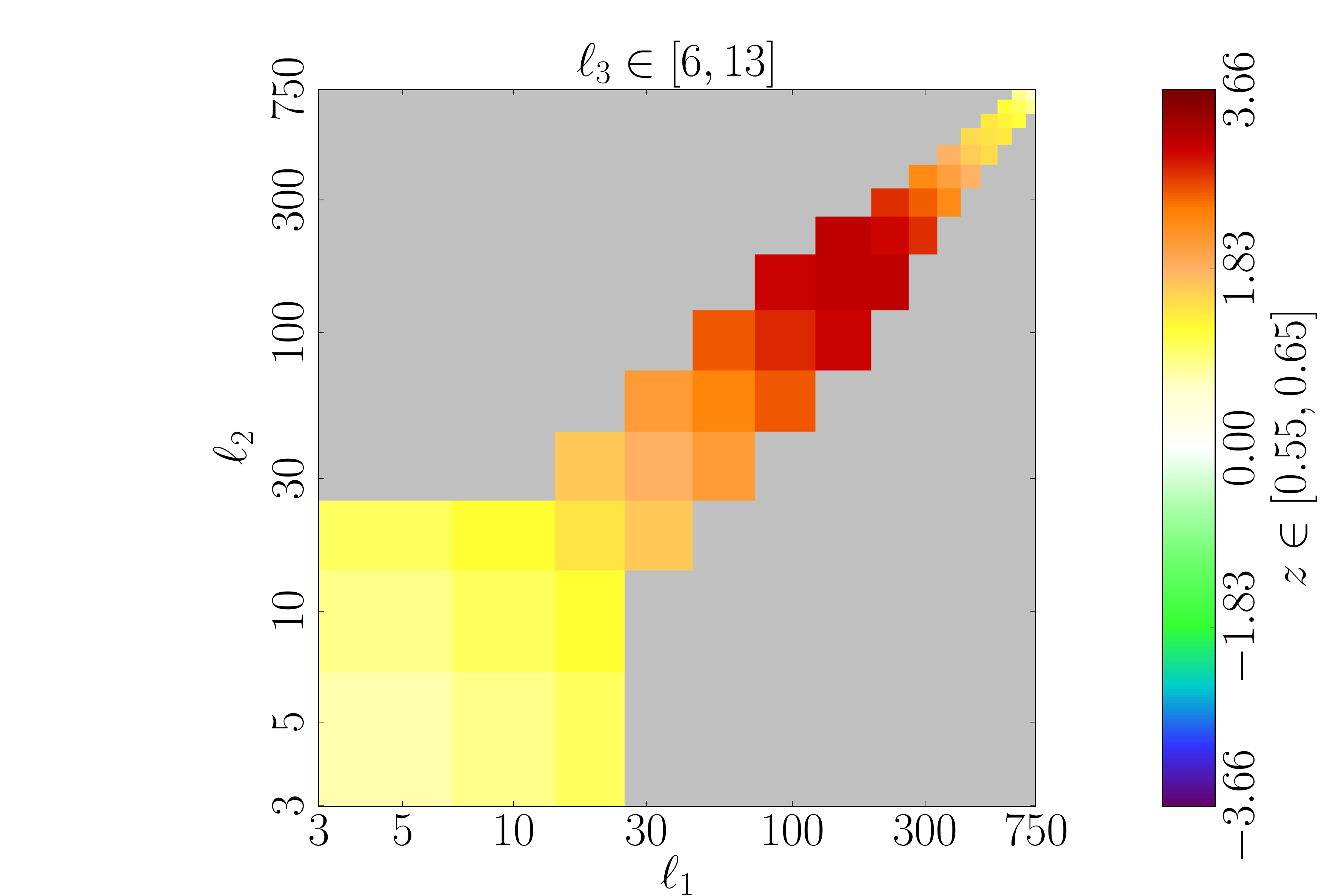} &
      \hspace{-0.45cm}\includegraphics[width=44mm, trim={7cm 0 6.5cm 0},clip]{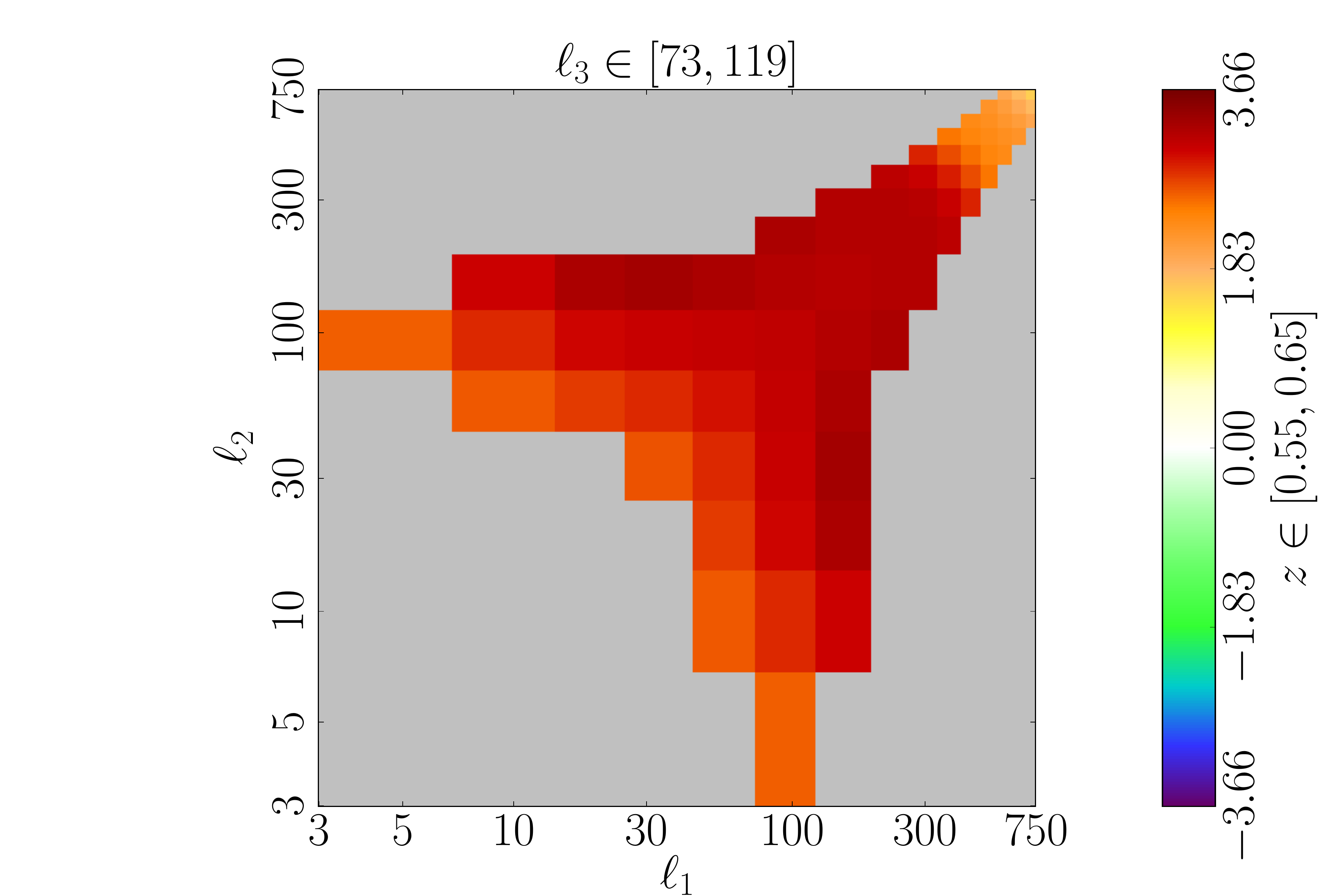} &
      \hspace{-0.45cm}\includegraphics[width=57mm, trim={7cm 0 0cm 0},clip]{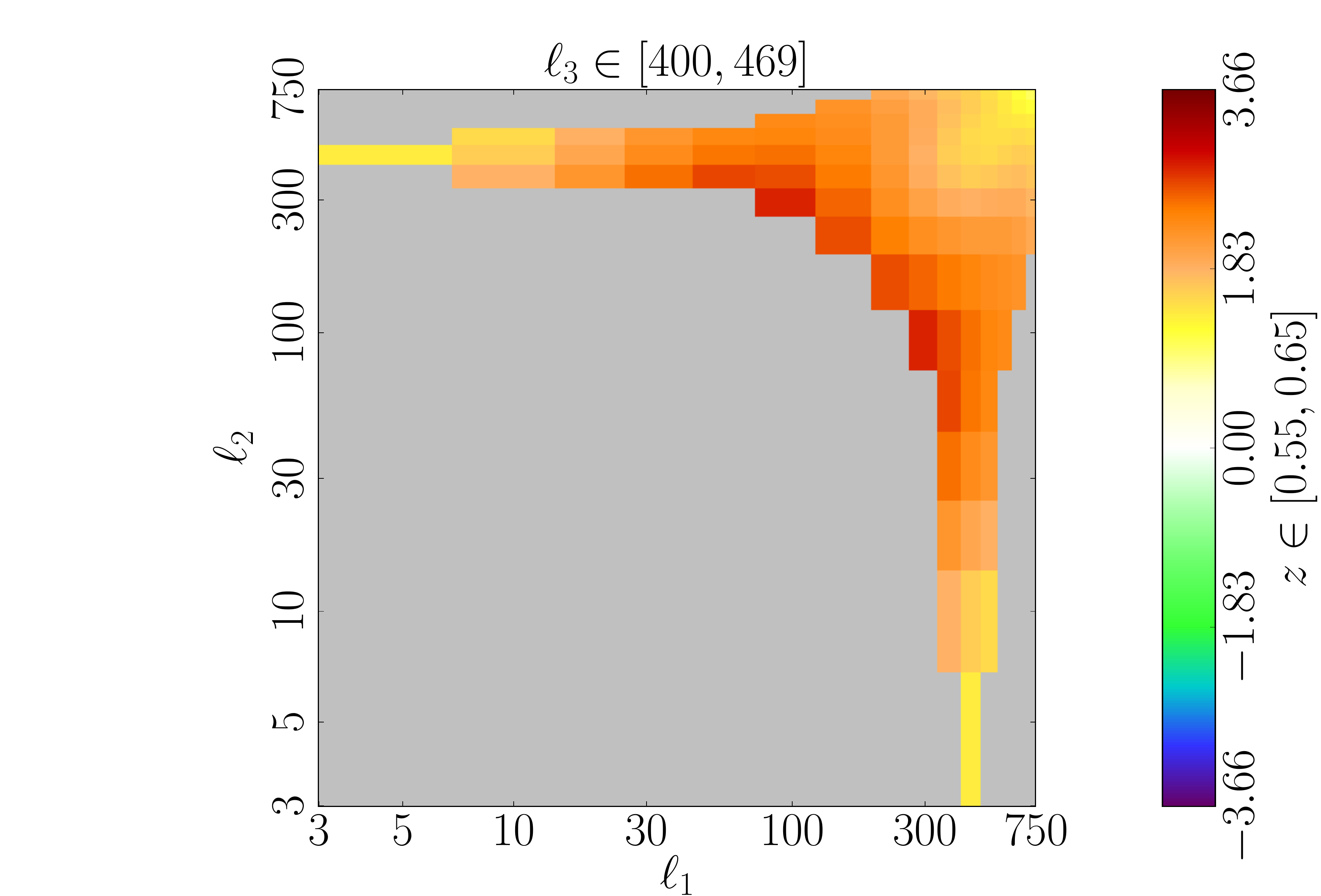}\\
      \includegraphics[width=50mm, trim={4cm 0 6.5cm 0},clip]{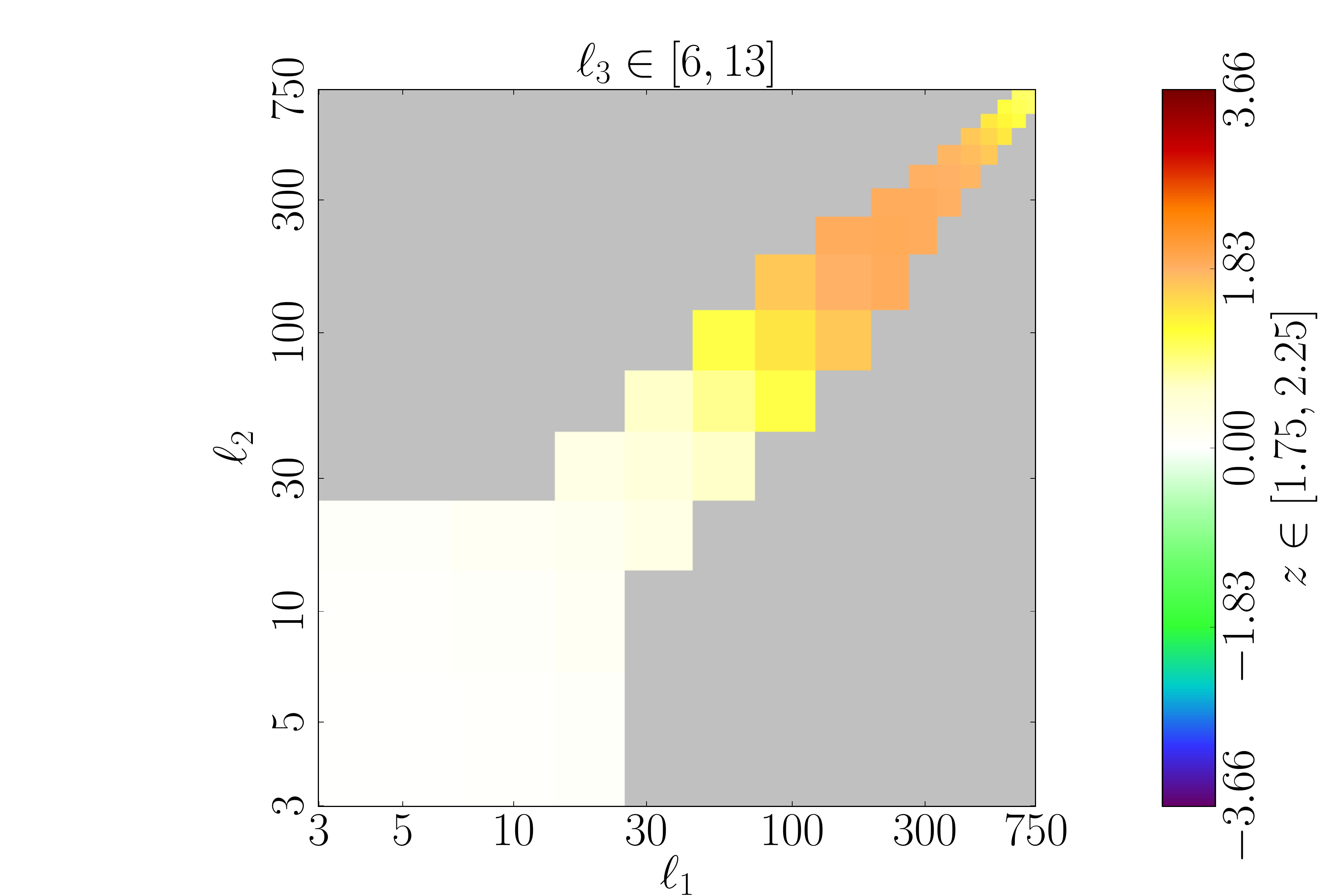} &
      \hspace{-0.45cm}\includegraphics[width=44mm, trim={7cm 0 6.5cm 0},clip]{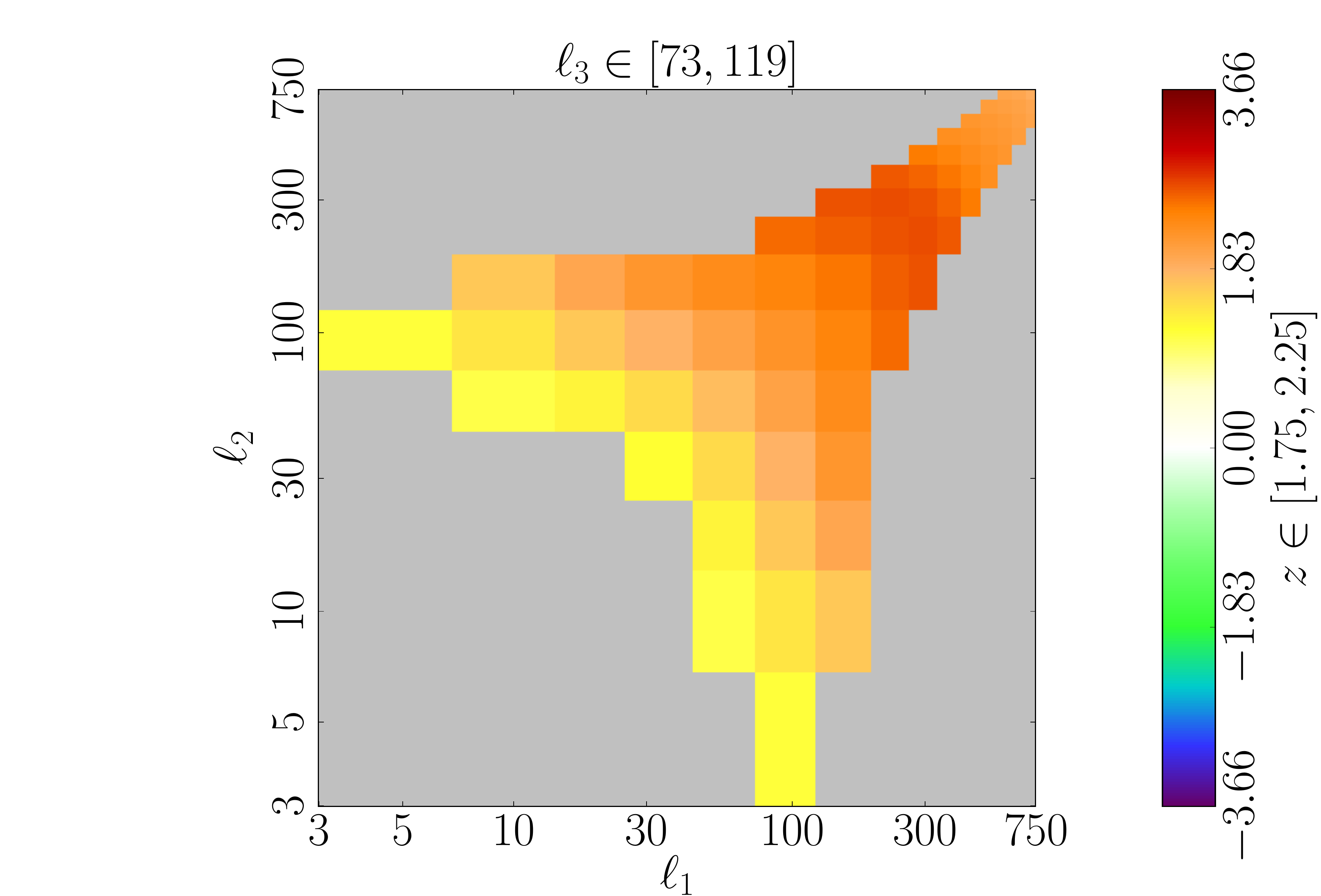} &
      \hspace{-0.45cm}\includegraphics[width=57mm, trim={7cm 0 0cm 0},clip]{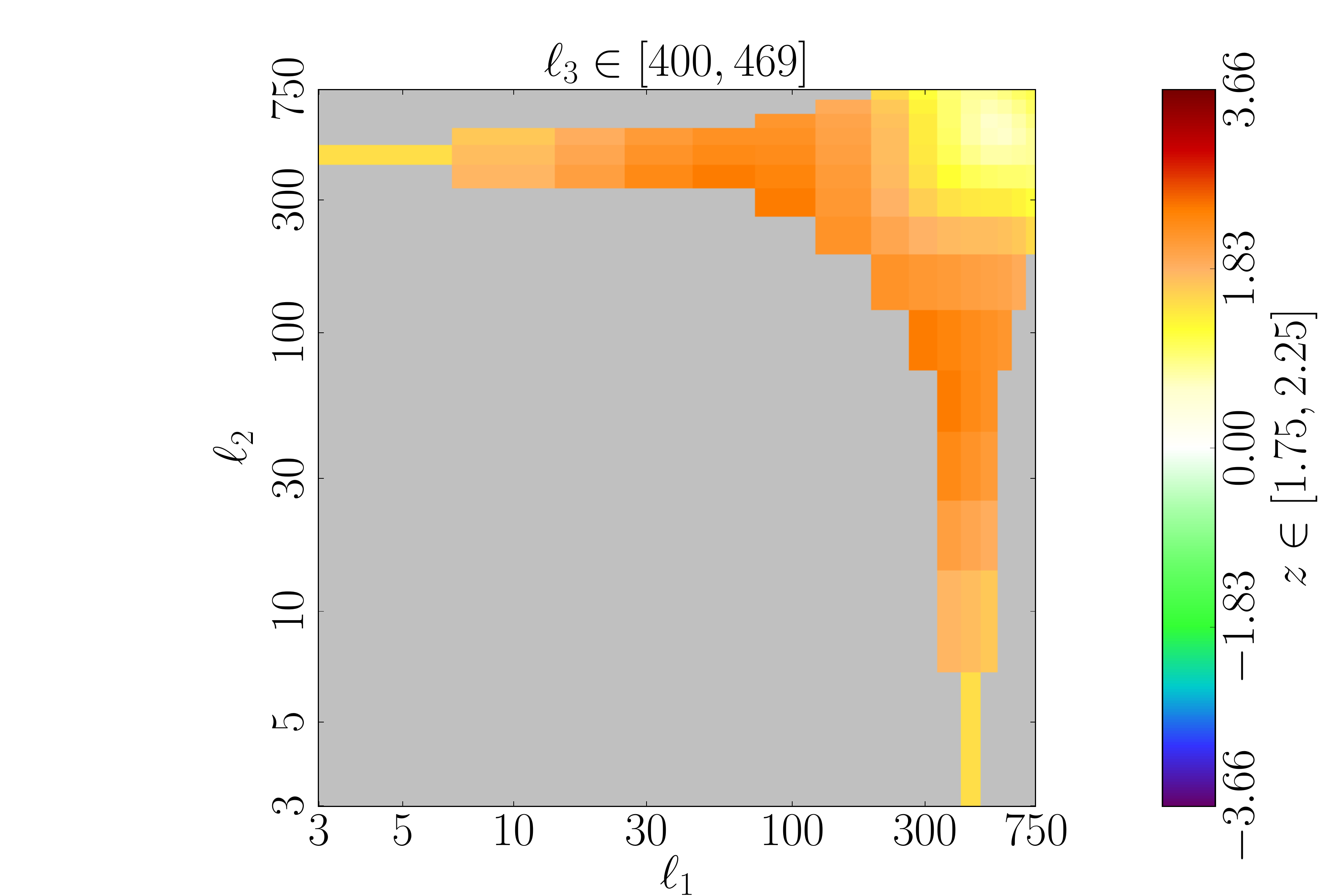}\\
\end{tabular}
    \caption{Similar to Fig.~\ref{fig:Newtonian_bisp}, but with the smoothed difference between the relativistic and the Newtonian bispectra divided by $\sigma_{\rm STD}$.}
    \label{fig:RmN}
\end{figure}

As explained in Section~\ref{sec:numbercount}, we use the binned bispectrum estimator developed initially for the \ac{CMB} and apply it to the maps of the number counts on the light cone. Currently, it takes as input not more than two different types of maps (originally the temperature and polarisation maps, but used here for up to two different redshift bins). It can then compute all the $8$ cross-correlated bispectra.\footnote{Enumerating all combinations of three ordered choices from two items with repetition.} However, we focus here on the auto-correlation bispectra only. We also leave the extension of the estimator to the case of three different maps to future work. 

In Fig.~\ref{fig:Bl_equi}, we plot in the first row the ensemble mean (averaged over ten simulations) of the auto-correlation bispectra ($B^{zzz}_{\ell_1\ell_2\ell_3}$) for each of the five redshift bins. In the left panel we show the equilateral configurations while in the right panel we fix $\ell_1$ to the second bin and vary $\ell_2 = \ell_3$. In the latter case, similarly to Fig.~\ref{fig:Bk_at_all_z}, the first point on the left of the horizontal axis represents an equilateral configuration while, as $\ell$ increases, the configuration becomes more squeezed. The error bars shown here are the standard errors $\sigma_{ \rm STE}$ defined in Eq.~\eqref{eq:ste}. 

The bottom panels show the bispectrum divided by $\sigma_{\rm STE}$ ($\sigma_{\rm STD}$) as solid  (dotted) lines. Focusing on the dotted lines, we see that in the equilateral configuration, only the lowest redshift can be detected at $\sim 5\,\sigma$ for $\ell > 100$. For the same multipole range, the redshift intervals $[0.7, 0.9]$ and $[0.95, 1.25]$ reach respectively $\sim 3\,\sigma$ and $2\,\sigma$. In the squeezed configuration, the redshift bins $[0.55, 0.65]$, $[0.7, 0.9]$ and $[0.95, 1.25]$ reach respectively $\sim 10\,\sigma $, $4\,\sigma $ and $3\,\sigma$ for $\ell > $100. This analysis therefore shows that for redshifts $\lesssim [0.7, 0.9]$, detection in ideal observations is possible for multipoles larger than $100$ and in the squeezed limit. Focusing now on the solid lines in the second row of Fig.~\ref{fig:Bl_equi}, which represent the best determination obtained by averaging our simulations, we see that the Newtonian bispectrum is well determined for all redshift bins and all multipoles except the first three bins.

As expected, the amplitude of the Newtonian bispectrum shown in the first row of Fig.~\ref{fig:Bl_equi} is growing with time. All angular bispectra have a positive slope at low-$\ell$, reach a peak, and then decay. For the equilateral configuration the peak for the largest redshift is at $\ell \sim 100$. This peak is shifted to lower $\ell$ for decreasing redshifts. For the lowest-redshift bin, centered on $z=0.6$, the peak is at $\ell \sim 20$. In the squeezed configuration, we notice that the decay occurs always at the same multipole as the decay of the angular power spectrum in Fig.~\ref{fig:Cl}. In perturbation theory, the tree-level squeezed bispectrum is expected to behave like the power spectrum, which also decays. Hence, the decay of the bispectrum is likely due to the combination of finite resolution and angular power spectrum decay. Finally, we note that the bispectrum is larger in its squeezed configurations than in the equilateral ones, as expected from perturbation theory.

As explained above, without smoothing, if we use the standard errors on the mean $\sigma_{\rm STE}$ instead of the standard deviations $\sigma_{\rm STD}$, we have a significant detection of the Newtonian bispectrum for all redshift bins. In real observations, however, where we do not have multiple realisations of the Universe and cannot use the pairing technique so that the errors will be significantly larger, almost none of these detections would be significant, as shown in Fig.~\ref{fig:Bl_equi}. This is where the smoothing methods, given by Eq.~\eqref{eq:smooth}, can help to increase the detection significance.

In Fig.~\ref{fig:Newtonian_bisp}, we show the smoothed Newtonian bispectrum averaged over the ten simulations. For the lowest redshift we recover a trend observed in Fig.~\ref{fig:Bl_equi}, namely that squeezed triangles have more power than the other configurations. However, by looking at the plots in the second row, we notice that the bispectrum has even more power for folded triangles, which is an expected feature of Newtonian gravity. This trend  is also confirmed through our results at intermediate redshifts (see Fig.~\ref{fig:Newtonian_bisp_annex}).

In order to study the significance of the detection, we define the signal-to-noise bispectrum
\begin{equation}\label{fig:significance}
            \mathcal B  =  \frac{  B  }{\sigma }\,.
\end{equation}
It is important to notice that, because of the smoothing procedure, the interpretation of a detection in terms of $\sigma$ is not completely straightforward, see Ref.~\cite{Bucher:2015ura}. However, these results can be interpreted in a qualitative way.

In Fig.~\ref{fig:Newtonian_detection_bisp}, we plot $S[\langle\mathcal B^{\rm N}_{\rm STD}\rangle]$, i.e.\ Eq.~\eqref{fig:significance} with $B=B^{\rm N}$ and $\sigma=\sigma_{\rm STD}^{\rm N}$ averaged over the simulations and smoothed, as a function of $\ell_1$ and $\ell_2$. For each column, we fix $\ell_3$ to a different bin. The first (second) row corresponds to the lowest (highest) redshift bin. The grey area is where the bispectrum is not defined because the triangle inequality cannot be satisfied. Note also that each panel is symmetric with respect to the first diagonal. The diagonal of the first panel on the left corresponds to the right panel of Fig.~\ref{fig:Bl_equi}.

As we can see, the smoothing procedure very significantly improves the detection of the Newtonian bispectrum. While cosmic variance meant it was only detected for the lowest-redshift bin and for high enough multipoles in Fig.~\ref{fig:Bl_equi}, it is now detected for all redshift bins and for all configurations at more than $4\,\sigma$, except for the equilateral configuration at the lowest $\ell$ at high redshift. Note that the colour bars are saturated at a significance of $10\,\sigma$ for readability. The best detection actually reaches $240\,\sigma$ for the lowest redshift and $140\,\sigma$ for the highest one. If we use the standard error $\sigma_{\rm STE}$ instead of the cosmic variance $\sigma_{\rm STD}$, the significance is multiplied by a factor 4 to 5 ($1100\,\sigma$ and $670\,\sigma$, respectively). This highlights the accuracy of the full bispectrum shown in Fig.~\ref{fig:Newtonian_bisp}, even at large scales and high redshift.

To calculate the ``pure'' relativistic bispectrum $B^{\mathrm R-\mathrm N}$, we subtract the Newtonian bispectrum from the full relativistic bispectrum,
\begin{equation}\label{eq:BRmN}
 B^{\mathrm R-\mathrm N} = B^{\mathrm R} - B^{\mathrm N} \,.  
\end{equation}
Similarly to Fig.~\ref{fig:Cl}, this difference comes from the combination of two effects: the relativistic nonlinearities which are consistently taken into account up to second order in the relativistic pipeline, and the gauge artifact induced by the relativistic ray tracing with particle positions in the wrong gauge. Unlike for the power spectrum, we do not estimate the amplitude of these effects separately. However, we can interpret $B^{\mathrm R-\mathrm N}$ as the \ac{RE} missing from the ``usual'' Newtonian simulated bispectrum, given that the gauge issue is commonly ignored in the construction of light cones from Newtonian simulations.

We again use the smoothing method to improve the significance of the pure relativistic bispectrum, which otherwise can only be determined by combining our simulations, i.e.\ by using~$\sigma_{\rm STE}$. In Fig.~\ref{fig:rRmN}, we show the smoothed difference between the relativistic and the Newtonian simulations, i.e.\ $S\left[ \left< B^{\mathrm R - \mathrm N} \right> \right]$. We recover the trend seen in the right panel of Fig.~\ref{fig:Bl_equi}, i.e.\ an increasing power in the squeezed limit until the power decays due to the combination of early radiation effects and numerical resolution at $\ell \sim 200$ for the highest-redshift bin, and at $\ell \sim 50$ for the lowest-redshift bin. Apart from these effects, the shape is more squeezed, especially for large redshift, than for the Newtonian bispectrum of Fig.~\ref{fig:Newtonian_bisp}. This ``bump'' in the squeezed limit is the dominant contribution in the bispectrum.

In Fig.~\ref{fig:RmN}, we plot $S[ \langle\mathcal B_{\rm STD}^{\mathrm R - \mathrm N }\rangle]$, which represents the significance of the pure relativistic bispectrum accounting for the cosmic variance. At high redshift (second row), we observe a $1\,\sigma$ to $2\,\sigma$ detection in squeezed and folded configurations. The efficiency of the detection for the lowest redshift (first row) is about twice as large, reaching $3.66\, \sigma$ in the best case. We recall that the smoothing procedure makes the interpretation of the significance of the detection of $\mathcal B$ difficult and would require a more sophisticated analysis (see e.g.\ \cite{Bucher:2015ura}). Hence, at such small significance, we cannot claim a detection because of the cosmic variance. 
Similarly to the Newtonian bispectrum, the significance of $B^{\mathrm R - \mathrm N}$ by combining all our simulations and using the pairing technique is estimated with $\sigma_{\rm STE}$ and is about five times larger than what is indicated by Fig.~\ref{fig:RmN}. Hence, except for the smallest-multipole equilateral configurations, we can claim  to have determined $B^{\mathrm{R}-\mathrm{N}}$ with a significance of at least $5\,\sigma$ and up to $14\,\sigma$ ($20\,\sigma$) for the highest (lowest) redshift. This high significance allows us to study the shape of the pure relativistic bispectrum.

\begin{figure}
    \centering
    \includegraphics[scale=0.4]{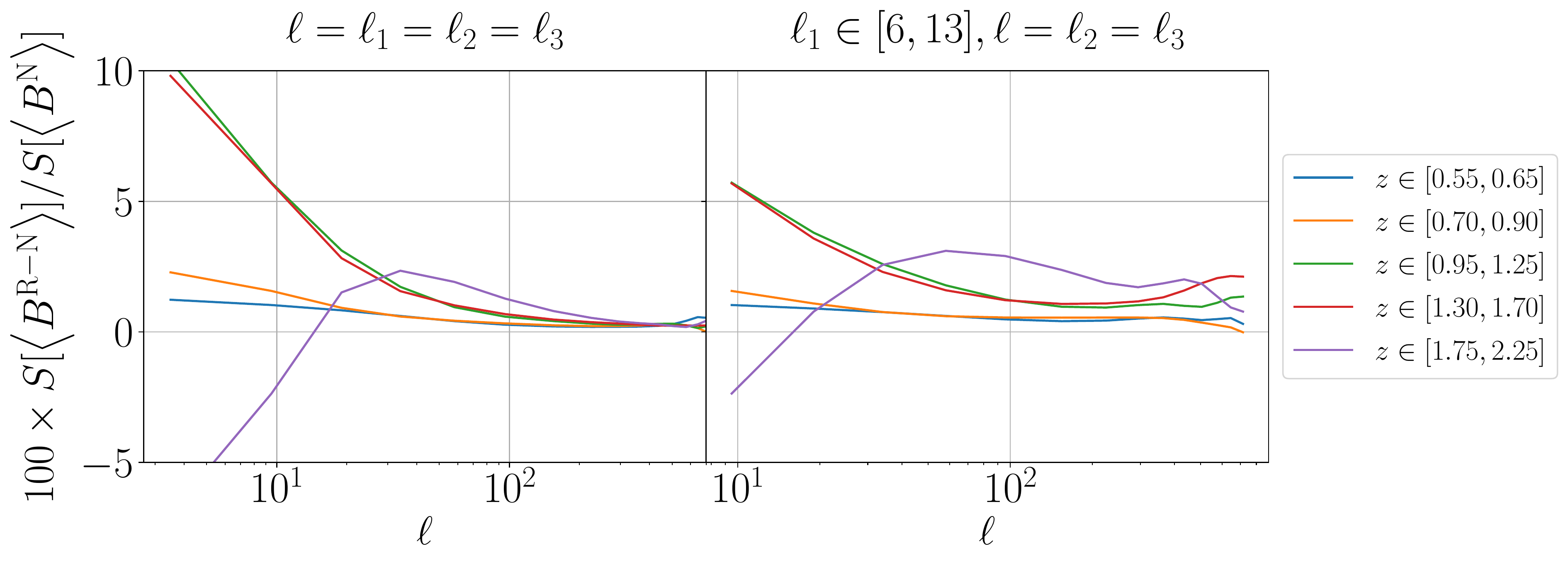}
    \caption{Ratio of the smoothed pure relativistic bispectrum, computed with eq.~\eqref{eq:BRmN}, and the smoothed Newtonian bispectrum. In the left (right) panel, we show the equilateral (squeezed) configuration.}
    \label{fig:ratio_BRmN_BN}
\end{figure}

Finally, in Fig.~\ref{fig:ratio_BRmN_BN}, we show the ratio between the smoothed pure relativistic bispectrum and the smoothed Newtonian bispectrum for the equilateral configuration and the squeezed configuration, respectively, in the left and right panels. 
As expected, relativistic effects increase for large-scale equilateral configurations and for higher redshifts. This behavior is expected because Newtonian nonlinearities grow faster than relativistic nonlinearities, see Section~\ref{sub:REIC}. Moreover, due to the relation \eqref{eq:lNyquist} linking the multipole $\ell$ and the mode $k$, a constant multipole corresponds to smaller physical scales, where the Newtonian nonlinearities are larger, as the redshift decreases. 

For the equilateral configuration (left panel in Fig.\,\ref{fig:ratio_BRmN_BN}), the ratio reaches $10\%$ for the largest scales considered for the redshifts $ z\geq 1$. It should be noted, however, that in this limit the Newtonian bispectrum in the denominator becomes very small (see Figs~\ref{fig:Newtonian_bisp} and~\ref{fig:Newtonian_bisp_annex}) so that this value might be less accurate. For all redshifts, the percentage of pure \ac{RE} falls to $0.1\%$ at $\ell \sim 400$. For the highest redshift bin, we notice a change of sign at large scales, which means that \ac{RE} decrease the bispectrum (this can also be seen in the bottom left panel of Fig.~\ref{fig:rRmN}).

In the right panel of Fig.\,\ref{fig:ratio_BRmN_BN}, we show the ratio in the squeezed limit. As expected, the ratio does not decay. We observe a plateau at $0.3\%$ for the lowest redshift bin, which increases by an order of magnitude, reaching $3\%$, for the highest redshift.

To conclude this analysis, thanks to the smoothing procedure, we have detected the Newtonian bispectrum for all redshift bins and for all the multipole bins, except for large-scale equilateral configurations formed with the first two bins at the highest redshift. %larger than $i = 2$. 
Our estimation of the cosmic variance shows that the Newtonian part should be detectable for an ideal experiment. We have cautioned that our measurements are numerically converged only in the range $\ell \ll 500$, but at the same time, low $\ell$ are also the regime where \ac{RE} are most important. We have measured, thanks to the combination of $10$ paired simulations, the pure relativistic part at high significance allowing us to study the phenomenological shape of the \ac{RE} in the bispectrum. The squeezed limit of the relativistic part of the bispectrum is dominant.
It is expected from perturbation theory that the bispectrum behaves like the power spectrum in the squeezed limit. It is however not guaranteed that perturbation theory holds for these scales at such late times. Moreover, lack of resolution also generates a decay of the power, similar to what is observed in the angular power spectrum. 
Our estimate of the cosmic variance shows that the pure relativistic part will be difficult to detect in real observations. However, the exact significance would require a more sophisticated statistical analysis. In addition, the combination  of multiple redshifts in a tomographic survey could change the detectability of \ac{RE} in future experiments.

Finally, we showed that at high redshift, the contribution of the \ac{RE} in the squeezed limit is about $3\%$ of the total amplitude, while it is $0.3\%$ for the lowest redshift probed. In the equilateral configuration, we reach $10\%$ for the smallest multipoles i.e. for $\ell \lesssim 5$.

\section{Conclusions}
\label{sec:conclusion}
The modelling of the relativistic bispectrum is a crucial step towards understanding the cosmological information encoded in the matter distribution of the Universe. Although the bispectrum is vastly suppressed with respect to the power spectrum at the time of decoupling, it quickly gains power during the late-time gravitational evolution. Hence, any probe of non-Gaussianity (primordial or otherwise) has to deal with it. As the standard model of cosmology provides a unique prediction for it, the relativistic bispectrum also has the potential to challenge the $\Lambda$CDM scenario. In this paper, we conduct a numerical study of this bispectrum. The squeezed limit of the bispectrum is particularly interesting. A detection of a primordial signal in this limit would have a profound impact on our understanding of the early Universe. However, this limit involves the coupling between large and small scales, which makes it challenging to model. On the one hand, the large scales are affected by \ac{RE}. On the other hand, the small scales are highly nonlinear and enter the horizon during radiation domination. The modelling of the squeezed limit must contain all these very different ingredients, i.e.\ relativistic and radiation effects as well as nonlinearities. In this paper, we develop a relativistic numerical pipeline for the observable number-count bispectrum by consistently taking into account all these effects at least up to second order in perturbation theory. To isolate the \ac{RE}, we perform two sets of simulations, one that is relativistic and includes radiation, and one that is Newtonian and takes into account radiation only through the \ac{IC}. Moreover, in order to reduce the variance due to phase correlation, each simulation is ``paired'' according to the method described in Ref.~\cite{Angulo:2016hjd}. 

First, we compute the Cauchy data of the relativistic N-body code \texttt{gevolution} up to second order by using standard perturbation theory including all relativistic and radiation effects. We remove the effect of aliasing and show explicitly that it would generate significant errors in our numerical setup. Then, we use a method based on discrete Lagrangian perturbation theory to initialise the particle positions. Running the N-body simulation with \texttt{gevolution}, we produce various snapshots at redshifts $\geq 0.5$ and compare the second-order power spectra and bispectra with the theoretical predictions as implemented in \texttt{MonofonIC}. This is, to our knowledge, the first time the second-order evolution was validated in a relativistic cosmological simulation. To access the power spectrum of the second-order density field, we subtract the linear field from the snapshots. Thanks to this method, we achieve a relative difference of order $\leq 10\%$ between $100 \geq z \geq 10$. In this same range, the average growth was slightly lower than expected from the second-order power spectrum for all scales in both the Newtonian and the relativistic simulations. For $z < 10$, the Newtonian growth is well recovered, but the residuals of the subtraction by the linear field contaminate the large scales. The bispectrum, which should be a cleaner probe of the nonlinear evolution, fits the theory well until redshift $0.5$ where the nonlinearities only pollute the small-scale equilateral configurations.

In order to compare the Newtonian and relativistic simulations, we construct the observed matter number counts on the light cone using a nonlinear ray tracer. Hence, in addition to the dynamical \ac{RE}, we also account for all the kinematical \ac{RE} (e.g.\ lensing and redshift-space distortions). Note that, while our Newtonian simulation has no dynamical \ac{RE} at second order, the kinematical \ac{RE} are accounted for in the same way as in the relativistic simulation. This means, following the standard approach in the literature (see \cite{Breton:2021htu} for example), that the ray tracer finds the particle coordinates in the wrong gauge when working with the output of a Newtonian simulation. This shortcoming has been discussed in the past, and remedies are available in principle, even though not commonly deployed. We choose to follow the common practice in analysing the Newtonian simulations, to represent the current state-of-the-art of a Newtonian treatment.

We first compare the angular power spectra to the relativistic linear prediction. We observe a difference between the relativistic and Newtonian simulation scaling like $\ell^{-1}$, and of the order $1\%$ to $10\%$ for $\ell \lesssim 5$ and for the highest-redshift bins. We also show that the error due to the gauge issue is one order of magnitude smaller, which indicates that genuine second-order \ac{RE} dominate the difference.

We then compute the bispectrum by using the binned bispectrum estimator that was originally developed for the Planck analysis. By combining all our paired simulations, the full bispectrum is extremely well resolved for $\ell > 20$ and for all the redshifts considered, i.e.\ $0.55 < z< 2.25$, with a significance between $3\,\sigma$ and $10\,\sigma$. The bispectrum peaks in the squeezed and folded configurations. However, our estimation of cosmic variance shows that only the low-redshift bispectrum can be cleanly detected in real observations. To improve the measurements, we use the smoothing method introduced in \cite{Bucher:2015ura} that was used in the Planck analysis to search ``blindly'' for any type of broad-band non-Gaussianity in the bispectrum, useful in particular if no theoretical template exists. The smoothing enhances the signal-to-noise ratio such that we can claim a highly significant detection, even with the cosmic variance, of the Newtonian bispectrum for all the redshifts studied and all triangle configurations.

We also study the difference between the relativistic and the Newtonian bispectra. Like in the case of the power spectrum, this difference is due to two effects: the nonlinear relativistic dynamics and the small gauge issue in the Newtonian treatment as mentioned above. The relativistic and Newtonian bispectra are very close to each other, such that the difference is not detectable in individual bispectrum bins if we account for the cosmic variance. Using the smoothing procedure we reach a maximum detection of $2\,\sigma$ to $3.7\,\sigma$ in the squeezed and folded configurations. With such low significance, detectability is not guaranteed and a more sophisticated statistical analysis would be required. However, note that future large-scale structure surveys will perform the analysis by combining different redshifts, see e.g.~\cite{Pardede:2023ddq}, which will increase the significance of the relativistic part of the bispectrum.
By combining all the paired simulations (i.e.\ ignoring the cosmic variance) we obtain a highly significant determination of the pure relativistic part of the bispectrum. This allows us to discuss its phenomenological shape, which seems to peak in the squeezed limit. Below the resolution limit of our simulations, i.e.\ at $\ell \ll 500$, the ``pure relativistic'' contribution amounts to up to $3\%$ of the total squeezed-limit bispectrum amplitude at redshift $z=2$, and falls to merely $0.3\%$ at redshift $z=0.6$. In the large-scale equilateral configuration, \ac{RE} amount for up to $10\%$ of the total amplitude. 

It is important to note that these percentages are expected to increase as we go further in the squeezed limit, or if one considers larger scales or redshifts. However, this would require to increase the box size while keeping the same resolution, which would increase the computational cost of the simulations.

Our new simulation pipeline provides a robust and self-consistent numerical framework for cosmological simulations that target second-order \ac{RE}, which are an important contribution to the intrinsically relativistic part in the bispectrum at very large scales. Our analysis shows that the combination of different redshifts might reveal the relativistic part at more than $3.7\,\sigma$. Hence, the power of future constraints on the local primordial amplitude $f_\mathrm{NL}$ coming from large-scale structure crucially depends on our ability to model these intrinsic effects accurately. In the absence of any primordial $f_\mathrm{NL}$, similar measurements from \ac{LSS} still amount to an interesting test of the infrared limit of our theory of gravity.

\acknowledgments
We thank Ruth Durrer, Martin Kunz, Ermis Mitsou, Enea Di Dio and Juan Calles for very helpful discussions. Some of the results in this paper have been derived using the HEALPix \cite{Gorski:2004by} package. TM and OH acknowledge funding from the European Research Council (ERC) under the European Union’s Horizon 2020 research and innovation programme, Grant Agreement No.\ 679145 (COSMO-SIMS). The work of JA is supported by the Swiss National Science Foundation. CS acknowledges funding from the European Research Council (ERC) under the European Union's Horizon 2020 research and innovation program (grant agreement No.\ 834148). We used high-performance computing resources provided by the Swiss National Super\-computing Centre (CSCS) under pay-per-use agreement (project ID ``uzh34''). TM thanks the Institute for Computational Science of the University of Zurich, and the University of Geneva for their hospitality. TM and BvT acknowledge the \href{https://cc.in2p3.fr}{IN2P3 Computer Center} for providing computing resources and services needed for the bispectrum analysis.

{\small\paragraph{Carbon footprint} 
We estimate that our numerical computing consumed up to $1120~\mathrm{kWh}$ of electrical energy. CSCS sources its electricity from emission-free hydropower.\footnote{According to \url{https://ethz.ch/en/news-and-events/eth-news/news/2022/11/at-cscs-energy-efficiency-is-a-key-priority-even-at-high-performance.html}, retrieved 21.\ November 2022.}
}

\appendix

\section{Second-order Einstein equations}\label{app:2EE}

In this Appendix we present the Einstein equations at second order which are relevant for our implementation in \texttt{Monofonic}.

Let us first notice that, in this paper, we have used two different definitions of  the density contrast. In Eqs.~\eqref{eq:5.54}, \eqref{eq:newtonian}, \eqref{eq:R1} and \eqref{eq:R2}, we give the expression of the second-order \ac{CDM} fluid density which is more convenient for the physical interpretation. This density is defined for a perfect fluid such that \cite{Adamek:2021rot}
\begin{equation}\label{delta_song}
T^{\mu\nu} = \bar{\rho} \left(1 + \delta^\mathrm{fl}\right) u^\mu u^\nu\,,
\end{equation}
where $u^\nu$ is the covariant matter $4$-velocity. However, when working with an N-body ensemble, it is more convenient to work with the density contrast in Poisson gauge, defined through Eq.~\eqref{eq:Tmunu_def4}, which we rewrite here in Poisson gauge
\begin{equation}\label{delta_gev}
 a^2 e^{2\psi} T^{00} = \bar{\rho} \left(1 + \delta \right)\,.
\end{equation}
In particular, this is the definition used by \texttt{gevolution}. At first order, $\delta^{\rm fl}_1$ and $\delta_1$ agree. At second order, the relation reads
\begin{equation}\label{eq:delta}
    \delta_2 = \delta^\mathrm{fl}_2 + (\nabla V_1)^2 \,.
\end{equation}
Once the density contrast is known, one can use the second-order Einstein equations to calculate the velocity and the potential.

At second order, the gravitational slip $\chi_{2} \equiv \phi_2 - \psi_2$ can be obtained via the traceless part of the Einstein equation
\begin{equation}
\label{eq:chi2}
    \Delta^2 \chi_{2} = \frac{3}{2} \left(\nabla_i \nabla_j - \frac{1}{3} \delta_{ij} \Delta\right) \left(2 \nabla^i \psi_{1} \nabla^j \psi_{1} + 3 \mathcal{H}^2 \Omega_\mathrm{m} \nabla^iV_1 \nabla^jV_1 \right)\,.
\end{equation}
The spatial trace of Einstein's equations yields a differential equation for $\phi_2'$ which can be solved in matter domination. The growing mode solution is given by
\begin{equation}
\label{eq:phiprime2}
    \phi'_{2} \simeq \frac{2 \Delta \chi_{2} - \left(\nabla \phi_{1}\right)^2 + 3 \mathcal{H}^2 \Omega_\mathrm{m} (\nabla V_{1})^2 }{21 \mathcal{H}}\,.
\end{equation}
The Hamiltonian constraint finally gives an equation for $\phi_{2}$,
\begin{multline}
\label{eq:phi2}
    \Delta \phi_{2} - 3 \mathcal{H}^2 \phi_{2} =\\ 3 \mathcal{H} \phi'_{2} - 3 \mathcal{H}^2 \chi_{2} - 3 \mathcal{H}^2 \psi^2_{1} - 2 \phi_{1} \Delta \phi_{1} + \frac{1}{2} \left(\nabla\phi_{1}\right)^2 + \frac{3}{2} \mathcal{H}^2 \Omega_\mathrm{m} \left[\delta_{2}^\mathrm{fl} + (\nabla V_{1})^2\right]\,.
\end{multline}

The second-order canonical momentum can be obtained from the longitudinal part of the momentum constraint, 
\begin{equation}
\label{eq:velocitydivergence}
    \Delta \left(\frac{1}{2}\mathcal{H} \psi_{1}^2 - \mathcal{H} \psi_{2} - \phi'_{2}\right) - \frac{3}{2} \mathcal{H}^2 \Omega_\mathrm{m} \nabla_i \left( \delta_{1}\nabla^i V_{1} \right) = \frac{3}{2} \mathcal{H}^2 \Omega_\mathrm{m} \Delta Q_{2}\,,
\end{equation}
where $Q_{2}$ is the ``canonical momentum field'' defined in Section~\ref{sub:REIC}. Finally, the curl part of the canonical momentum vanishes, as argued in \cite{Lu:2008ju,Adamek:2021rot}. 

\section{Second-order power spectrum}\label{app:2ps}

Similarly to Eq.~\eqref{eq:second-order-stuff}, we can write any second-order field as the convolution integral
\begin{equation}
    \label{eq:second-order-stuff_A}
    \I_{2} (\tau,\boldsymbol{k}) = \int \frac{d^3k_1 d^3k_2}{(2\pi)^3} \delta_\mathrm{D}^{(3)}(\boldsymbol{k} - \boldsymbol{k}_1 -\boldsymbol{k}_2) T^{(2)}_\I(k_1,k_2,k) \zeta(\boldsymbol{k}_1) \zeta(\boldsymbol{k}_2)\,,
\end{equation}
where $T^{(2)}_\I$ is the second-order transfer function of the field $\I$ and $\zeta$ is the primordial curvature perturbation. From Eq.~\eqref{eq:second-order-stuff_A}, one can show that the second-order power spectrum, defined as 
\begin{equation}
    \label{eq:second-order_P}
    \left< \I_{2}(\boldsymbol{k}_1) \I_{2} (\boldsymbol{k}_2) \right> = (2\pi)^3 \delta_\mathrm{D}^{(3)}(\boldsymbol{k}_1 +\boldsymbol{k}_2) P^{\I}_{22}(k_1) \,,
\end{equation}
reads 
\begin{equation}
\label{parimordialsp2_final}
    P^\I_{22}(k) =  \int \frac{ d^3 k_1}{(2\pi)^3} \left[ T^{(2)}_{\I}(k_1, |\boldsymbol k - \boldsymbol k_1| ,k)\right]^2 P_{\zeta}(k_1) P_{\zeta}(|\boldsymbol{k}-\boldsymbol{k}_1|) \,.
\end{equation}
The primordial power spectrum $P_{\zeta}$ is given by the usual power law
\begin{equation}
\label{parimordial}
    P_{\zeta}(k) = \frac{2\pi^2 A_\mathrm{s}}{k^3} \left( \frac{k}{k_\mathrm{p}} \right)^{n_\mathrm{s}-1} \,.
\end{equation}

The second-order density transfer function can be obtained by Fourier transform of Eq.~\eqref{eq:5.54}. It reads \cite{Tram:2016cpy}
\begin{equation}
    \label{eq:kernel}
      T^{(2)}_{\delta^{\text{fl}}}(k_1,k_2,k) = k_1^2k_2^2 T^{(1)}_{\phi \phi} \left( \frac{2}{3\H^2\Omega_{m}}\right)^2 
      \left[ \beta-\alpha + \frac{\beta}{2} \mu \left(\frac{k_1}{k_2}+\frac{k_2}{k_1}\right) + \alpha \mu^2 + \gamma \left(\frac{k_1}{k_2}-\frac{k_2}{k_1}\right)^2\right]\,,  
\end{equation}
where we introduce the shorthand
\begin{equation}
    \label{eq:def_tij}
T^{(1)}_{\I\mathcal J} = T^{(1)}_{\I} (k_1) T^{(1)}_{\mathcal J} (k_2) \,,
\end{equation}
and where $\mu = \boldsymbol{k}_1 \cdot \boldsymbol{k}_2 / ( k_1 k_2)$ and 
\begin{equation}
    \label{eq:VR}
\begin{split}
    \alpha =& \frac{7-3v}{14} + \left( 4f +\frac{3}{2} \Om-\frac{9}{7}w \right)\frac{\H^2}{k^2} + \left( 18f^2+9f^2\Om-\frac{9}{2}f\Om \right)\frac{\H^4}{k^4}\,,\\ 
    \beta =& 1 + \left( -2f^2 + 6f-\frac{9}{2}\Om \right) \frac{\H^2}{k^2} + \left( 36f^2+18f^2\Om \right)\frac{\H^4}{k^4}\,,\\
    \gamma =& \frac{1}{2}\left( -f^2 + f-3 \Om\right) \frac{\H^2}{k^2} + \frac{1}{4} \left( 18f^2+9(f^2-f)\Om \right)\frac{\H^4}{k^4}\\ &-\frac{1}{2} \left( f + \frac{3 \Om}{2} \right) \left(\frac{\H^2}{k^2}+3 f\frac{\H^4}{k^4}\right) \frac{\partial\log{T_\phi}}{\partial\log{k}}\,.
\end{split}
\end{equation}

In the equilateral limit, $k_1=k_2=k$, the last term of \eqref{eq:kernel} vanishes while in the squeezed limit, $k_1 \ll k_2,k$ or  $k_2 \ll k_1,k$, this last term becomes dominant. The function $\gamma$ is therefore the contribution to the squeezed limit. The Newtonian limit is recovered when $\mathcal H^2 / k^2 \rightarrow 0$. This means that relativistic corrections are important near the horizon where $k \sim \H$. The time dependence is recovered by multiplying each coefficient of \eqref{eq:VR} by $\D^2$, see Eq.~\eqref{eq:kernel}. In the Einstein-de Sitter limit $\D = a \propto \tau^2$ and $\H=2/\tau$. Hence the terms $\propto \H^4 / k^4$ are constant in time and come from the second-order \ac{IC} that are dropped in the Newtonian limit. The terms $\propto \H^2/k^2$, including the squeezed limit, grow like $a$, hence do not decay with respect to the linear solution. The Newtonian terms $\propto \H^0/k^0$, excluding the squeezed limit, grow like~$a^2$.

Following \cite{Adamek:2021rot}, we use the four scalar Einstein equations to compute the canonical momentum at second order, as well as the metric field $\phi$. For these equations, one can compute the second-order transfer functions. For $\chi_{2}$ it reads (see also appendix A of \cite{Adamek:2017grt})
\begin{equation}
\label{eq:Tchi2}
    T^{(2)}_\chi(k_1,k_2,k) = \frac{k_1^2k_2^2}{k^4} \left[ 3(1+\frac{k_1}{k_2}\mu)(1+\frac{k_2}{k_1}\mu) -\frac{k^2}{k_1k_2}\mu \right]    \left(\frac{3\H^2}{2}\Om T^{(1)}_{vv} + T^{(1)}_{\psi\psi}  \right)\,.
\end{equation}

The transfer function of the time derivative of the potential is 
\begin{equation}
\label{eq:Tphiprime2}
    T^{(2)}_{\phi'}(k_1,k_2,k) = \frac{k_1k_2}{21\H}\left( \mu T^{(1)}_{\phi\phi} -3\H^2 \Om \mu T^{(1)}_{vv} - 2 \frac{k^2}{k_1k_2} T^{(2)}_\chi \right) \,.
\end{equation}
The second-order transfer function for the potential itself is given by
\begin{multline}
\label{eq:Tphi2}
    -(3\H^2+k^2) T^{(2)}_{\phi}(k_1,k_2,k) =   \frac{3\H^2\Om}{2} \left(T^{(2)}_{\delta^\mathrm{fl}} - k_1 k_2 \mu T^{(1)}_{vv} \delta \right) -  3\H^2 T^{(2)}_\chi -3\H^2 T^{(1)}_{\psi\psi}\\
    +3\H T^{(2)}_{\phi'} + \left( k_1^2+k_2^2 - \frac{1}{2} k_1k_2\mu \right)T^{(1)}_{\phi\phi} \,.
\end{multline}

\begin{figure}
    \centering
    $S[ \left< B^{\rm N} \right>]$
\begin{tabular}{c c c}
% \includegraphics[width=50mm, trim={4cm 1.1cm 6.5cm 1cm},clip]{binned_bispec_z0_NpNm_logaxis_T+E_1024_1024_2_2000_2_2000_log2linbinsv3_dt10_0_nomask__0_01_sml2.0.pdf} &
      % \hspace{-0.45cm}\includegraphics[width=44mm, trim={7cm 1.1cm 6.5cm 1cm},clip]{binned_bispec_z0_NpNm_logaxis_T+E_1024_1024_2_2000_2_2000_log2linbinsv3_dt10_0_nomask__0_05_sml2.0.pdf} &
      % \hspace{-0.45cm}\includegraphics[width=57mm, trim={7cm 1.1cm 0cm 1cm},clip]{binned_bispec_z0_NpNm_logaxis_T+E_1024_1024_2_2000_2_2000_log2linbinsv3_dt10_0_nomask__0_10_sml2.0.pdf}\vspace{-0.1cm}\\
      \includegraphics[width=50mm, trim={4cm 1.1cm 6.5cm 1cm},clip]{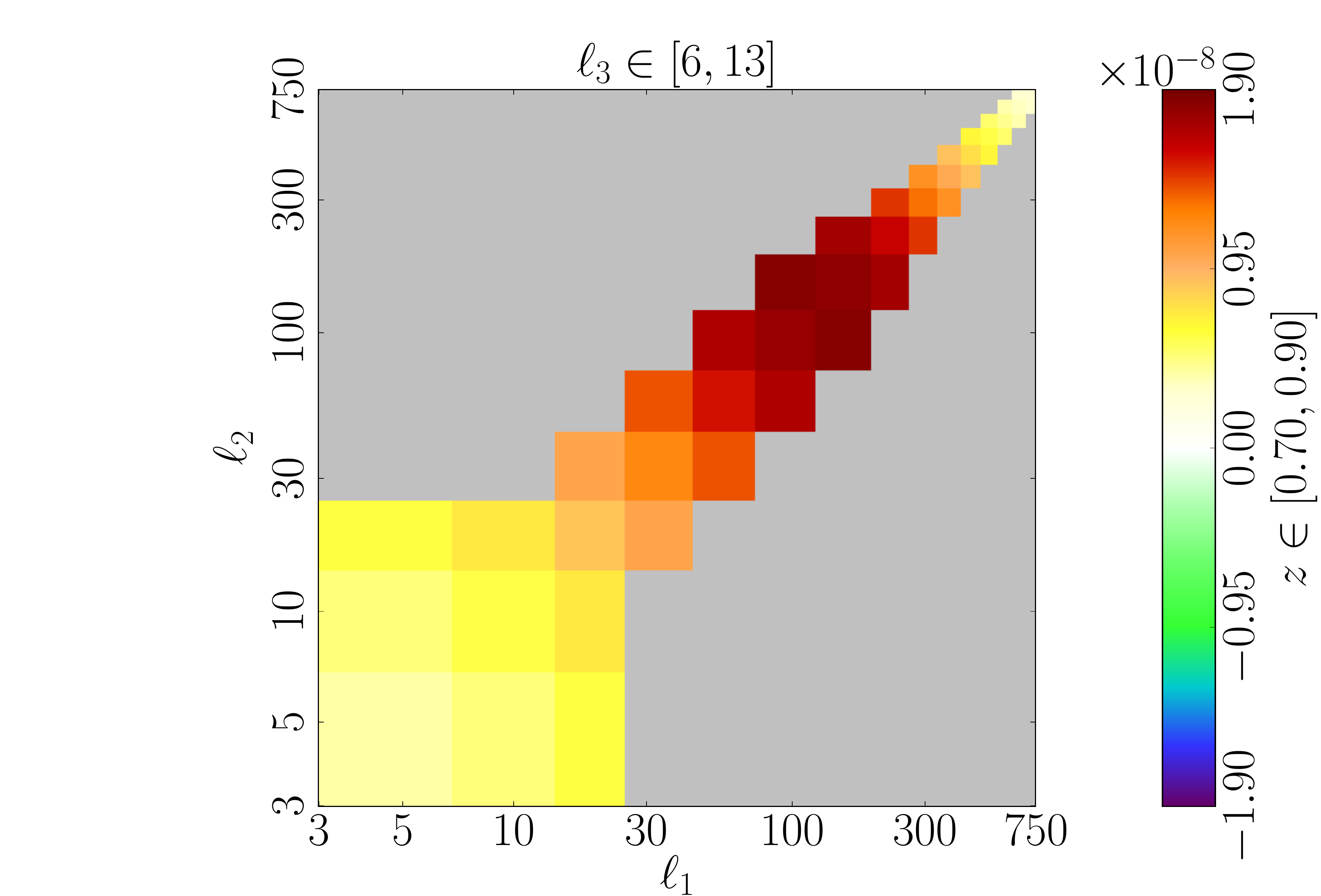} &
      \hspace{-0.45cm}\includegraphics[width=44mm, trim={7cm 1.1cm 6.5cm 1cm},clip]{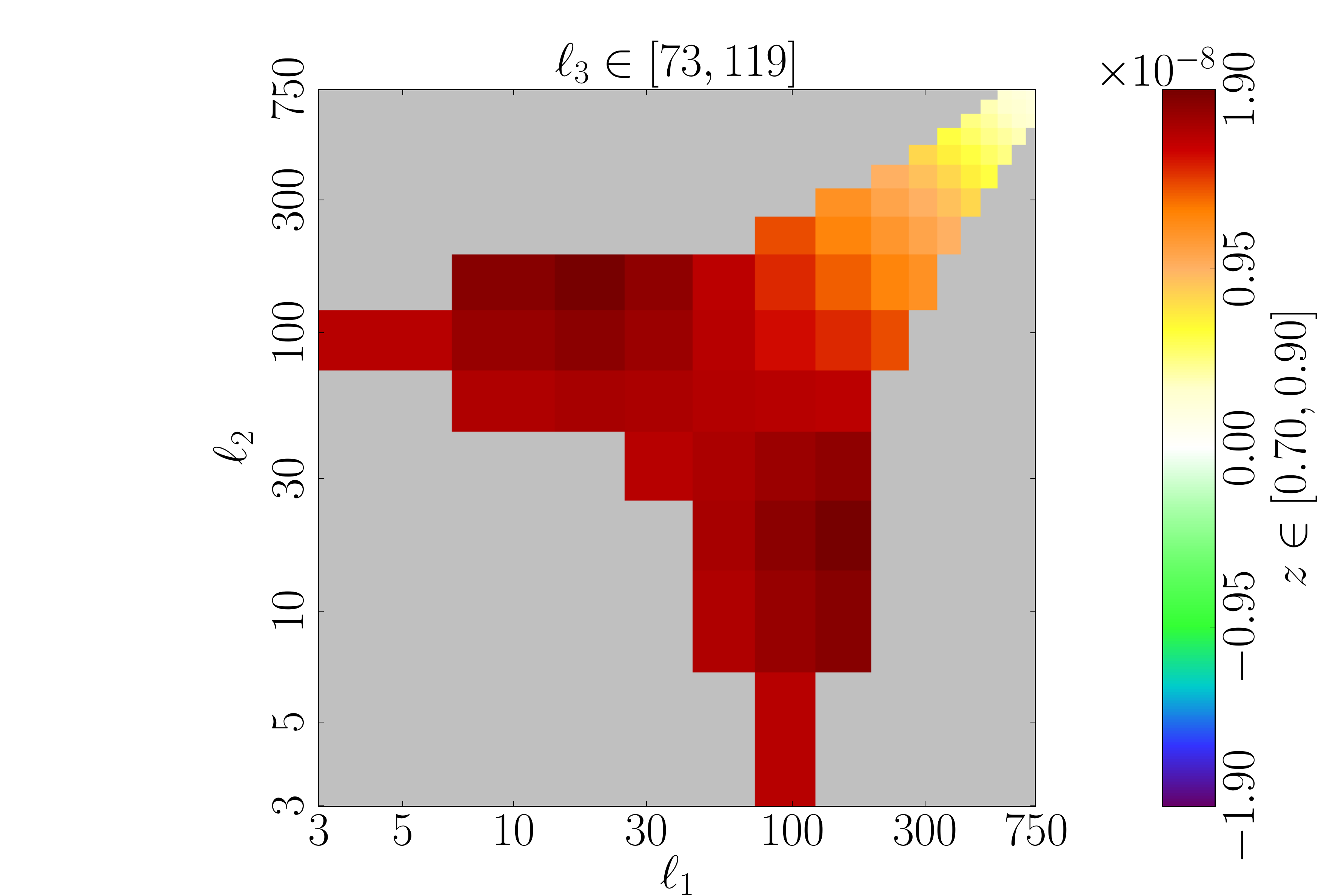} &
      \hspace{-0.45cm}\includegraphics[width=57mm, trim={7cm 1.1cm 0cm 1cm},clip]{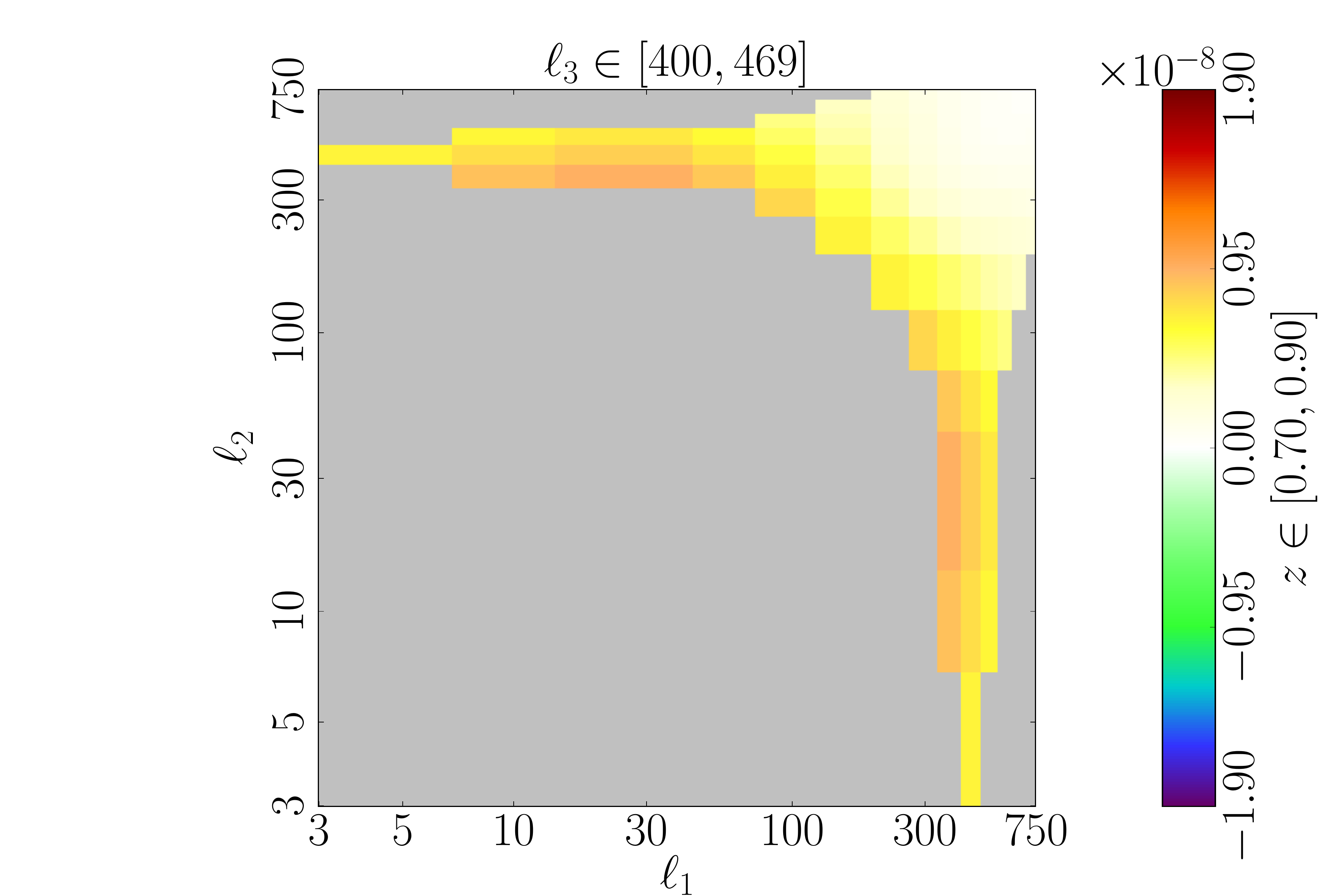}\vspace{-0.1cm}\\
      \includegraphics[width=50mm, trim={4cm 1.1cm 6.5cm 1cm},clip]{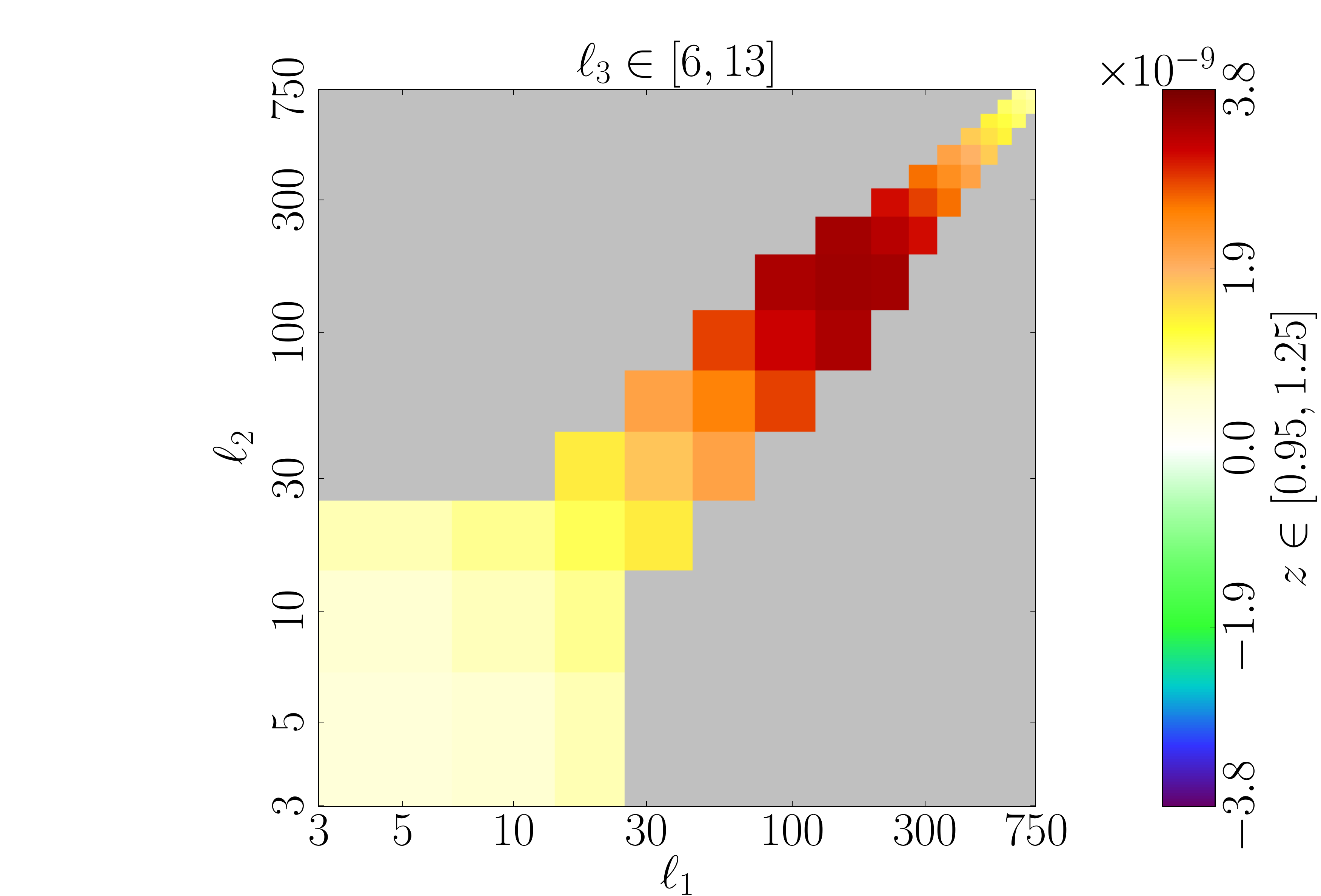} &
      \hspace{-0.45cm}\includegraphics[width=44mm, trim={7cm 1.1cm 6.5cm 1cm},clip]{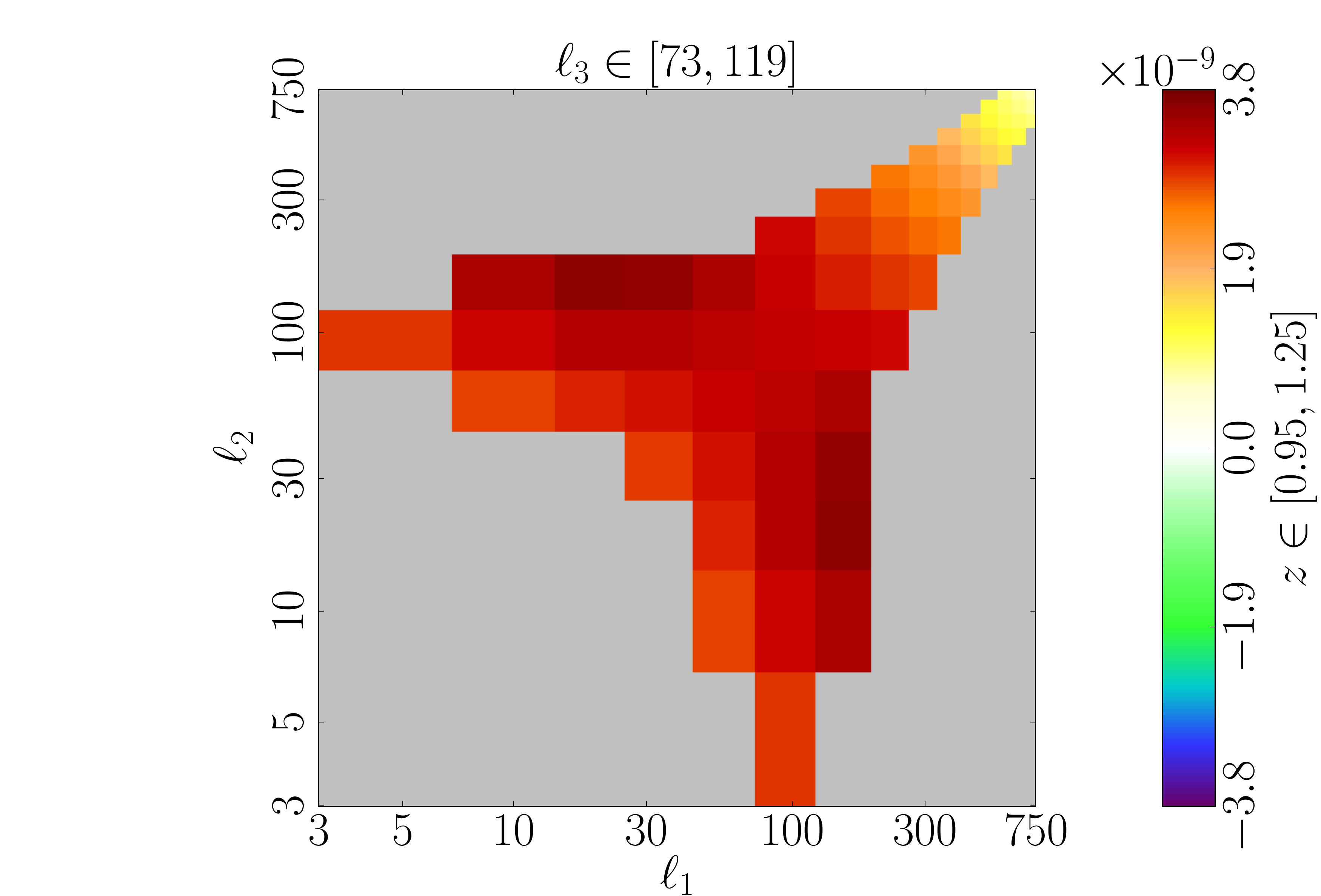} &
      \hspace{-0.45cm}\includegraphics[width=57mm, trim={7cm 1.1cm 0cm 1cm},clip]{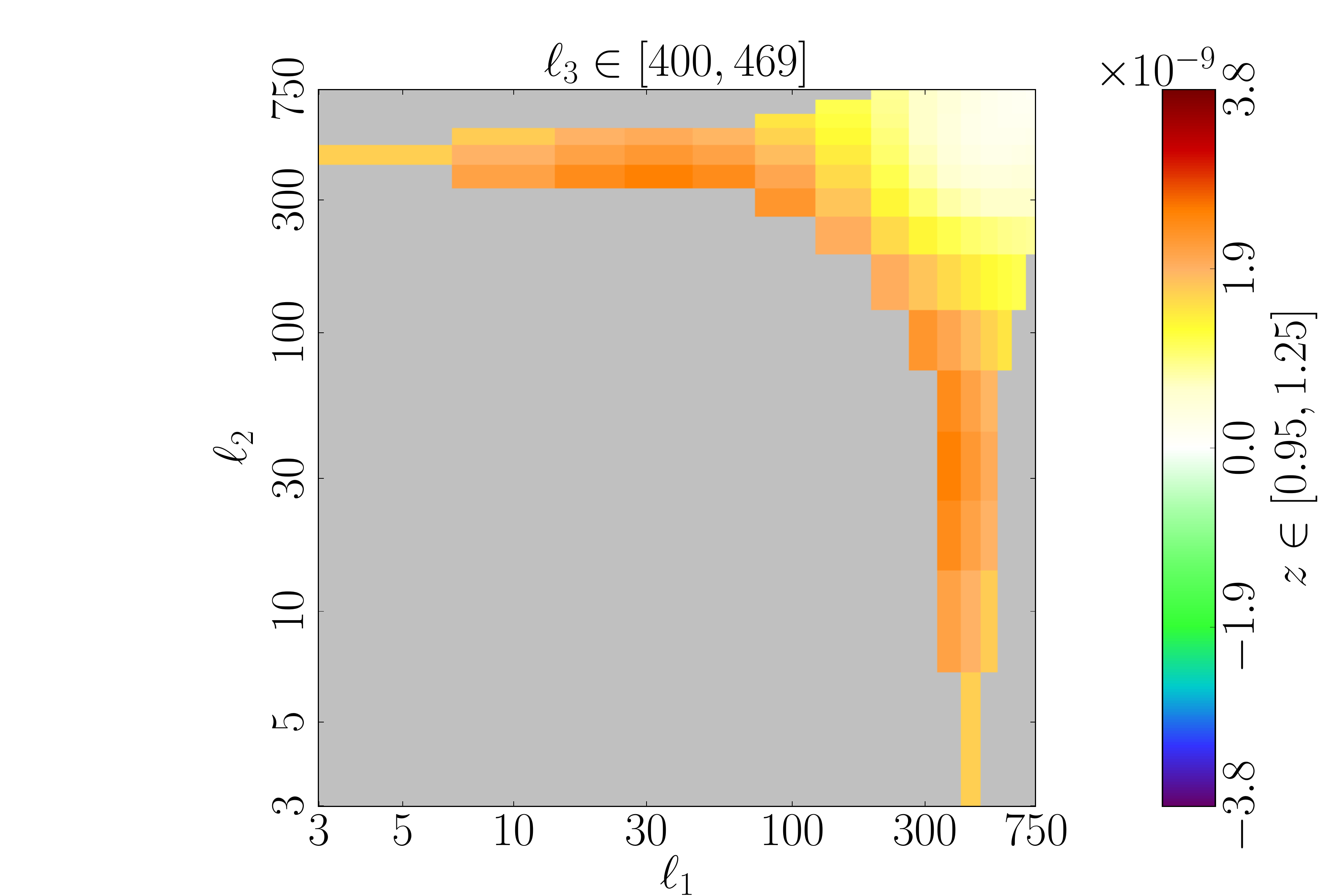}\vspace{-0.1cm}\\
    \includegraphics[width=50mm, trim={4cm 1.1cm 6.5cm 1cm},clip]{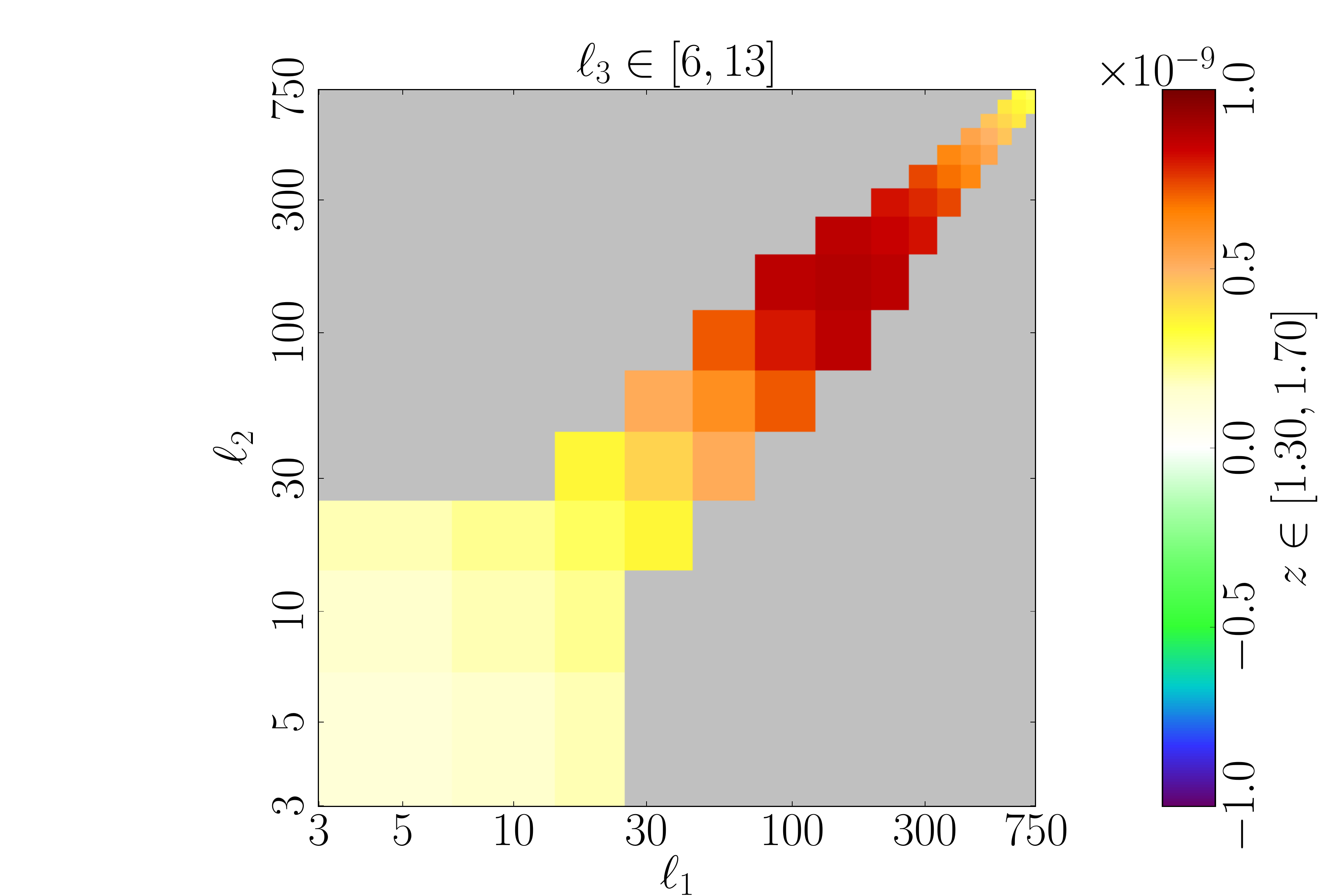} &
      \hspace{-0.45cm}\includegraphics[width=44mm, trim={7cm 1.1cm 6.5cm 1cm},clip]{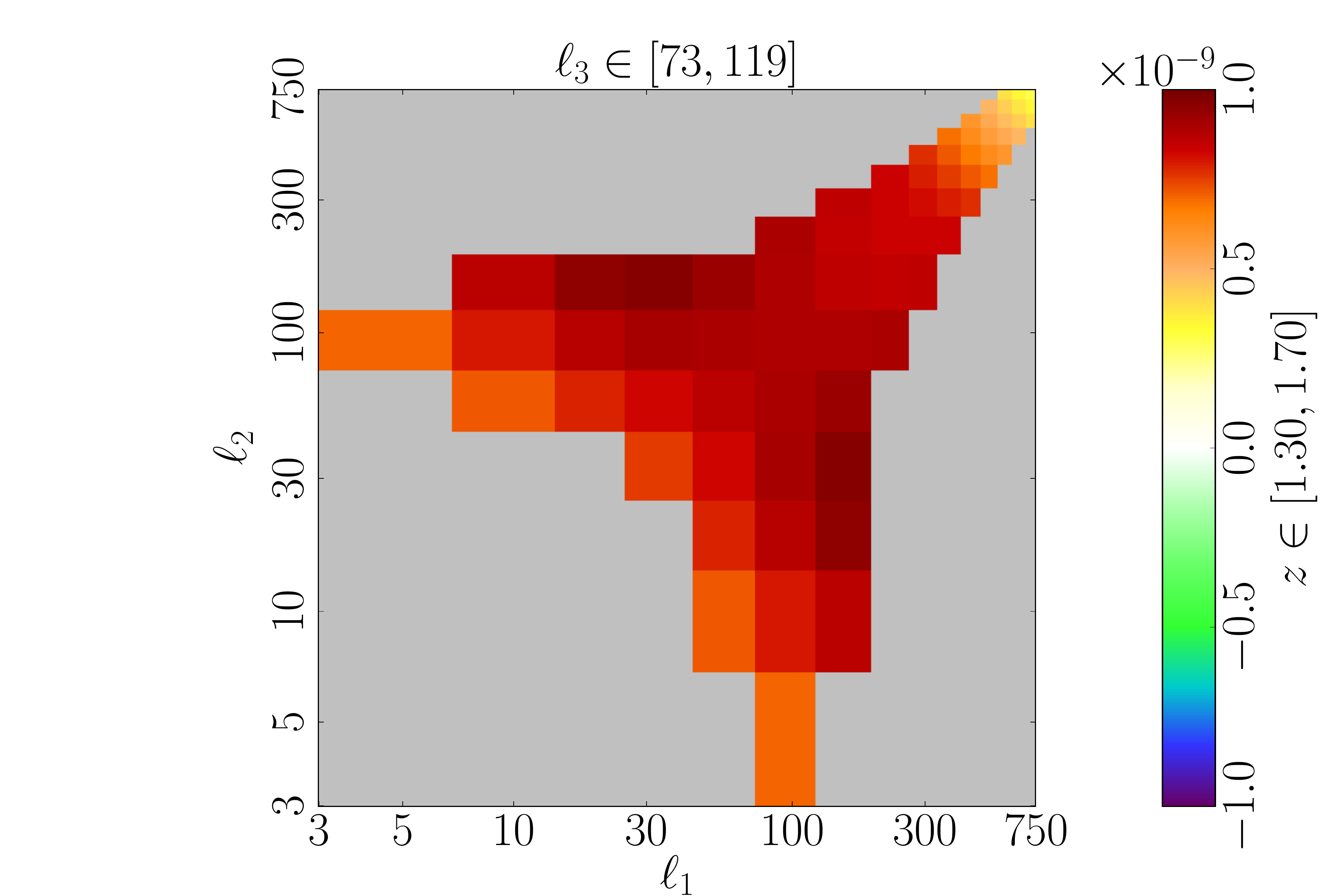} &
      \hspace{-0.45cm}\includegraphics[width=57mm, trim={7cm 1.1cm 0cm 1cm},clip]{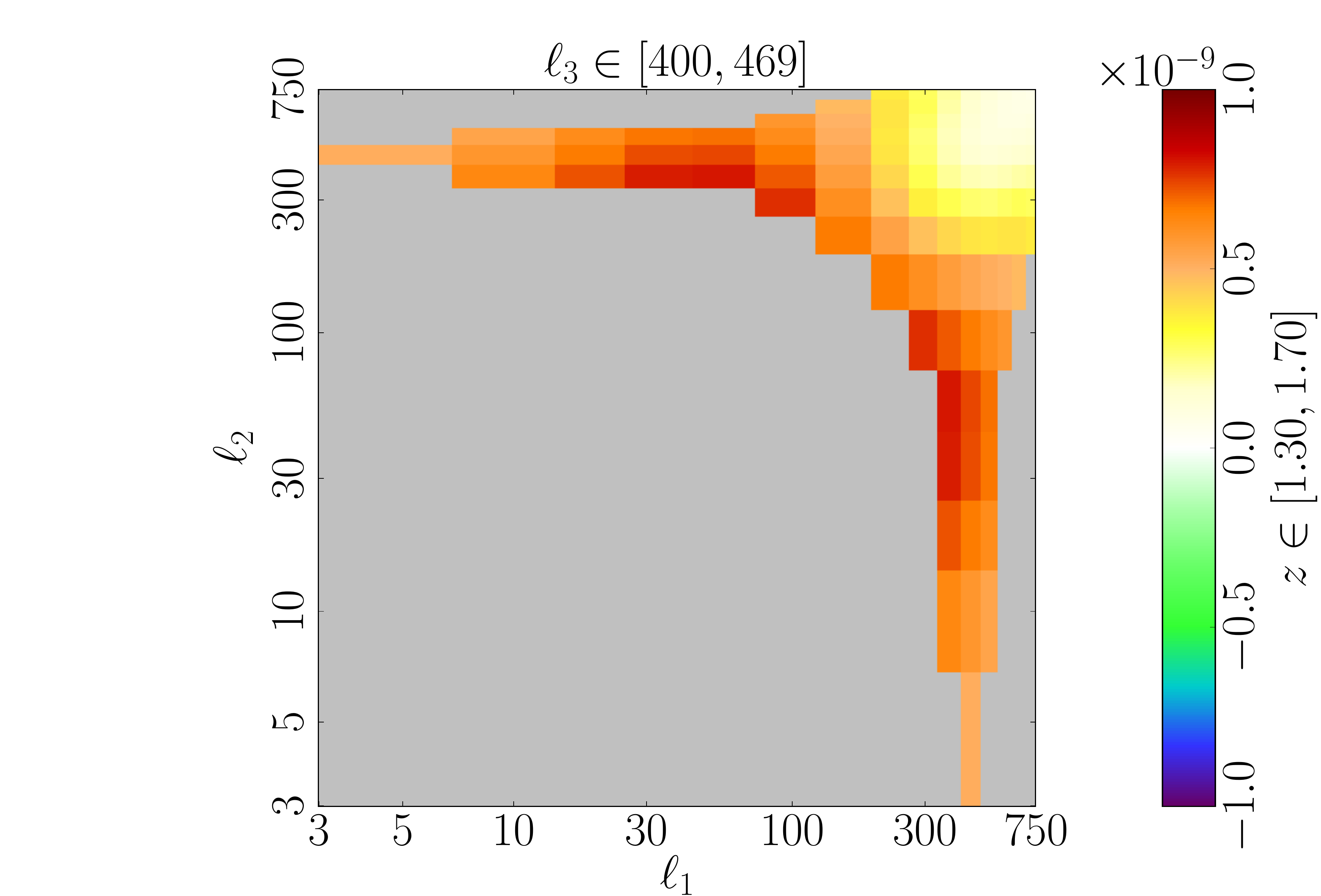}\\
      %   \includegraphics[width=50mm, trim={4cm 0 6.5cm 1cm},clip]{binned_bispec_z4_NpNm_logaxis_T+E_1024_1024_2_2000_2_2000_log2linbinsv3_dt10_0_nomask__7_01_sml2.0.pdf} &
      % \hspace{-0.45cm}\includegraphics[width=44mm, trim={7cm 0 6.5cm 1cm},clip]{binned_bispec_z4_NpNm_logaxis_T+E_1024_1024_2_2000_2_2000_log2linbinsv3_dt10_0_nomask__7_05_sml2.0.pdf} &
      % \hspace{-0.45cm}\includegraphics[width=57mm, trim={7cm 0 0cm 1cm},clip]{binned_bispec_z4_NpNm_logaxis_T+E_1024_1024_2_2000_2_2000_log2linbinsv3_dt10_0_nomask__7_10_sml2.0.pdf}\\
\end{tabular}
    \caption{Similar to Fig.~\ref{fig:Newtonian_bisp}, but for all intermediate redshift bins. Note the different colour scales used on each row.
    }
    \label{fig:Newtonian_bisp_annex}
\end{figure}

\begin{figure}
    \centering
    $S[ \left< B^{\mathrm R-\mathrm N} \right>]$
\begin{tabular}{c c c}
% \includegraphics[width=50mm, trim={4cm 1.1cm 6.5cm 1cm},clip]{binned_bispec_z0_NpNm_logaxis_T+E_1024_1024_2_2000_2_2000_log2linbinsv3_dt10_0_nomask__0_01_sml2.0.pdf} &
%       \hspace{-0.45cm}\includegraphics[width=44mm, trim={7cm 1.1cm 6.5cm 1cm},clip]{binned_bispec_z0_RpRm-NpNm_logaxis_T+E_1024_1024_2_2000_2_2000_log2linbinsv3_dt10_0_nomask__0_05_sml2.0.pdf} &
%       \hspace{-0.45cm}\includegraphics[width=57mm, trim={7cm 1.1cm 0cm 1cm},clip]{binned_bispec_z0_RpRm-NpNm_logaxis_T+E_1024_1024_2_2000_2_2000_log2linbinsv3_dt10_0_nomask__0_10_sml2.0.pdf}\vspace{-0.1cm}\\
      \includegraphics[width=50mm, trim={4cm 1.1cm 6.5cm 1cm},clip]{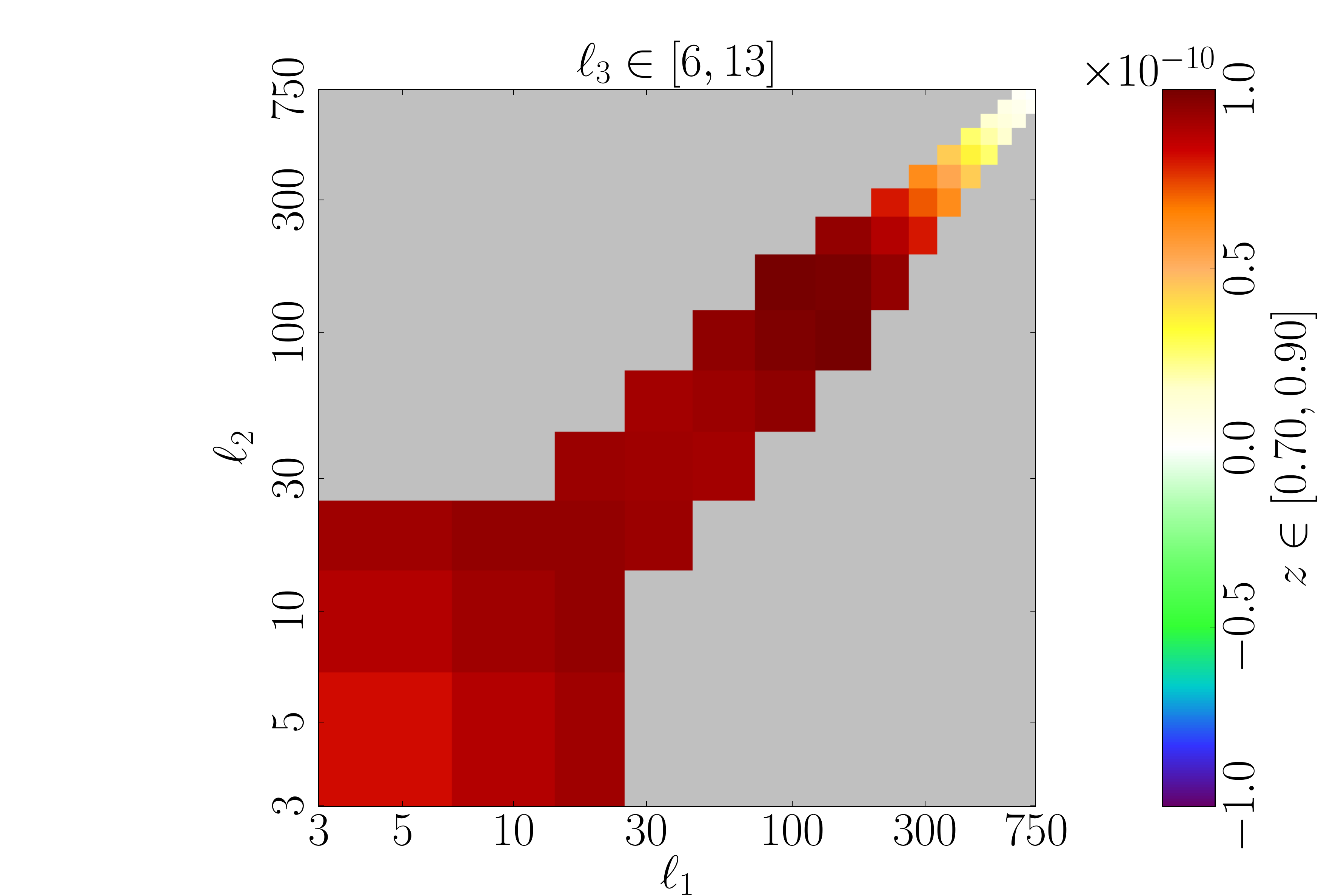} &
      \hspace{-0.45cm}\includegraphics[width=44mm, trim={7cm 1.1cm 6.5cm 1cm},clip]{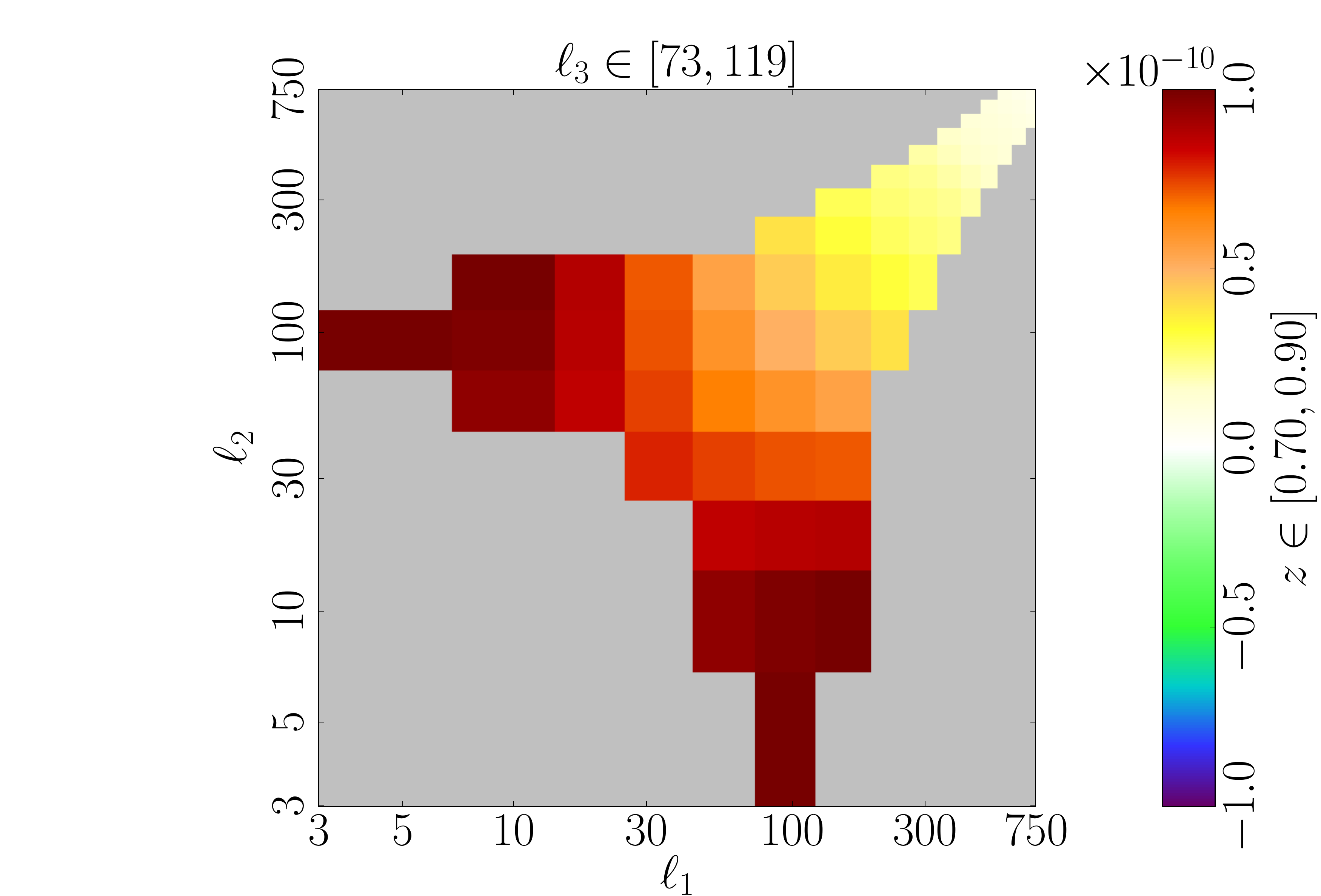} &
      \hspace{-0.45cm}\includegraphics[width=57mm, trim={7cm 1.1cm 0cm 1cm},clip]{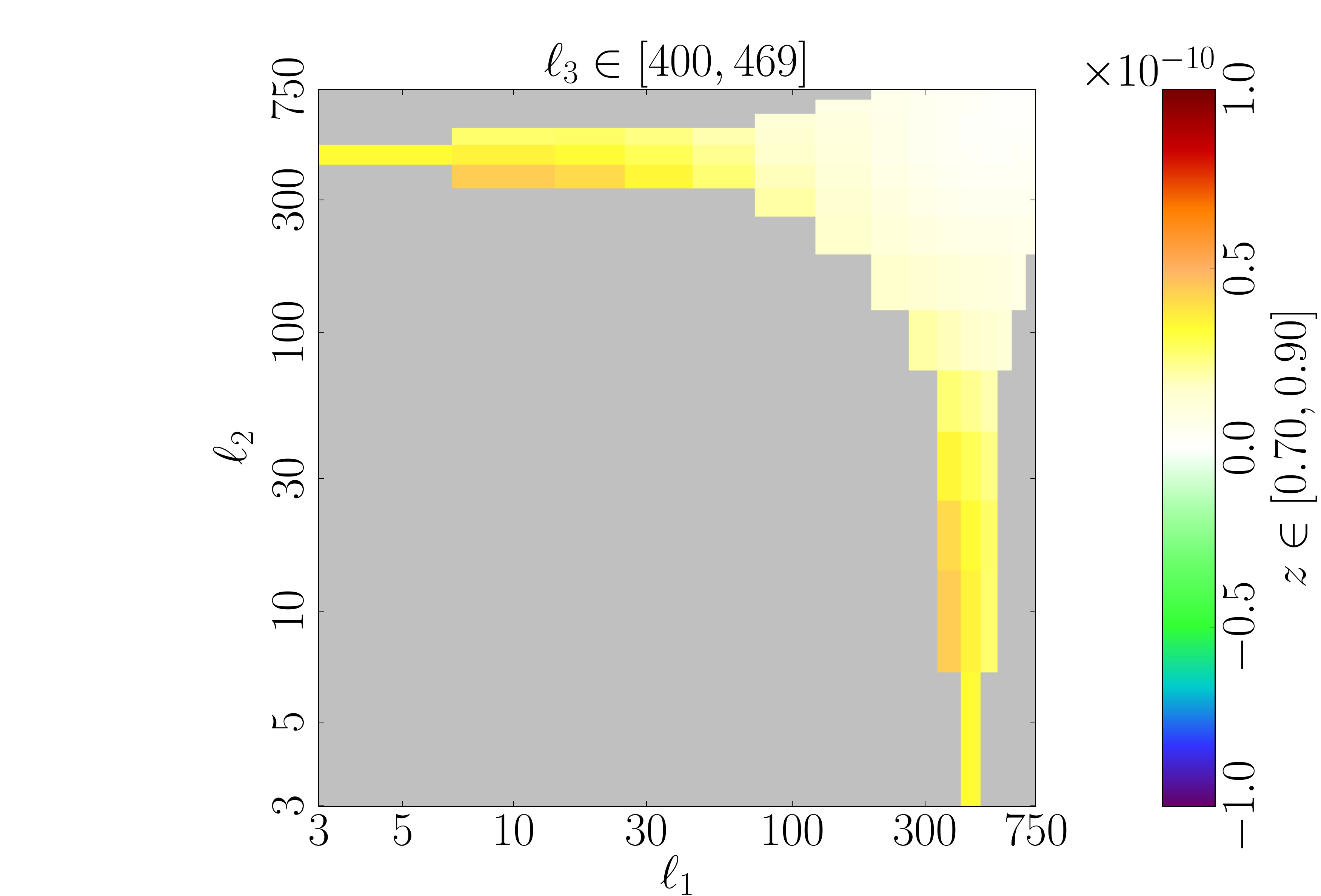}\vspace{-0.1cm}\\
      \includegraphics[width=50mm, trim={4cm 1.1cm 6.5cm 1cm},clip]{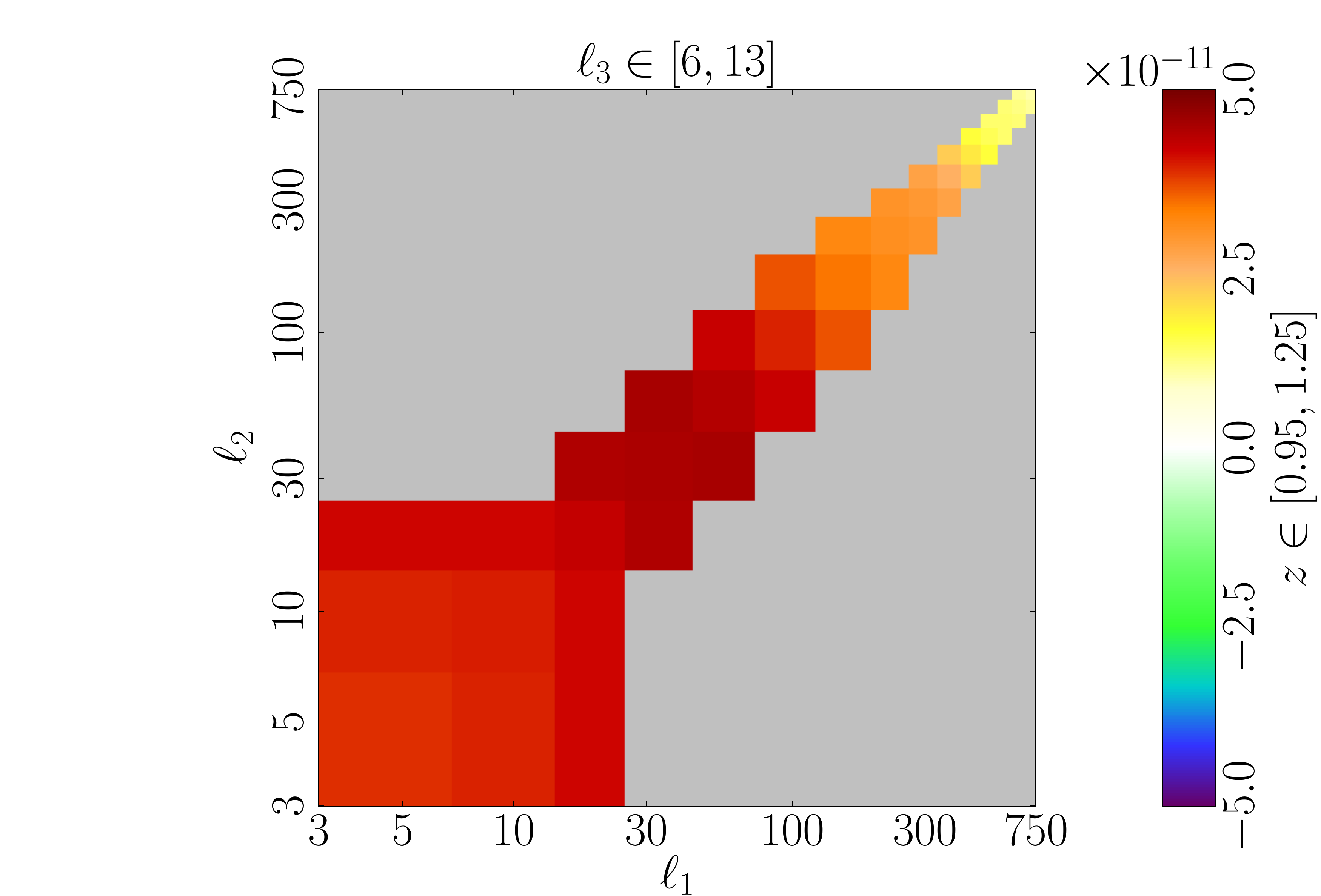} &
      \hspace{-0.45cm}\includegraphics[width=44mm, trim={7cm 1.1cm 6.5cm 1cm},clip]{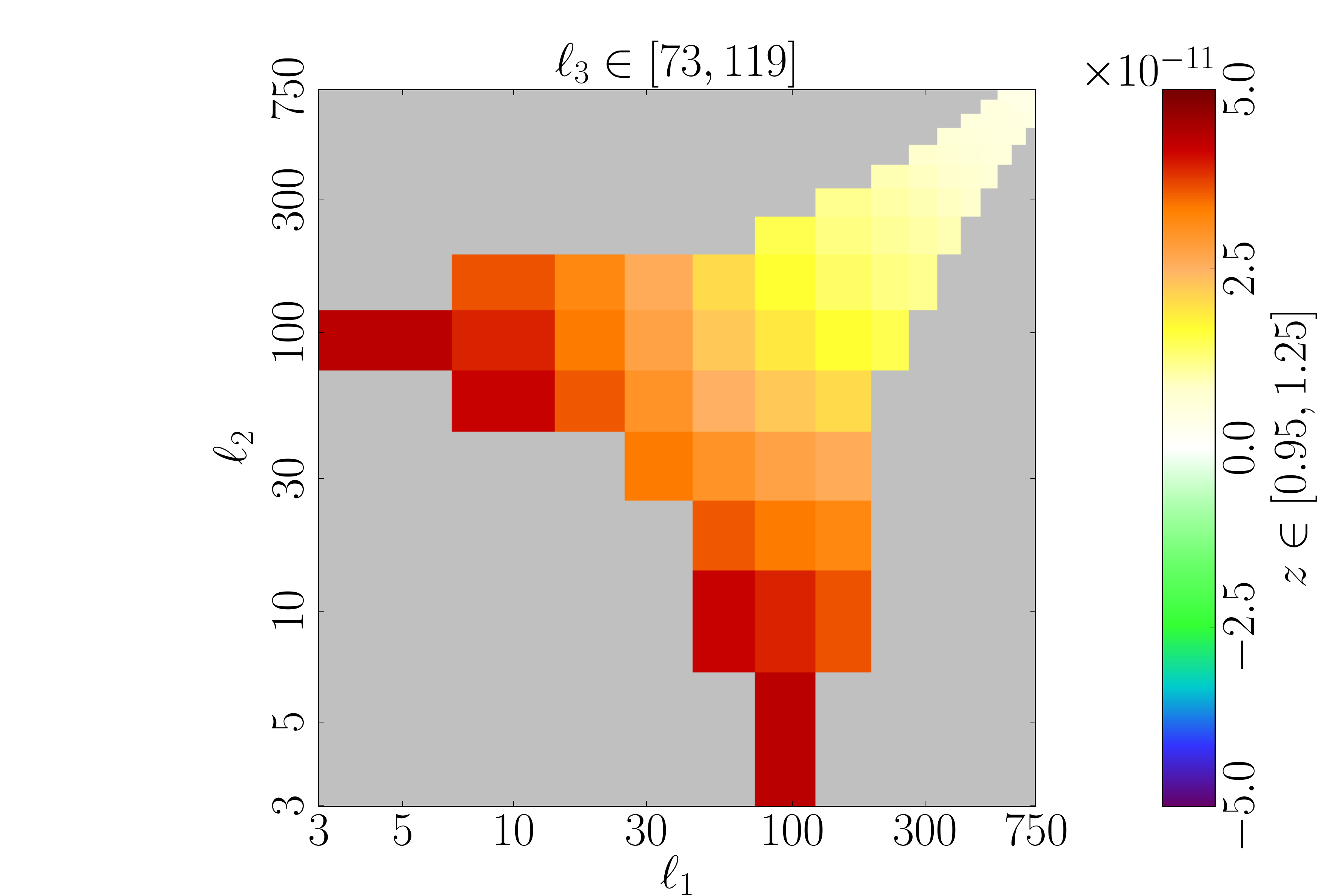} &
      \hspace{-0.45cm}\includegraphics[width=57mm, trim={7cm 1.1cm 0cm 1cm},clip]{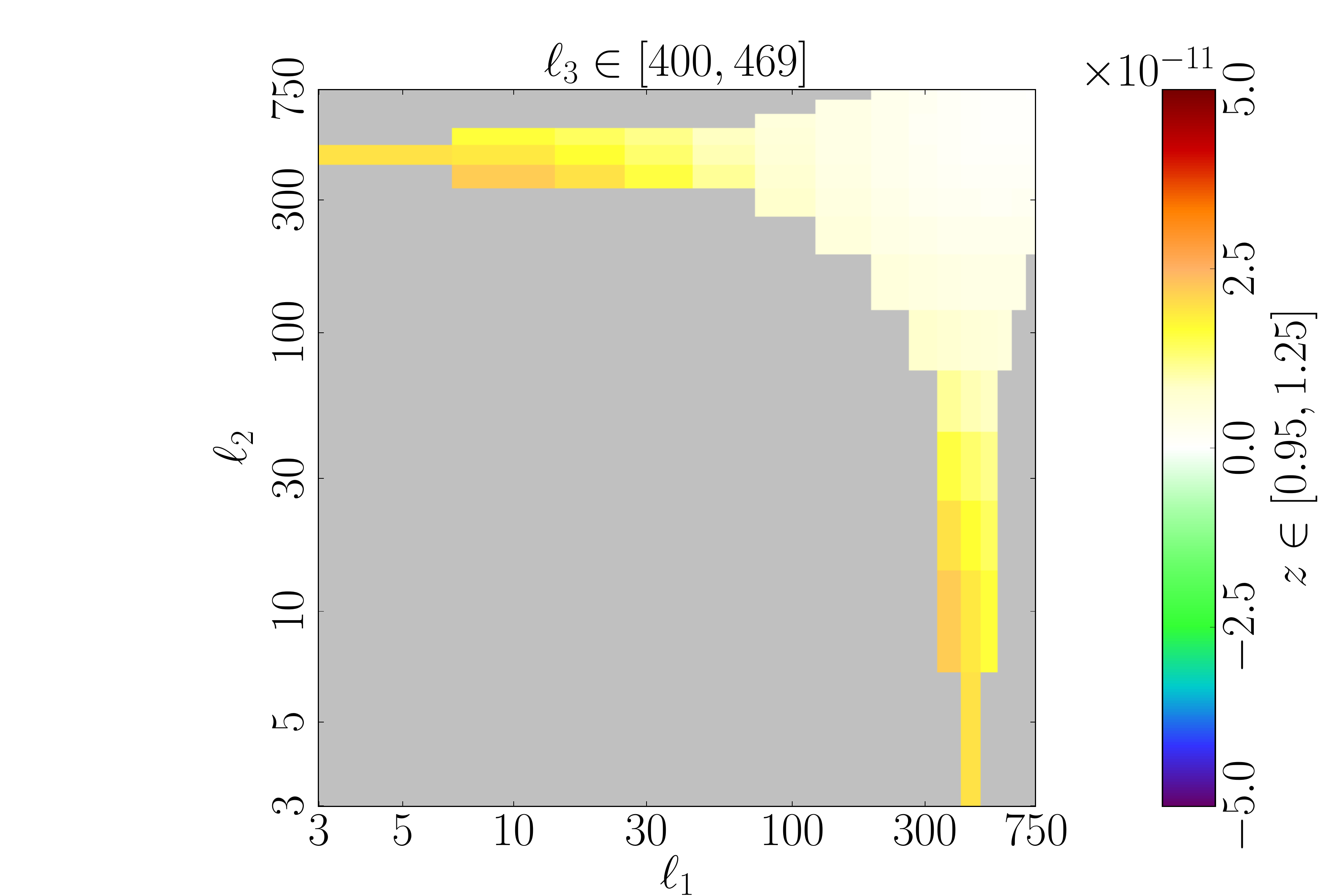}\vspace{-0.1cm}\\
    \includegraphics[width=50mm, trim={4cm 1.1cm 6.5cm 1cm},clip]{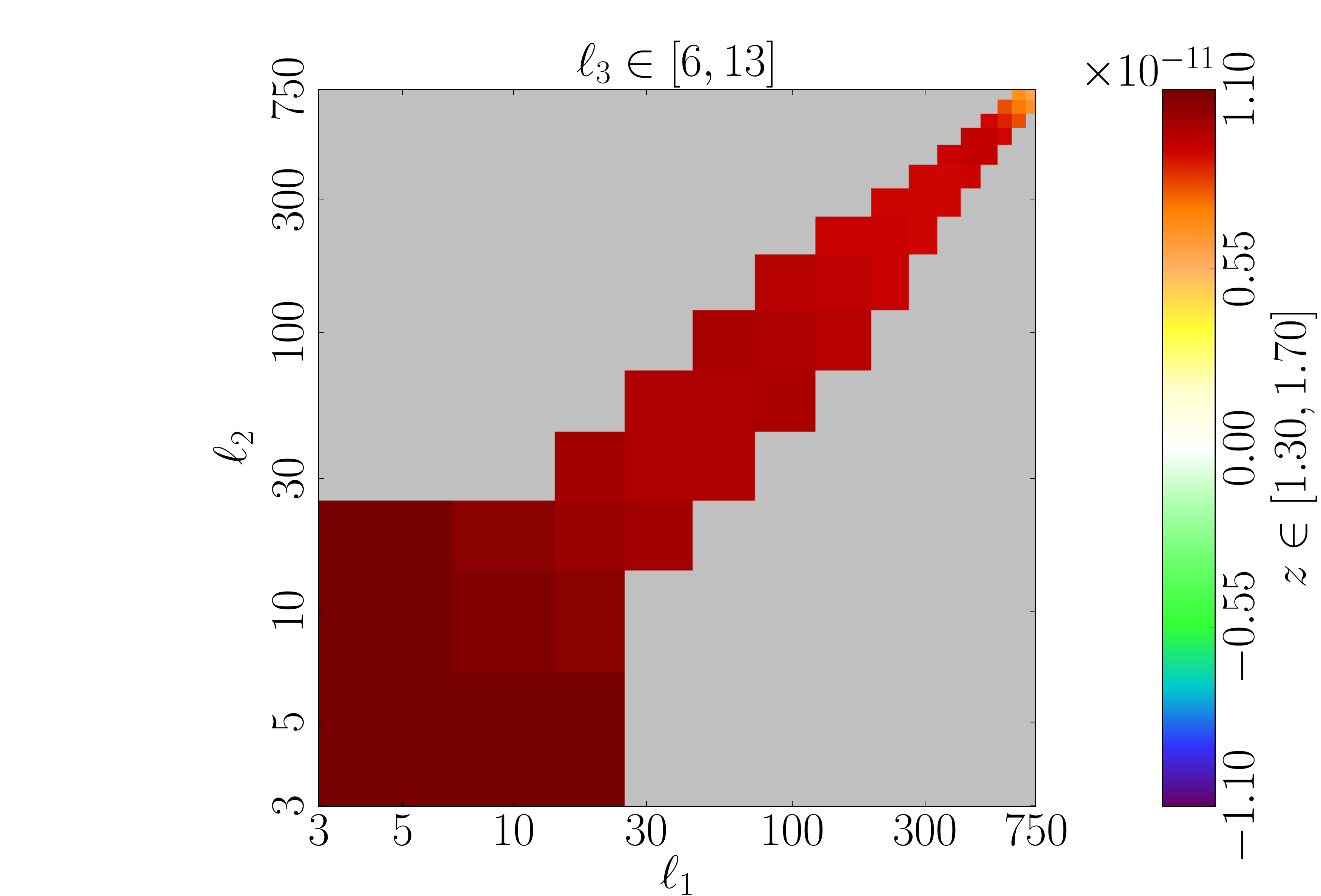} &
      \hspace{-0.45cm}\includegraphics[width=44mm, trim={7cm 1.1cm 6.5cm 1cm},clip]{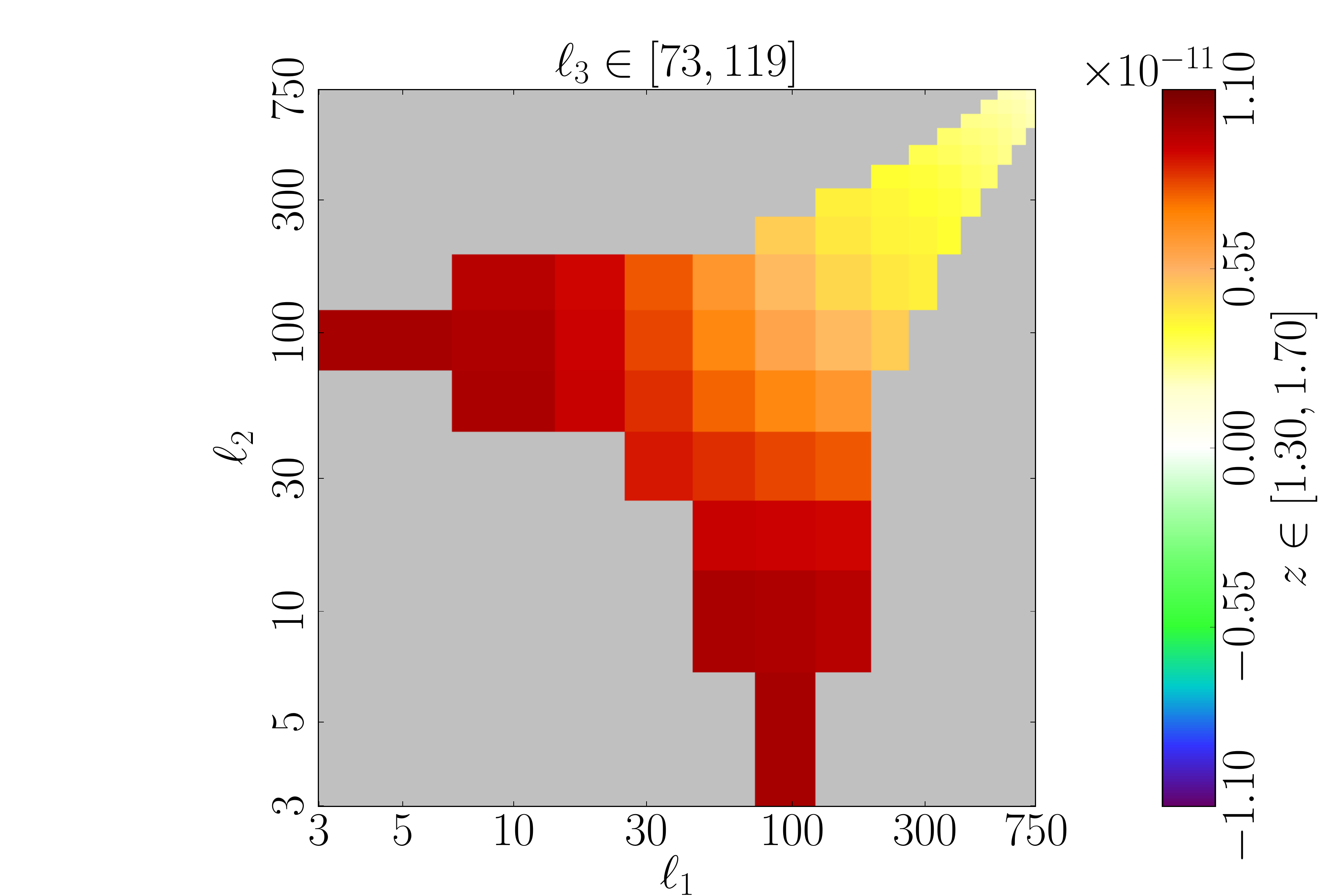} &
      \hspace{-0.45cm}\includegraphics[width=57mm, trim={7cm 1.1cm 0cm 1cm},clip]{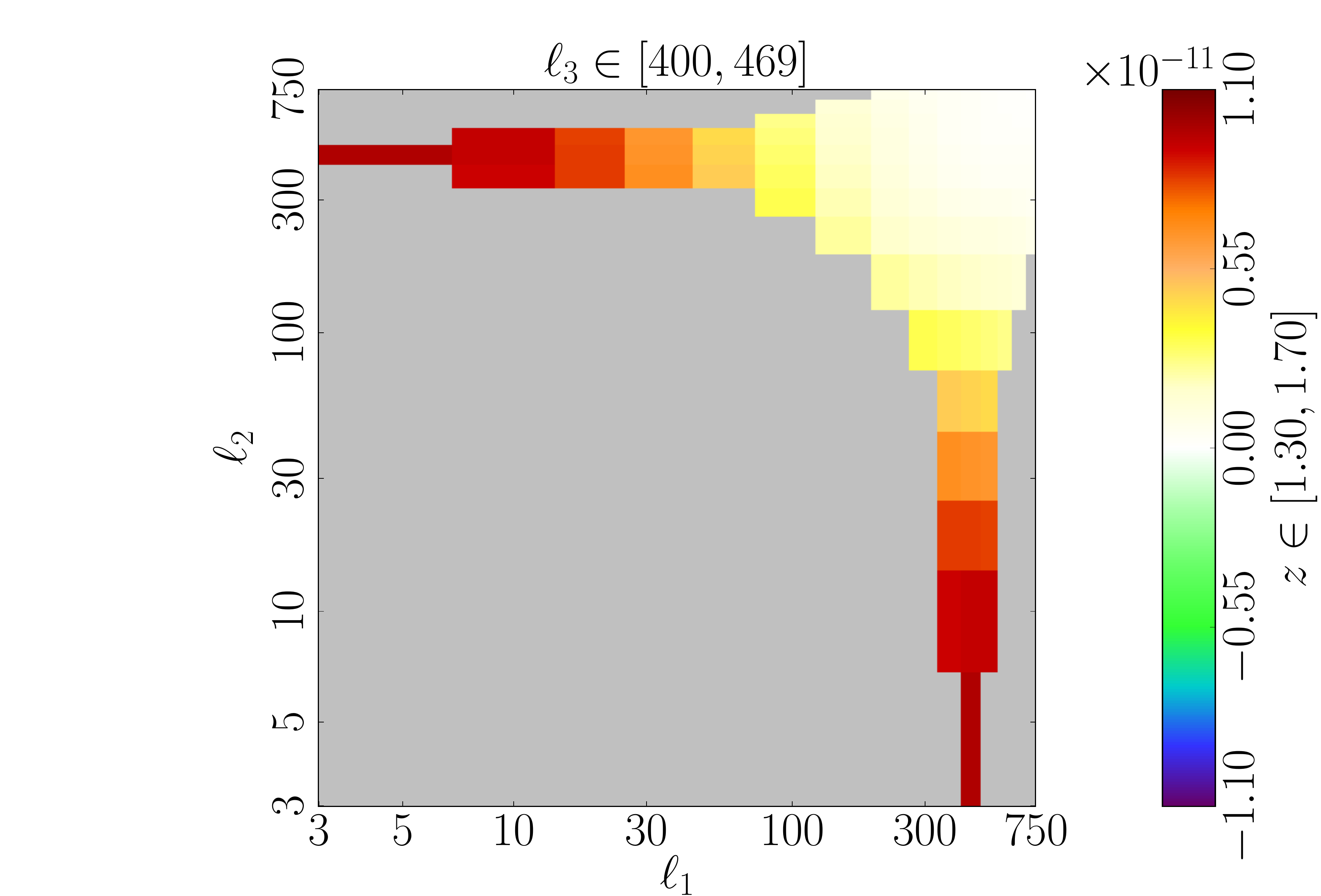}\\
      %   \includegraphics[width=50mm, trim={4cm 0 6.5cm 1cm},clip]{binned_bispec_z4_RpRm-NpNm_logaxis_T+E_1024_1024_2_2000_2_2000_log2linbinsv3_dt10_0_nomask__7_01_sml2.0.pdf} &
      % \hspace{-0.45cm}\includegraphics[width=44mm, trim={7cm 0 6.5cm 1cm},clip]{binned_bispec_z4_RpRm-NpNm_logaxis_T+E_1024_1024_2_2000_2_2000_log2linbinsv3_dt10_0_nomask__7_05_sml2.0.pdf} &
      % \hspace{-0.45cm}\includegraphics[width=57mm, trim={7cm 0 0cm 1cm},clip]{binned_bispec_z4_RpRm-NpNm_logaxis_T+E_1024_1024_2_2000_2_2000_log2linbinsv3_dt10_0_nomask__7_10_sml2.0.pdf}\\
\end{tabular}
    \caption{Similar to Fig.~\ref{fig:rRmN}, but for all intermediate redshift bins. Note the different colour scales used on each row.
    }
    \label{fig:Rela_bisp_annex2}
\end{figure}

Finally, the divergence of the momentum constraint gives us the second-order momentum 
\begin{multline}
\label{eq:Tvelocitydivergence}
     T^{(2)}_{Q}(k_1,k_2,k) =   \frac{2}{3\mathcal{H}^2 \Omega_m} \left(\frac{1}{2}\mathcal{H} T^{(1)}_{\psi\psi} + \mathcal{H} T^{(2)}_\psi +  T_{\phi'}^{(2)}\right) \\ + \frac{1}{2} \frac{k_1k_2}{k^2} 
     \left( \left[\frac{k_1}{k_2}+\mu \right] T^{(1)}_{v\delta}+ \left[\frac{k_2}{k_1}+\mu \right] T^{(1)}_{\delta v} \right)\,.
\end{multline}

\section{Bispectra for all redshift bins}\label{app:bispectre}
We show in Fig.~\ref{fig:Newtonian_bisp_annex} the smoothed Newtonian bispectrum for all intermediate redshift bins, see Fig.~\ref{fig:Newtonian_bisp} for the highest and lowest redshift bins. It confirms that the Newtonian bispectrum has most power in the squeezed and folded configurations. Similarly, in Fig.~\ref{fig:Rela_bisp_annex2} we show the purely relativistic bispectrum for all intermediate redshift bins (see Fig.~\ref{fig:rRmN} for the highest and lowest redshifts bins), confirming that it peaks in the squeezed limit.

\bibliographystyle{JHEP.bst}
\bibliography{bib.bib}

\end{document}